\documentclass{aa}
\usepackage{graphicx}
\usepackage{txfonts}
\usepackage{hyperref}
\hypersetup{colorlinks = true, citecolor = {blue}, urlcolor= {blue}}

\newcommand{\msun}{M_{\odot}}

\def\PGPU{$\varphi-$GPU }

\def\gapprox{\;\rlap{\lower 3.0pt                       
        \hbox{$\sim$}}\raise 2.5pt\hbox{$>$}\;}
\def\lapprox{\;\rlap{\lower 3.1pt                       
        \hbox{$\sim$}}\raise 2.7pt\hbox{$<$}\;}

\newcommand{\be}{ \begin{equation} }
\newcommand{\ee}{\end{equation}}

\newcommand{\ben}{\begin{enumerate}}
\newcommand{\een}{\end{enumerate}}

\renewenvironment{thebibliography}[1]{\begin{oldthebibliography}{#1}\setlength{\parskip}{0ex}\setlength{\itemsep}{0ex}}{\end{oldthebibliography}}

\defcitealias{Ishchenko2023a}{Paper~I}
\defcitealias{Ishchenko2023b}{Paper~II}
\defcitealias{Ishchenko2024}{Paper~III}

\usepackage{diagbox}
\usepackage{siunitx}
\newcommand{\orcid}[1]{\href{https://orcid.org/#1}{\protect\includegraphics[width=8pt]{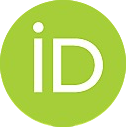}}}
\makeatletter
\renewcommand*\aa@pageof{, page \thepage{} of \pageref*{LastPage}}
\makeatother

\begin{document}

\title{Evolution of the disky second generation of stars in globular clusters on cosmological timescale}

\author{Peter~Berczik
\inst{1,2,3,4}\orcid{0000-0003-4176-152X}
\and
Taras~Panamarev
\inst{5,4}\orcid{0000-0002-1090-4463}
\and
Maryna~Ishchenko
\inst{1,2,4}\orcid{0000-0002-6961-8170}
\and
Bence~Kocsis
\inst{5}
}

\institute{
Nicolaus Copernicus Astronomical Centre Polish Academy of Sciences, ul. Bartycka 18, 00-716 Warsaw, Poland
  \and
Main Astronomical Observatory, National Academy of Sciences of Ukraine, 27 Akademika Zabolotnoho St, 03143 Kyiv, Ukraine  \email{\href{mailto:berczik@mao.kiev.ua}{berczik@mao.kiev.ua}}
  \and
Konkoly Observatory, Research Centre for Astronomy and Earth Sciences, HUN-REN CSFK, MTA Centre of Excellence, Konkoly Thege Mikl\'os \'ut 15-17, 1121 Budapest, Hungary
  \and
Fesenkov Astrophysical Institute, Observatory 23, 050020 Almaty, Kazakhstan
  \and
Rudolf Peierls Centre for Theoretical Physics, Parks Road, OX1 3PU, Oxford, UK
  }
   
\date{Received xxx / Accepted xxx}


\abstract
{Many Milky Way globular clusters (GCs) host multiple stellar populations, challenging the traditional view of GCs as single-population systems. It has been suggested that second-generation stars could form in a disk from gas lost by first-generation stars or from external accreted gas. Understanding how these multiple stellar populations evolve under a time-varying Galactic tidal field is crucial for studying internal mixing, the rotational properties, and mass loss of GCs over cosmological timescales.}
{We investigate how the introduction of a second stellar generation affects mass loss, internal mixing, and rotational properties of GCs in a time-varying Galactic tidal field and different orbital configurations.}
{We conducted direct $N$-body simulations of GCs on three types of orbits derived from the observed Milky Way GCs using state-of-the-art stellar evolution prescriptions. We evolved the clusters for 8 Gyr in the time-varying Galactic potential of the IllustrisTNG-100 cosmological simulation. After 2 Gyr, we introduced a second stellar generation, comprising 5\% of the initial mass of the first generation, as a flattened disk of stars. For comparison, we ran control simulations using a static Galactic potential and isolated clusters.}
{We present the mass loss, structural evolution, and kinematic properties of GCs with two stellar generations, focusing on tidal mass, half-mass radii, velocity distributions, and angular momentum. We also examine the transition of the second generation from a flattened disk to a spherical shape.}
{Our results show that the mass loss of GCs depends primarily on their orbital parameters, with tighter orbits leading to higher mass loss. The growth of the Galaxy led to tighter orbits, implying that the GCs lost much less mass than if the Galaxy had always had its current mass. The initially flattened second-generation disk became nearly spherical within one relaxation time. However, whether its distinct rotational signature was retained depends on the orbit: for the long radial orbit, it vanished quickly; for the tube orbit, it lasted several Gyr; but for the circular orbit, rotation persisted until the present day.}

\keywords{Galaxy: globular clusters: general - Galaxy: center - Methods: numerical}

\titlerunning{Evolution of the second generation in the globulat cluster}
\authorrunning{P.~Berczik et al.}
\maketitle

\section{Introduction}\label{sec:Intr}

According to the standard $\Lambda$CDM model, the Milky Way (MW) globular clusters (GCs) are among the first bound stellar systems that formed in the early Universe, with typical ages of about 10-12~Gyr \citep{Weinberg1993}. Originally, globular clusters were described as spherical systems formed during a single star formation episode. However, over the last two decades, it has become clear that most globular clusters show evidence of multiple stellar populations \citep[and references therein]{Bastian_Lardo2018,Milone_Marino2022}, and at least 20-30 of the globular clusters in the Milky Way exhibit internal rotation \citep{Bellazzini2012, Bianchini2013, Fabricius2014, Kacharov2014, Kimmig2015, Lardo2015, Boberg2017, Jeffreson2017, Ferraro2018, Kamann2018, Lanzoni2018,Bianchini2018,Sollima2019,Cordoni+2020,Vasiliev_Baumgardt2021,Petralia2024}. Modelling the evolution of multiple stellar populations over cosmic time is crucial for understanding globular cluster formation, as well as the observed anomalous element distributions and internal anisotropic structures \citep[e.g.][]{Henault-Brunet2015,Cordero+2017,Cordoni+2020,Scalco+2023,Martens+2023,Libralato+2023,McKenzie_Bekki2021,Livernois+2021,Tiongco+2021,Mastrobuono-Battisti_Perets2021,Szigeti+2021,Vesperini+2021,Lacchin+2022,Cordoni+2023}.

The origin of the observed multiple population is not clear \citep[see e.g.][]{Renzini+2015,Bastian_Lardo2018,Gratton+2019}. Second generation ({\tt g2}) stars could have formed from gas lost from earlier first generation ({\tt g1}) stars and/or from external accreted gas. It is generally expected that ({\tt g2}) stars form from gas which collapsed into a cooling flow possibly forming a disk \citep{D'Ercole+2008,Calura+2019,McKenzie_Bekki2021,Lacchin+2022}. 
In particular, \citet{Bekki2010} demonstrated that {\tt g2} stars, formed from asymptotic giant branch (AGB) ejecta of  {\tt g1} stars, could develop significant rotation with a rotation rate depending on the initial kinematics of the {\tt g1} stars. More recently, \citet{Lacchin+2022} found that under specific conditions — such as low external gas density — {\tt g2} stars can form a rapidly rotating disk-like structure, often with higher rotational speeds than the {\tt g1} stars. Additionally, \citet{McKenzie_Bekki2021} highlighted the role of pristine gas accretion from the host galaxy in enhancing the rotational characteristics of {\tt g2} stars.
The present day rotation is likely a remnant of a higher degree of earlier rotation \citep{Mapelli+2017,Bianchini2018,Kamann2018,Sollima2019,Tiongco+2021,Livernois+2021}.  

Stronger evidence for star formation in disks comes from nuclear star clusters (NSCs), where young stellar generations can be observed today \citep{LevinBeloborodov2003, PaumardEtAl2006, Neumayer2020}. In NSCs, stars may form due to fragmentation of a gaseous accretion disk \citep{LevinBeloborodov2003, Levin2007}, the collapse of a giant molecular cloud \citep{Bonnell2008, Generozov2022}, or through the capture of stars by the gaseous accretion disk \citep{Panamarev2018}.

Simulations have shown that the initial differences between different subpopulations decrease over time in GCs due to spatial and kinematic mixing on the two-body relaxation timescale \citep{Mastrobuono-Battisti_Perets2013,Vesperini+2013,Henault-Brunet2015,Mastrobuono-Battisti_Perets2016,Meiron_Kocsis2018,Tiongco+2019,Vesperini+2021}. Additionally, mixing may occur more rapidly in angular momentum space due to vector resonant relaxation \citep{Rauch_Tremaine1996,Kocsis_Tremaine2015,Meiron_Kocsis2019}, leading to anisotropic mass segregation, where more massive objects produce a larger anisotropy \citep{Szolgyen+2019,Tiongco+2019,Tiongco+2021,Livernois+2021,Livernois+2022,Wang_Kocsis2023}. Similarly, numerical studies of such stellar disks in galactic nuclei indicate that mixing in angular momentum space occurs on vector resonant relaxation time scales, which are of the order of 10 Myrs for the Milky Way NSC \citep{Kocsis_Tremaine2015,Panamarev2022}. However, full mixing occurs on two-body relaxation time scales, which are much longer (see e.g. \citealt{Meiron_Kocsis2018,Panamarev2019}). As a result, due to the conservation of angular momentum, a fully relaxed system will exhibit net rotation.
 
To study the dynamical evolution of multiple stellar populations in different environments, such as globular clusters, it is essential to perform realistic $N$-body simulations that incorporate stellar evolution and the effects of a realistic galactic potential. As previously mentioned, globular clusters have been extensively studied over their relaxation time scales. Some of the most realistic simulations, which include both single and binary stellar evolution, are the DRAGON simulations \citep{Wang2016}. These simulations were recently enhanced to incorporate up-to-date stellar evolution prescriptions \citep{dragon2a2024,dragon2b2023,dragon2c2024}. However, these models assumed an initially spherically symmetric and isotropic distribution of stars, considered only a single generation of stars, and modelled the clusters on circular orbits within a static galactic potential. 

The main idea of this work is to carry out direct N-body simulations of globular clusters with two stellar components within a time-varying Galactic tidal field, exploring different orbital configurations. We begin with an initially spherically symmetric and isotropic stellar distribution and then introduce a second, initially geometrically flattened stellar generation. Using state-of-the-art stellar evolution prescriptions, we study the internal mixing of these systems and assess the impact of the external potential.

The paper is organised as follows. In Section \ref{sec:met} we describe the initial conditions and data for the first and second stellar generations. We also present the description of the integration procedure in external Galactic potentials. In Section \ref{sec:gen-g1-g2} we present the dynamical evolution of the mass loss for single and multiple generation systems. In Sections \ref{sec:gen-g2} we focus on the second generation mainly in a context of shape and angular momentum evolution of the generations. In Section \ref{sec:conc}, we summarise our results and outcomes.

\section{Initial conditions and models}\label{sec:met}

This work builds on our previous studies, where we analysed the orbits of 159 Galactic GCs using \textit{Gaia} Data Release 3, integrated them backward in time to reconstruct their orbital evolution on cosmological timescales 
(\citealt{Ishchenko2023a}, hereafter \citetalias{Ishchenko2023a}), examined the interaction rates of the GCs with the Galactic centre 
(\citealt{Ishchenko2023b}, hereafter \citetalias{Ishchenko2023b}),
and investigated their mass loss due to the tidal field in the central region of the Galaxy 
(\citealt{Ishchenko2024}, hereafter \citetalias{Ishchenko2024}). In this study, we select three different orbits from those obtained in \citetalias{Ishchenko2023a} and initialise a GC on each orbit in time variable external potential at a lookback time of 8 Gyr. After 2 Gyr of evolution, i.e. at $T$ = -6 Gyr, we introduce a disk of stars representing the second stellar generation. We then evolve the system to the present day. For comparison, we ran the same simulations of the same GCs within a static external potential and one control simulation of an isolated cluster. In the following, we describe the details of this procedure.

\subsection{Galactic potentials and integration procedure}\label{sybsec:integr-TNG}

To add more realistic dynamical orbital evolution into our simulations, we extended the standard $N$-body code (described below) to incorporate time-dependent external potentials derived from the IllustrisTNG-100 cosmological simulation database \citep{Nelson2018, Nelson2019, NelsonPill2019}. From this database, we selected a time-variable potential TNG, denoted as {\tt 411321}, which closely matches that of the MW at redshift $z = 0$ in terms of halo and disk masses, as well as their characteristic scales. 
The sampling and fitting procedures for this potential, as well as other MW-like potentials are discussed in detail in \citetalias{Ishchenko2023a} and \citet{Mardini2020}. The key parameters for this potential are summarised in Table~\ref{tab:pot}. The code routines for sampling and fitting the selected potentials are also publicly available on GitHub\footnote{The ORIENT: \\~\url{ https://github.com/Mohammad-Mardini/The-ORIENT}}.

In addition to the time-variable potential, we performed comparative simulations using a static potential, known as {\tt FIX}. This potential has constant parameters throughout the simulation and using the three-component (bulge-disk-halo) axisymmetric Plummer-Kuzmin model \citep{Miyamoto1975}. The parameters for both the static potential {\tt FIX} and the TNG are provided in Table~\ref{tab:pot}.

\begin{table}[htbp!]
\caption{Parameters of the external potentials at redshift zero.} 
\centering
\begin{tabular}{llrr}
\hline
\hline
\multicolumn{1}{c}{Parameter} & Unit & {\tt 411321} & {\tt FIX} \\
\hline
\hline
Bulge mass, $M_{\rm d}$         & $10^{10}~\rm M_{\odot}$ & -- & 1.4  \\
Disk mass,  $M_{\rm d}$         & $10^{10}~\rm M_{\odot}$ & 7.110 & 9.0  \\
Halo mass,  $M_{\rm h}$         & $10^{12}~\rm M_{\odot}$ & 1.190 & 0.72  \\

Bulge scale length, $a_{\rm d}$ & 1~kpc                     & -- & 0.0  \\
Bulge scale height, $b_{\rm d}$ & 1~kpc                     & -- & 0.3 \\

Disk scale length, $a_{\rm d}$ & 1~kpc                     & 2.073 & 3.3  \\
Disk scale height, $b_{\rm d}$ & 1~kpc                     & 1.126 & 0.3  \\
Halo scale height, $b_{\rm h}$ & 10~kpc                    & 2.848 & 2.5  \\
\hline
\end{tabular}
\label{tab:pot}
\end{table} 

For the dynamical evolution of GCs, we used the high-order parallel dynamical $N$-body code \PGPU\footnote{$N$-body code \PGPU: \\~\url{ https://github.com/berczik/phi-GPU-mole}}. This code is based on the fourth-order Hermite integration scheme with hierarchical individual block time steps \citep{Berczik2011,BSW2013}. The current version of the code also incorporates the recently updated stellar evolution models \citep{Banerjee2020}. The most important updates were made in stellar evolution: metallicity dependent stellar winds;  metallicity dependent core-collapse supernovae, new fallback prescription and their remnant masses;  electron-capture supernovae accretion, induced collapse and merger-induced collapse; remnant masses and natal kicks; BH natal spins. For more details on the current stellar evolution library, please refer to \citet{Kamlah2022, Kamlah2022MNRAS}.

We apply a Plummer softening parameter
for inner spatial resolution to $\epsilon=0.01$~pc. Additionally, we ignore the effects of dynamical friction, which, as shown by \citet{Just2011}, significantly affects only clusters very close to the Galactic centre\footnote{The possible influence of dynamical friction on clusters with close approaches to the Galactic centre is analysed in Sec.~4 of \citetalias{Ishchenko2024}}.

\subsection{First stellar generation}\label{subsec:fr-stel-pop}

As a basis for our current study, we use the set of reconstructed orbits for real MW GCs obtained in  \citetalias{Ishchenko2023a}. In that study, we integrated orbits up to 10 Gyr lookback time for the 159 GCs with initial positions, proper motions, radial velocities, and geocentric distances from \citet{Baumgardt2021} which is based on the \textit{Gaia} DR3 catalogue. Visualisation for all simulated GCs can also be found online\footnote{Visualisation for the 159 GCs orbits: \url{https://sites.google.com/view/mw-type-sub-halos-from-illustr/GC-TNG?authuser=0}}. 

\begin{figure*}[ht]
\centering
\includegraphics[width=0.90\linewidth]{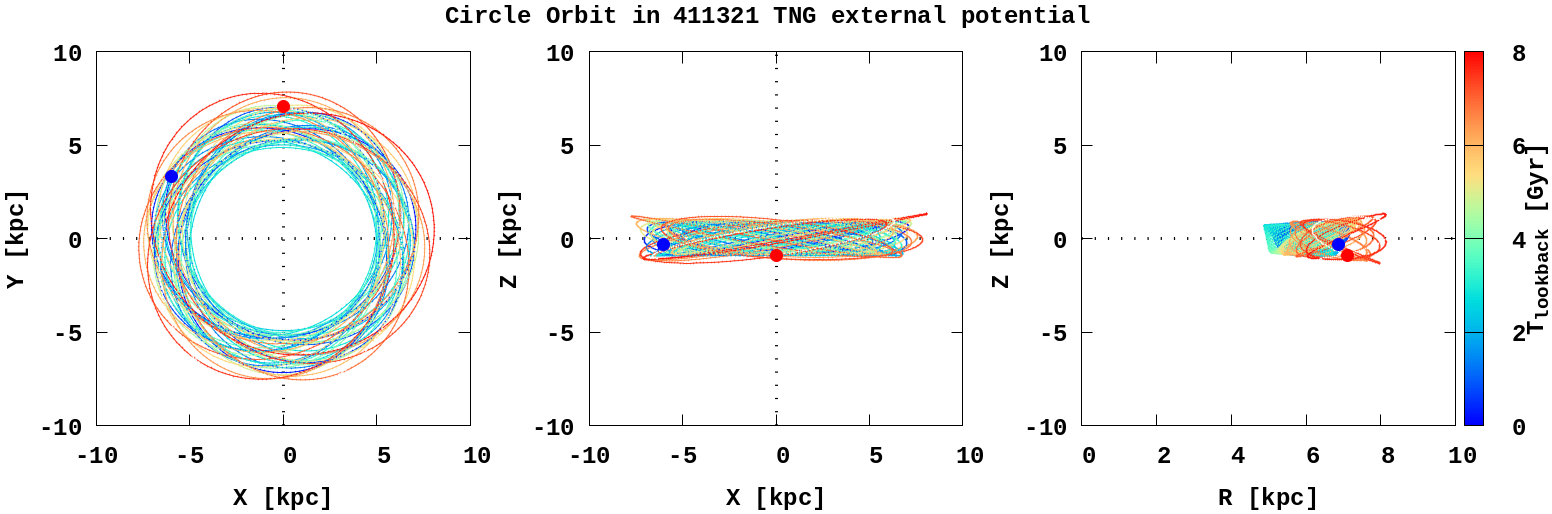}
\includegraphics[width=0.90\linewidth]{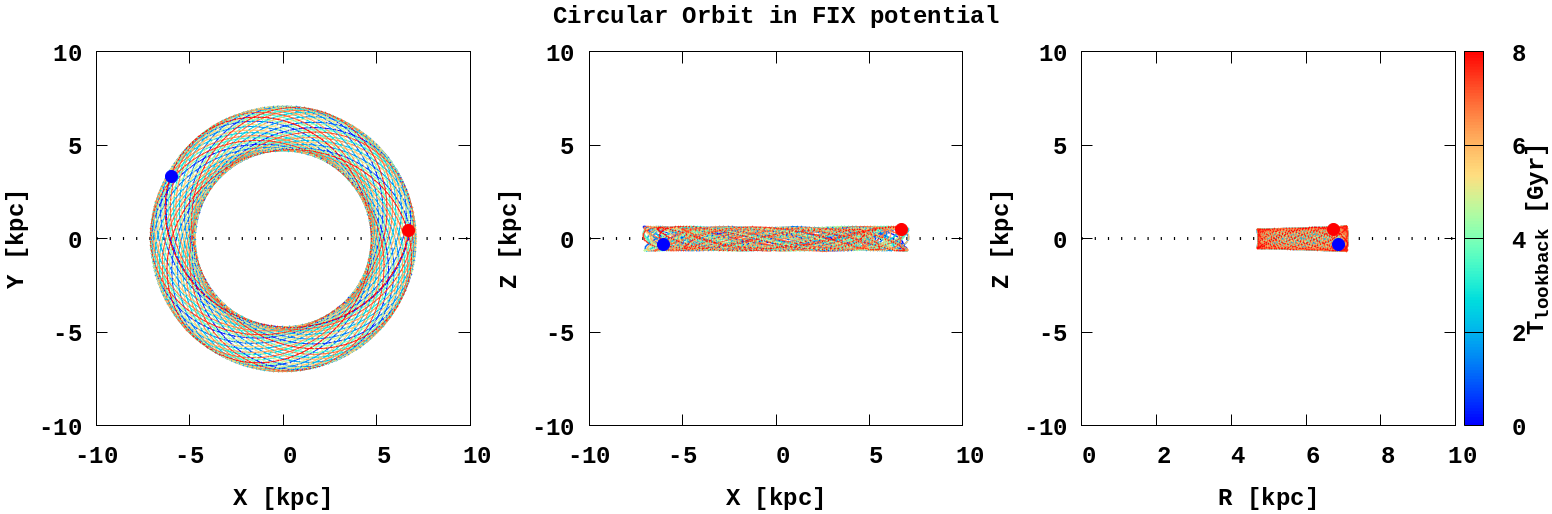}
\caption{Circular orbit in {\tt 411321} TNG (top row) and {\tt FIX} external potentials (bottom row). Time represents by colour palette. Red dots represent the GC's position up to 8 Gyr in lookback time, blue - position for today.}
\label{fig:orb4}
\end{figure*}

For the multi-generation GC models, we examine the evolution of GCs on three types of orbit families in the host galaxy: a circular (CR) orbit, a tube (TB) orbit, and a long radial (LR) orbit. As prototypes for these type of MW GC orbits we selected three particular clusters: NGC6838, NGC6401, and NGC1904 on CR, TB, and LR orbits, respectively. In \citetalias{Ishchenko2023a} the clusters were integrated in the TNG potential, however, for this work we carried out the orbital integration also in {\tt FIX} static potential for these GCs. Eventually, we have two numerical models for each of these three clusters in both potentials. As an illustration, Fig.~\ref{fig:orb4} shows the orbital visualisation for circular orbit in TNG and {\tt FIX} potentials over the span of 8 Gyr up to the present time. Orbital visualisation for tube and long radial orbits in both potentials can be found in Appendix~\ref{app:det-orb} in Figures~\ref{fig:tube-orb}~and~\ref{fig:lr-orb}. 

To carry out the simulations, we assume the following initial conditions. The first generation stars in each cluster are initially set in a state of dynamical equilibrium with individual positions and velocity vectors, which were obtained from the backward orbital integration at $T = -8\,$Gyr, see Table \ref{tab:init-param}. 

For the first stellar generation, we adopt the initial mass for the total mass M = 1.5$\times 10^{5}\rm\;M_{\odot}$, Kroupa initial mass function (IMF) \citep{Kroupa2001} in the range 0.08--100 $\msun$. The total number of particles is $N$ = 261 384. We use the distribution function of the most concentrated King model \citep{King1966} with the dimensionless central potential $W_{0}$ = 9.0. The initial half-mass radius $r_{\rm hm}$ is assumed to be 2 pc. Because we use a possible lower limit for our GCs initial masses, we compensate this using a quite small initial half mass radius for the GC first generation. With our current parameters the model GC has a average mass density around $\sim$19{\it k} M$_\odot$/pc$^3$. Such a density is quite compatible with our earlier GC numerical simulations 
in \citetalias[][see Table 2]{Ishchenko2024}. Also we use the classical Solar value for the initial stellar metallicity: $Z = Z_\odot = 0.02$ \citep{Grevesse1998}. These conditions are the same for all our GC models, see Table~\ref{tab:init-param}. The first stellar generation is labelled as {\tt g1}. 

\begin{table}[tbp]
\setlength{\tabcolsep}{4pt}
\centering
\caption{Initial kinematical and physical characteristics of GCs at $T = -8\,\mathrm{Gyr}$.}
\label{tab:init-param}
\begin{tabular}{llcccccc}
\hline
\hline 
Pot. & Orbit & $X$ & $Y$ & $Z$ & $V_x$ & $V_y$ & $V_z$ \\
 & & pc & pc & pc & km~s$^{-1}$ & km~s$^{-1}$ & km~s$^{-1}$ \\
\hline
\hline
TNG   & CR & 4 & 7108 & -871 & 189 & -39 & 28  \\
      & TB & -1493 & -3213 & 398 & 5 & -3 & -57 \\
      & LR & 3597 & 13451 & -3767 & -43 & -167 & 23  \\
 FIX  & CR & 6725 & 413 & 498 & -30 & -201 & 25 \\
      & TB & 566 & 265 & -602 & 170 & 139 & -104 \\
      & LR & -16851 & -4979 & 7933 & 23 & 8 & 12 \\
 ISO  & -- & 0 & 0 & 0 & 0 & 0 & 0 \\
\hline 
\end{tabular}
\tablefoot{Initial conditions (King models with $W_0$ = 9.0) for our theoretical GC models: $M = 1.5 \times 10^{5}\,$M$_{\odot}$ with corresponding particle numbers $N$ = 261 384 and half mass radius $r_{\rm hm} = 2.0\,$pc.}
\vspace{6pt}
\end{table}
 
\begin{figure*}[ht]
\centering
\includegraphics[width=0.33\linewidth]{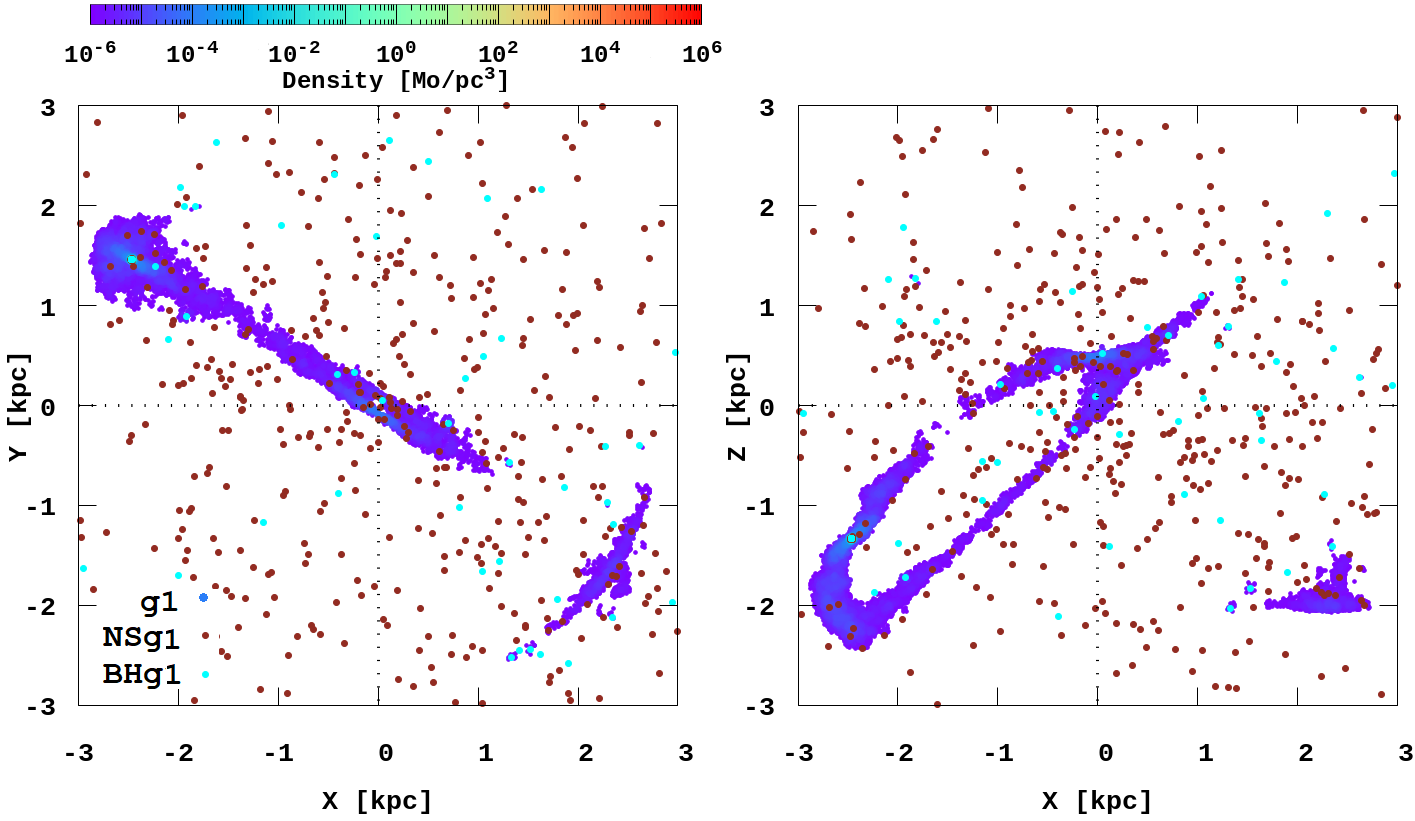}
\includegraphics[width=0.33\linewidth]{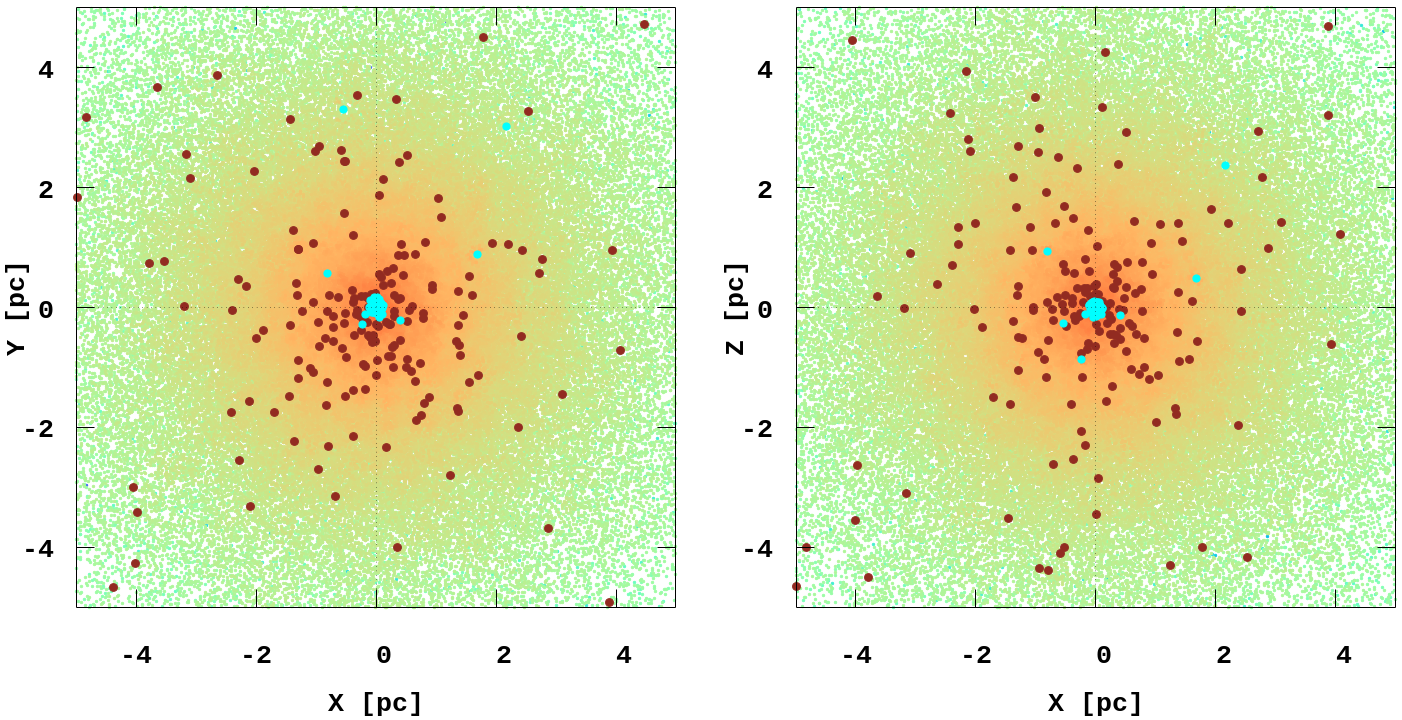}
\includegraphics[width=0.33\linewidth]{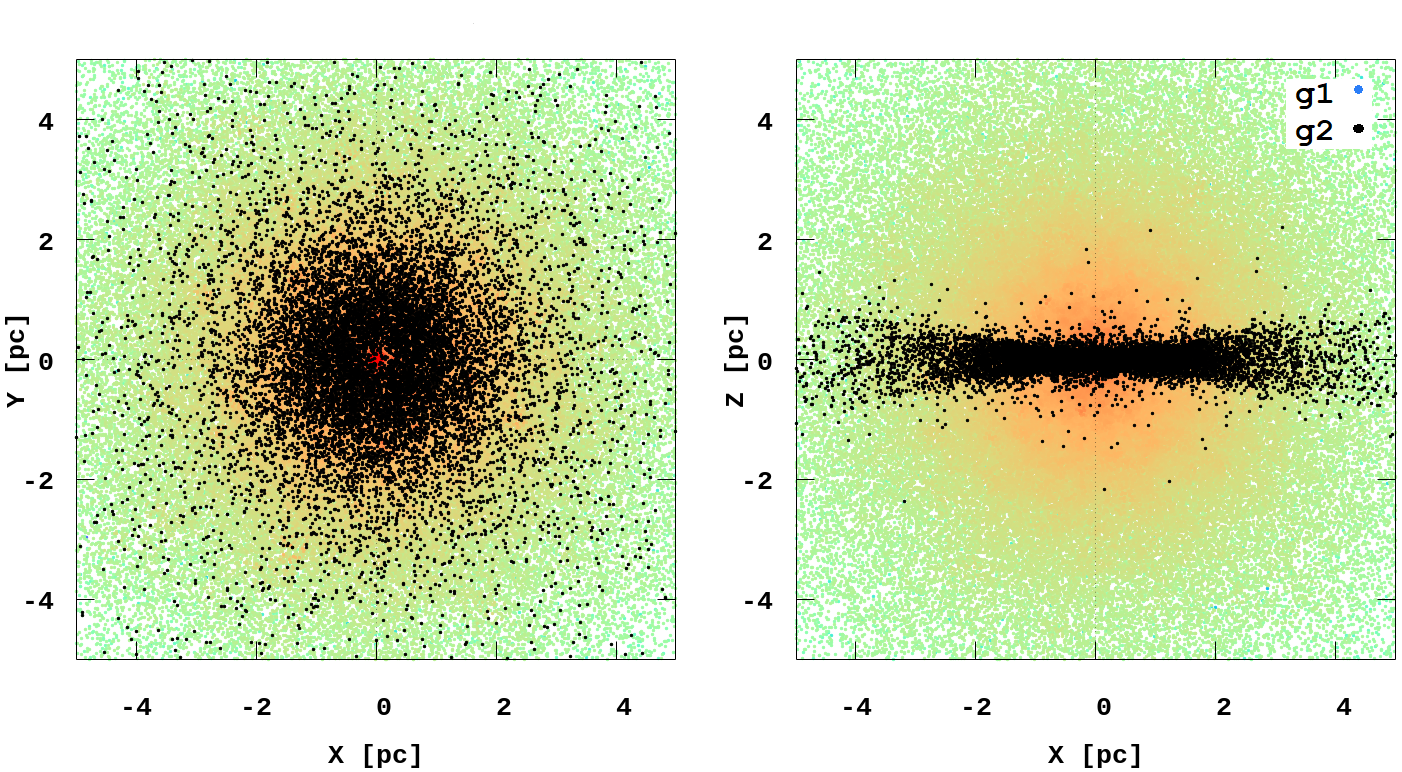}
\includegraphics[width=0.33\linewidth]{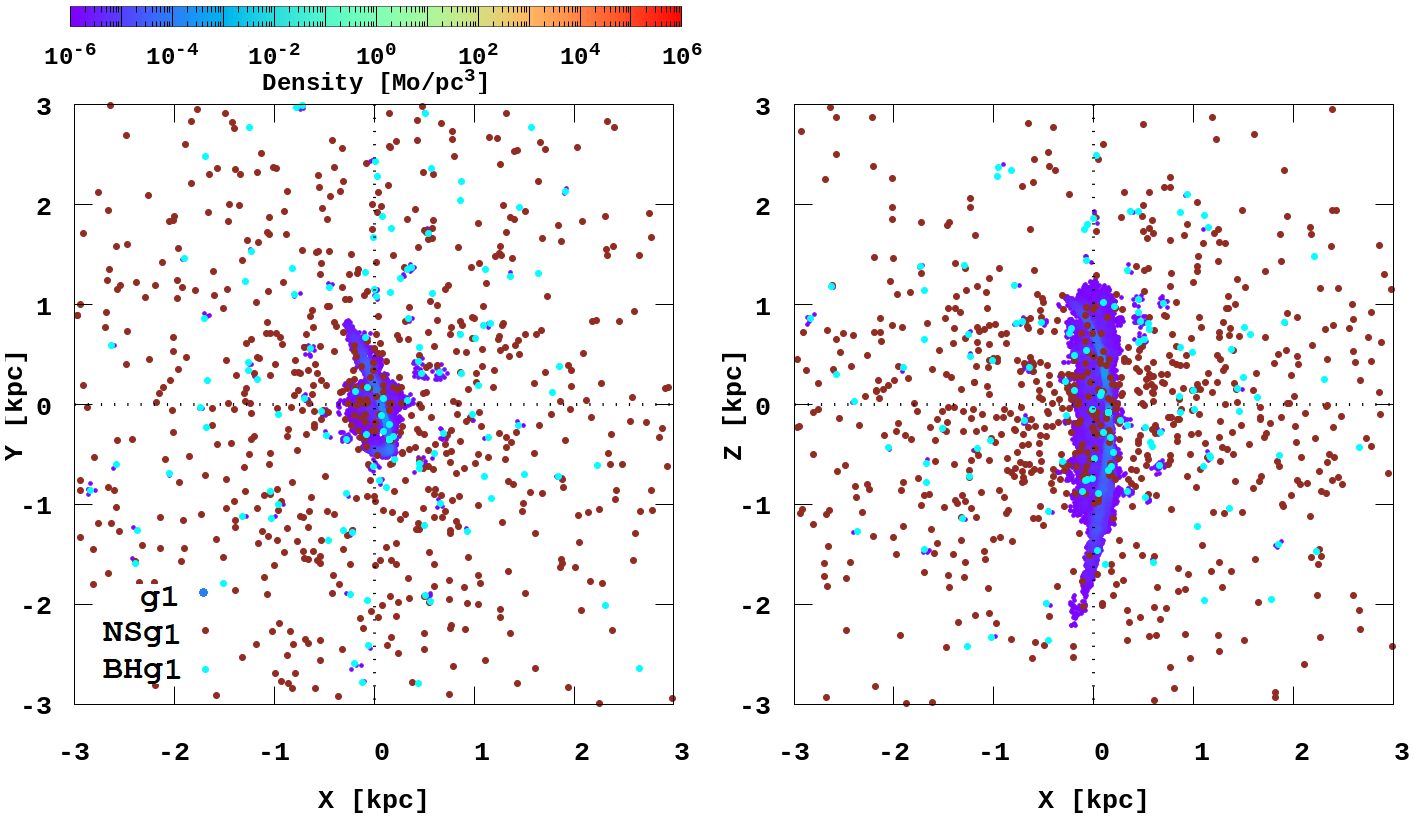}
\includegraphics[width=0.33\linewidth]{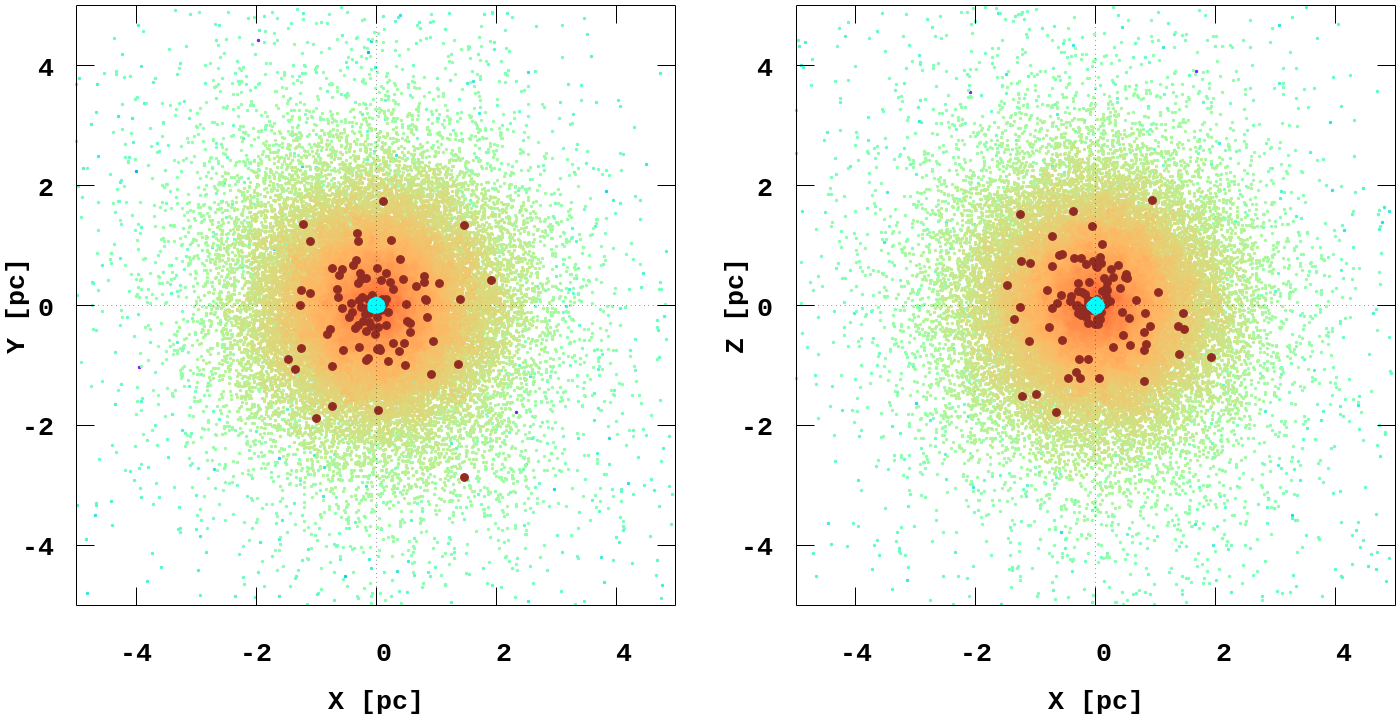}
\includegraphics[width=0.33\linewidth]{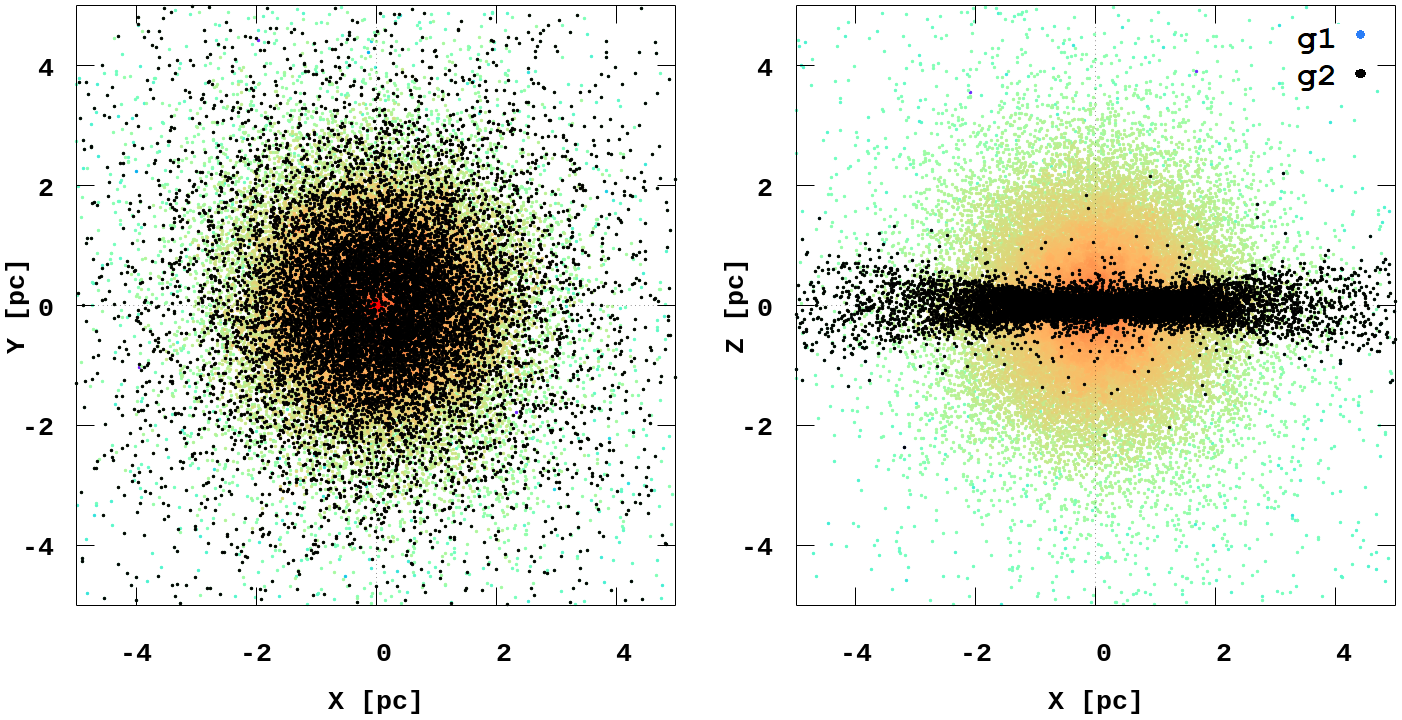}
\caption{GC density distribution at -6.0 Gyr for the tube orbit at the moment when the second generation stars ({\tt g2}) are added to the GC. The \textit{two left panels}  show the GC in global Galactocentric coordinates. Cyan and brown dots show BHs and NSs in the GC. The \textit{two middle panels} show the density of the central GC’s core (5×5 pc). The \textit{two right panels} show the initial spatial distributions of the {\tt g2} stars in the GC with black dots which form a disk. The \textit{upper} and \textit{bottom panels} show the density distribution inside the GC in the {\tt 411321} and the {\tt FIX} potentials, respectively.}
\label{fig:tb-init}
\end{figure*}

Fig.~\ref{fig:tb-init} presents the density distribution at -6.0 Gyr for the GC on a tube orbit in the TNG (upper panels) and {\tt FIX} potentials (lower panels). The four left panels show densities in the Galactocentric coordinate system, while the four middle panels show density distribution in the central core of the cluster (5x5 pc). Additionally, cyan and brown dots indicate black holes and neutron stars, respectively. Finally, the four right panels show the second stellar generation (described in Sec.~\ref{subsec:DR3}). Density distributions for other orbits can be found in Appendix~\ref{app:global}, in Figures \ref{fig:cr-init} and \ref{fig:lr-init}. 

By the time we insert the second generation stars, after 2 Gyr of evolution, the first generation stars already experienced significant mass loss, see Table \ref{tab:2gyr-evol}. As we see from the table, TB and LR models in {\tt FIX} potential almost lost $\sim$80--70\% of the initial masses. In the case of the TNG the TB model lost $\sim$50\%, see black vertical line in Fig.~\ref{fig:mass-loss}. This is evident for the TB and LR orbits,  where the clusters undergo close approaches to the Galactic centre, leading to enhanced tidal stripping. The mass loss is more prominent in the {\tt FIX} potential. 

\begin{table}[htbp!]
\caption{Physical parameters of the GCs after 2\,Gyr of evolution.} 
\centering
\begin{tabular}{llccccc}
\hline
\hline
Pot. & Orbit & $r_{\rm hm}$ & $r_{\rm tid}$ & $M (r_{\rm hm})$ & $M (r_{\rm tid})$ & $N_{\rm bound}$ \\
   &      & pc & pc & {$10^{4}\rm\;M_{\odot}$} & {$10^{4}\rm\;M_{\odot}$} & {$10^{3}$} \\
\hline
\hline
TNG & CR & 3.4 & 59 & 4.8 & 9.7 & 259  \\
    & TB & 2.5 & 15 & 3.7 & 7.3 & 194  \\
    & LR & 3.2 & 47 & 4.5 & 9.1 & 241  \\
FIX & CR & 3.3 & 40 & 4.7 & 9.5 & 252  \\
    & TB & 0.9 & 30 & 1.2 & 2.5 & 54   \\
    & LR & 1.6 & 38 & 2.2 & 4.3 & 104  \\
ISO & -- & 3.1 & --  & 4.5 & -- & 238  \\
\hline
\end{tabular}
\label{tab:2gyr-evol}
\end{table} 

\subsection{Second stellar generation}\label{subsec:DR3}

After 2 Gyr of dynamical evolution of the cluster we add to the system the {\tt g2} stellar sub-system. For each simulation, we add the same mass of the  {\tt g2} population namely 8000 M$_\odot$ with the same Kroupa IMF as the first generation {\tt g1}. We add to the {\tt g1} stars additional 13 940 {\tt g2} stars, which corresponds to 5\% of the initial mass of the {\tt g1}. The main assumption in our models is that the g2 stars formed from the AGB ejecta of g1 stars, which imposes an upper limit on the initial mass of g2 stars relative to the initial mass of g1 stars. For an IMF of the stellar system given by $dN = \xi(m)\, dm$, assuming that, g2 stars form from AGB stars with initial masses between $m_1$ and $m_2$, the mass fraction of AGB stars relative to the total stellar mass is:

\begin{equation}
f = \frac{\int_{m_1}^{m_2} m\, \xi(m)\, dm}{\int_{m_{\text{low}}}^{m_{\text{high}}} m\, \xi(m)\, dm},
\end{equation}
where $m_{\text{low}}$ and $m_{\text{high}}$ are the overall lower and upper mass limits of the IMF. Following the arguments of \citet{Bekki2010}, who proposed that {\tt g2} stars form from AGB stars with initial masses between $5$--$8\,M_\odot$, the mass fraction of AGB stars relative to the total stellar mass is approximately $0.06$ for the Kroupa IMF $ \xi(m) =\xi_0 m^{-\alpha}$ with the power law index $\alpha=1.3$ for $0.08 < m < 0.5 M_\odot$ and $\alpha=2.3$ for $m > 0.5 M_\odot$ (with the mass limits between $0.08\,M_\odot$ and $100\,M_\odot$). Considering that $83\%$--$86\%$ of the mass of these AGB stars can be ejected \citep{Gnedin2002}, and assuming that all of this material converts into {\tt g2} stars, we estimate that approximately $5\%$ of the total mass of the {\tt g1} stars may end up in {\tt g2} stars. We allow the {\tt g1} stars to relax dynamically before introducing the {\tt g2} stars, resulting in mass fractions of {\tt g2} stars ranging between $8\%$ and $32\%$ at the time of their addition.

In our setup the main difference between {\tt g1} and {\tt g2} stars was the disk-like distribution of the {\tt g2}  stars. There are no solid constraints on the initial thickness of the possible {\tt g2} stars in globular clusters. However, star formation in disks which is observed in galactic nuclei, e.g. in the central half-parsec of the MW, the so-called clockwise stellar disk, appears to have the orbital angular momentum vectors distributed in the range of 10-15 degrees \citep{vonFellenberg2022, Jia2023}. In our case, the ``disk'' distribution was generated using the rejection method from the initial spherical Plummer particle distribution, selecting stars which have positive angular momentum with a polar angle less than 10 degrees from the $Z$-axis:
\begin{equation}\label{eq:Lz}
\arccos\left(\frac{L_{\rm z}}{L_{\rm tot}}\right) \leq 10^{\circ},
\end{equation}
where the angular momentum is calculated with respect to the centre of mass of the cluster.

The original Plummer model was generated with an initial half-mass radius $r_{\rm hm}$ of 2 pc. These {\tt g2} stars are initialised at the centre of the evolved {\tt g1} stellar system. As the {\tt g1} stars experienced stellar evolution and dynamical (Galactic tidal) mass loss over the first 2 Gyr, the insertion of 8000 M$_\odot$ of {\tt g2} stars results in different mass fractions of {\tt g2} relative to {\tt g1}, depending on the orbit and potential. In Table~\ref{tab:mass-rat}, we present the mass and number ratios of {\tt g1} and {\tt g2} stars at the moment when the {\tt g2} stars are inserted, highlighting how the additional {\tt g2} mass contributes different percentages relative to the current stellar mass of the {\tt g1} population for each orbit/potential. 


\begin{table}[htbp!]
\caption{Initial mass ratio of the {\tt g2} to the {\tt g1} current mass.} 
\centering
\begin{tabular}{llcc}
\hline
\hline
Pot. & Orbit & $M_{\rm ini~ {\tt g2}}/M_{\rm cur~ {\tt g1}}$ & $N_{\rm ini~ {\tt g2}}/N_{\rm cur~ {\tt g1}}$ \\
 & & in \% & in \% \\
\hline
\hline
TNG & CR  &  8.2 & 5.4 \\
    & TB  & 10.9 & 7.2 \\
    & LR  &  8.8 & 5.8 \\
FIX & CR  &  8.4 & 5.5 \\
    & TB  & 32.0 & 25.8 \\
    & LR  & 18.6 & 13.4 \\
ISO & --  &   -- & 5.8 \\
\hline
\end{tabular}
\label{tab:mass-rat}
\end{table} 

\begin{figure}[ht]
\centering
\includegraphics[width=0.98\linewidth]{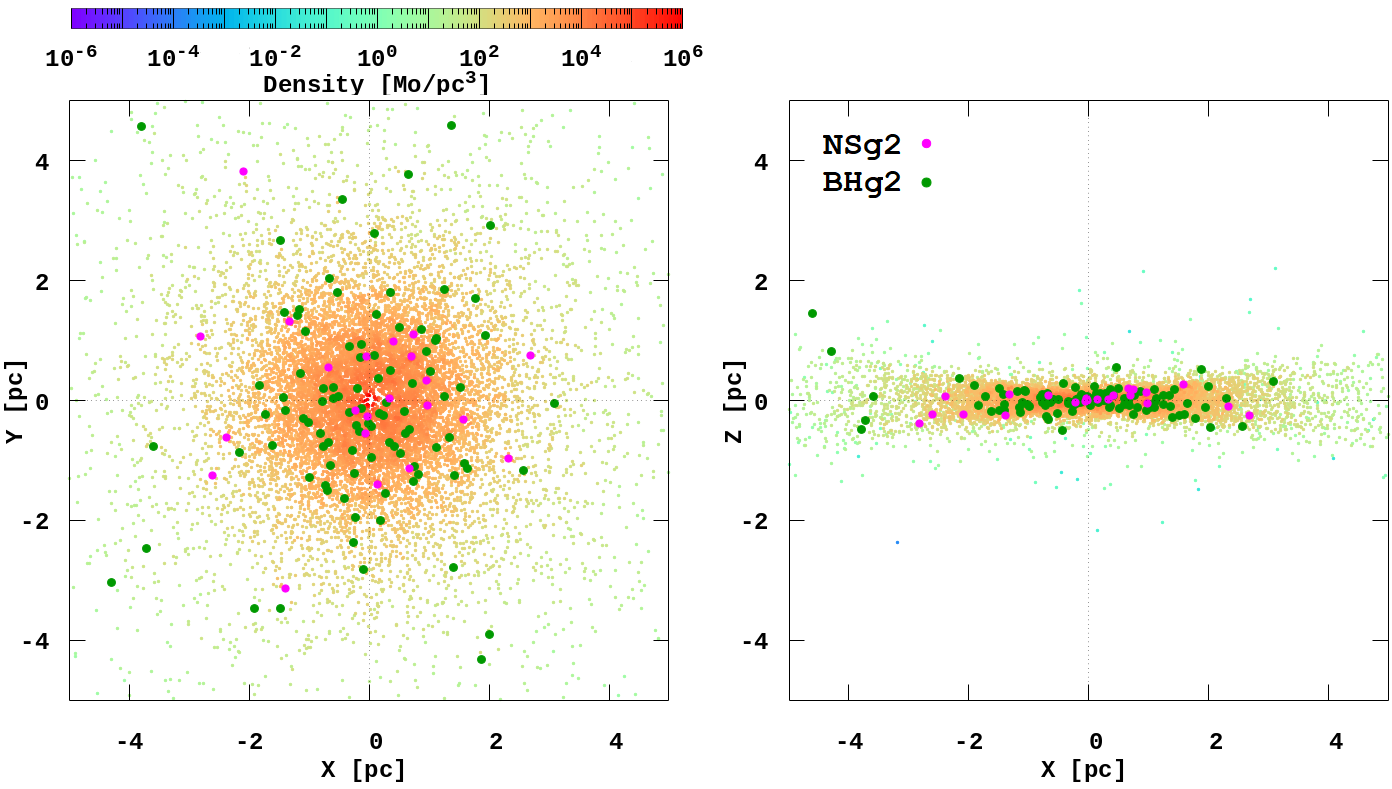}
\caption{The initial density distribution for second generation stars (colour-codded dots). Dark green and magenta dots show the positions of the NS and BH progenitors, respectively.}
\label{fig:init-stardisk}
\end{figure}

On Fig. \ref{fig:init-stardisk} we shows the initial positions of {\tt g2} stars. The 3D mass density values within the star cluster (in M$_\odot$/pc$^3$) is represented by colour coding within the central 5$\times$5 pc region. 
The two right panels in Fig. \ref{fig:tb-init} show the initial positions for the {\tt g2} stars in the central $5\,\mathrm{pc}\times 5\,\mathrm{pc}$ of the GCs on TB orbits. Similarly, Fig. \ref{fig:cr-init} and \ref{fig:lr-init} in  Appendix \ref{app:global} show the initial positions for GCs on CR and LR orbits, respectively. 

\subsection{Isolated system}\label{sybsec:iso-run}

For comparison, we also simulate the star clusters in complete isolation, i.e. no external galactic potential. These models are labelled {\tt ISO}. For these set of runs we use exactly the same initial particle distributions as described in the previous sections. 

\begin{figure}[ht]
\centering
\includegraphics[width=0.98\linewidth]{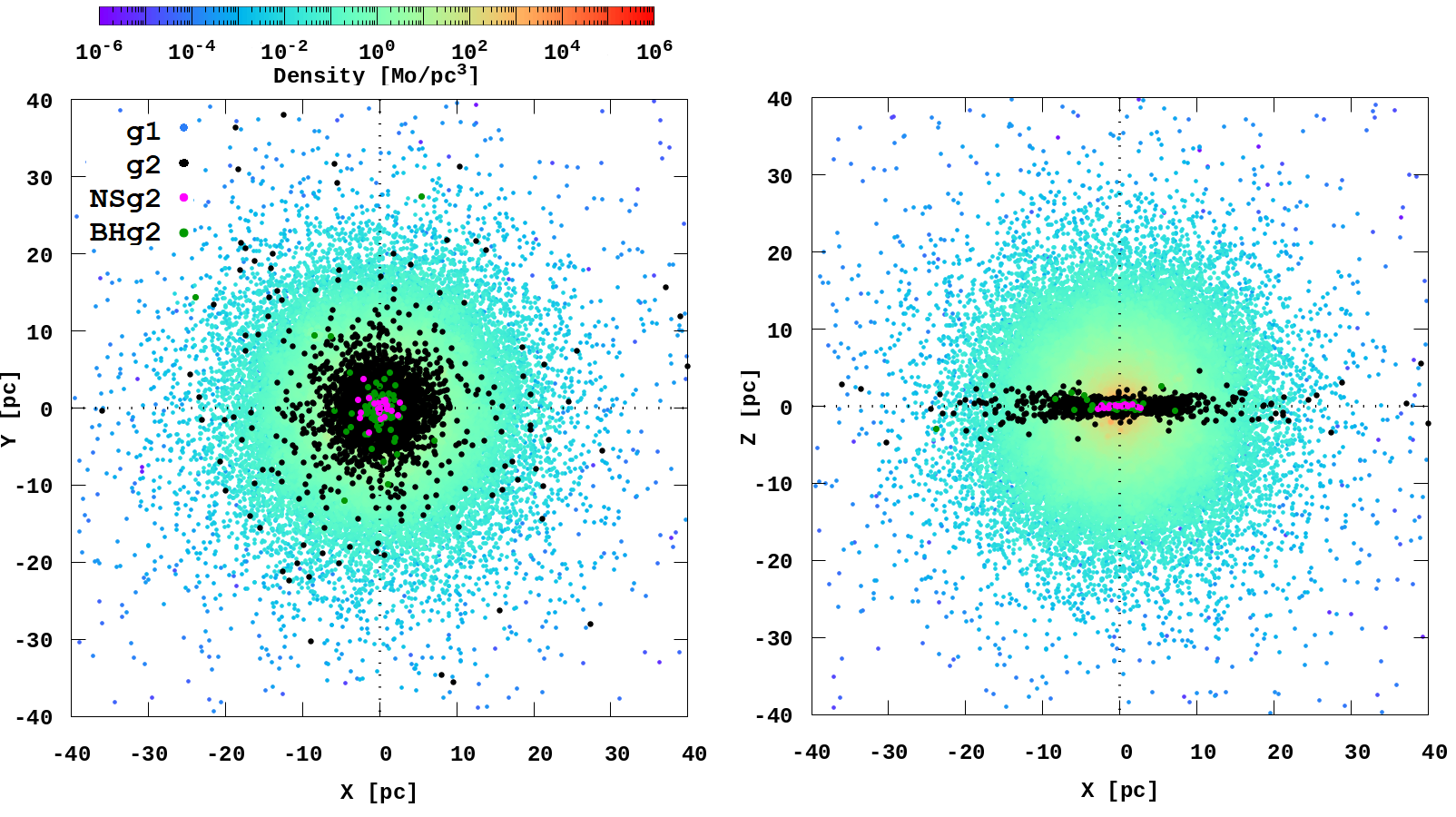}
\caption{GC's density distribution at -6.0 Gyr lookback time for the {\tt ISO} model. Dark green and magenta dots show the BHs and NSs, respectively. Black dots represent the initial positions of the second generation stars {\tt g2}.}

\label{fig:iso-init}
\end{figure}

To illustrate the {\tt g1} and {\tt g2} stars' distribution, Fig. \ref{fig:iso-init} shows the central {\tt g1} density distributions and black dots show the {\tt g2} initial positions within the GC. 

The initial half-mass relaxation time for this system on which the orbital energies mix for different stars is (\citealt{BinneyTremaine2008}, eq. 7.106) 
\begin{equation}
\label{eq:2-body}
t_\mathrm{relax} = 0.34\frac{\sigma^3(r)}{G^2\rho(r) \langle m\rangle\ln\Lambda} \approx 900\,\mathrm{Myr} = \frac{1}{\alpha^2} N(r) P(r),
\end{equation}
where $\sigma$ is one-dimensional velocity dispersion, $\rho$ is the stellar density, $\langle m\rangle\approx0.574 M_\odot$ is the average mass and $\ln\Lambda\approx11.0$ is the Coulomb logarithm, and the final equality \citep{Meiron_Kocsis2019} has $\alpha=7.32\pm0.15$, $P(r)$ is the period and $N(r)$ is the enclosed number of stars for which $P=0.37\,\mathrm{Myr}$ at the half mass radius.  
The vector resonant relaxation (VRR) timescale on which the orbital inclinations mix is 
\begin{equation}
\label{eq:vrr}
t_\mathrm{vrr} = \frac{\sqrt{N(r)} P(r)}{\beta_v} = \frac{\alpha^2}{\beta_v } \frac{t_\mathrm{relax}(r)}{\sqrt{N(r)}} \approx 100\,\mathrm{Myr} ,
\end{equation}
where $\beta_\nu=1.33 \pm 0.50$ \citep{Meiron_Kocsis2019}. The VRR timescale is prolonged for a rotating cluster \citep{Kocsis_Tremaine2011}.

\section{General dynamical evolution of the combined systems}\label{sec:gen-g1-g2}

\subsection{Mass loss}\label{sybsec:glo}

\begin{figure}[ht]
\centering
\includegraphics[width=0.98\linewidth]{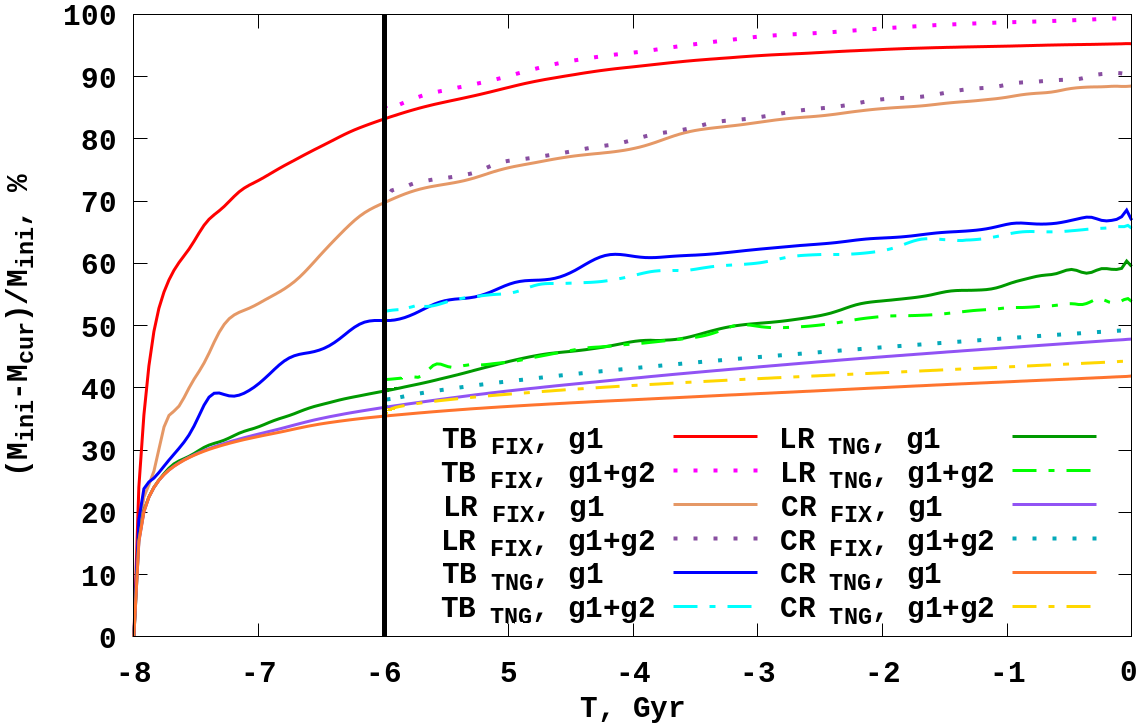}
\caption{Evolution of the mass loss in different GCs. Solid lines -- single generation  {\tt g1}.  Dotted and dot dashed lines -- multiple generations {\tt g1+g2}. Black vertical line represents the time when the second generation of stars {\tt g2} was added into the GCs.}

\label{fig:mass-loss}
\end{figure}

Our numerical simulations follow the stellar mass loss of the GCs due to both stellar evolution and tidal stripping by the external potential. First we compare the mass loss of GCs with one and two stellar generations in Fig. \ref{fig:mass-loss}, respectively. We compare the mass loss for simple {\tt g1}-only (solid lines) and two stellar population models {\tt g1+g2} (dashed lines). The figure shows that in the first $\sim$10 Myr the single generation GCs (i.e. only {\tt g1} stars) lost roughly $\sim$25\% of their initial masses. 

Let us now examine the influence of the external potential on the clusters' mass loss with one- and two stellar generation models. In the time evolving TNG potential, the GCs on TB and LR orbits lost $\sim$40--60\% of their initial masses by today. In the opposite case with the static {\tt FIX} potential, the GCs' mass loss is significantly higher, i.e. $\sim$60--95\%. GC models in the {\tt FIX} external potential have lost typically two times more mass compared to the corresponding GC model in the time variable TNG external potential. 
Fig. \ref{fig:mass-loss} shows that adding a second stellar generation {\tt g2} into the cluster leads to somewhat different mass loss behaviours. The GCs dissolve faster for the {\tt FIX} potential for long radial and tube orbits. In the time evolving TNG potentials the models show a slower cluster dissolution rate. 

\begin{figure*}[ht]
\centering
\includegraphics[width=0.98\linewidth]{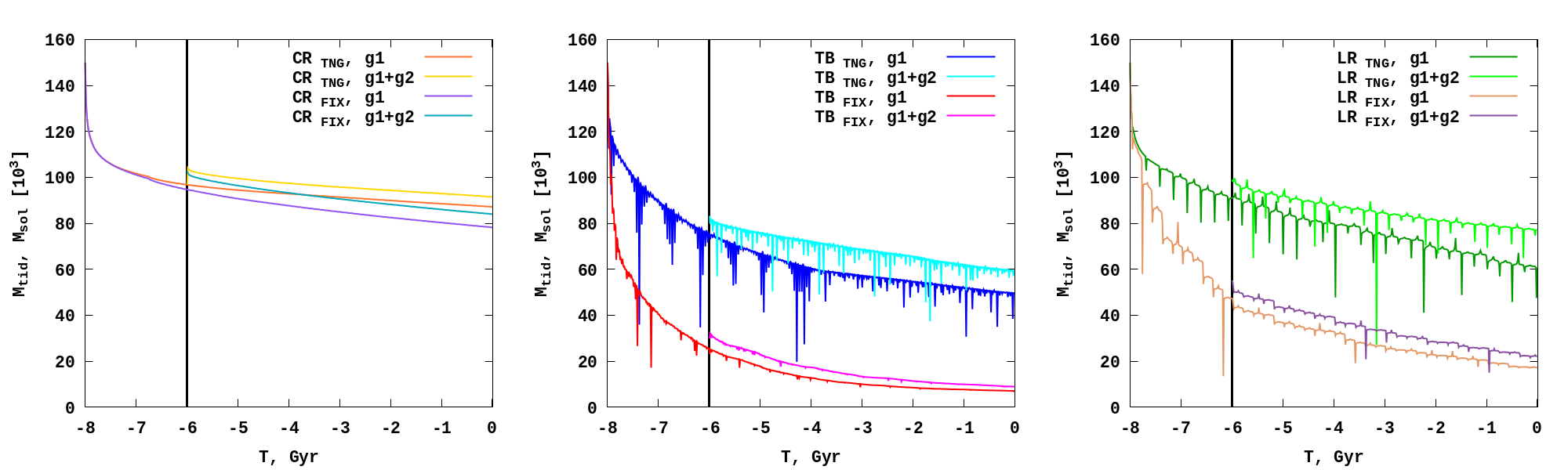}
\includegraphics[width=0.98\linewidth]{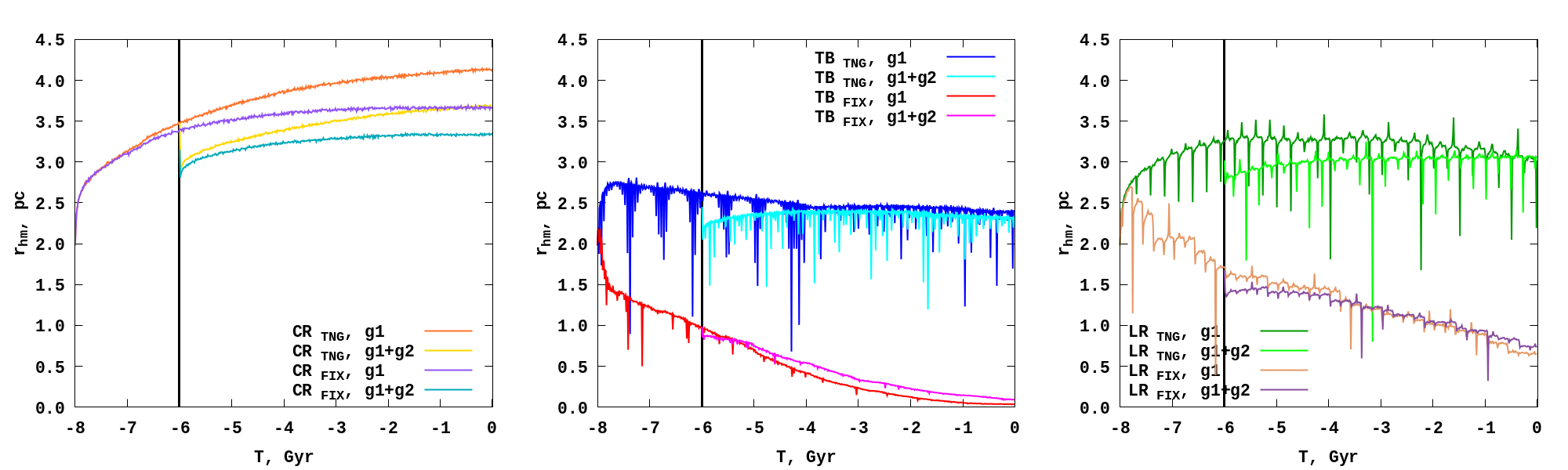}
\caption{Evolution of the GC's physical parameters as a function of time: the tidal mass \textit{(upper panels)}, and half-mass radii \textit{(bottom panels)}. The left, middle, and right columns show GCs on circular, tube, and radial orbits, respectively. Black vertical line represents the time when the {\tt g2} generation was added to the GCs.}
\label{fig:mtid-rhm-rtid}
\end{figure*}

Let us now discuss in detail the dynamical evolution of the single generation {\tt g1} and mixed generation {\tt g1+g2} clusters. Fig. \ref{fig:mtid-rhm-rtid} presents the evolution of the GC's global physical parameters including the tidal mass ($M_{\rm tid}$) and half-mass radii ($r_{\rm hm}$). We obtained the cluster tidal mass based on the iterative procedure, described in detail in \citet{King1962, Just2009, Ernst2011}.  As a result, we simultaneously obtain the cluster tidal mass -- $M_{\rm tid}$, which is currently bound to the cluster and the corresponding cluster tidal or Jacobi radius -- $r_{\rm J}$. After identifying the cluster members based on the simple distance criteria $r < r_{\rm J}$ we derive the half-mass radius of the system. We repeat this procedure for each simulation snapshot.

Generally we find that clusters on TB and LR orbits show significantly fluctuating, non-monotonic behaviour for the tidal mass and half mass radii during their whole orbital evolution both for the {\tt g1} and also for the {\tt g1+g2} stellar populations. These short time variations have a quite significant range up to $\sim$ 20-30\% for the TNG external potential. This kind of behaviour is much less prominent, nearly non-existent for the {\tt FIX} potential. In the case of TB orbits, such a behaviour can be explained by the much more compact extent of the orbit in the {\tt FIX} potential, see Figures~\ref{fig:tube-orb}). The orbits are initially much wider in the TNG potential in order to yield the same final orbit observed today, since the halo and galaxy masses were initially much smaller in the TNG case. In the case of the LR orbit the differences in the orbital shape is much less prominent, see Figures~\ref{fig:lr-orb}. 

Also, Table \ref{tab:2gyr-evol} shows that the GCs in the {\tt FIX} potential have lost more stellar mass than their counterparts in the {\tt TNG} potential. The GC on tube orbit in the {\tt FIX} potential is almost completely destroyed, the GC on circular orbit was more resilient. This effect can be easily explained by the strong influence of the mass in the {\tt FIX} bulge potential. At the same time, models with circular orbits, which have almost no interaction with the bulge during their evolution, show higher survival rates.  

For the clusters on CR orbits we practically see a smooth evolution for both the TNG and {\tt FIX} potentials for both {\tt g1} and also {\tt g1+g2} populations as expected since in this case the potential does not fluctuate rapidly along the GC orbit. 

For all models the addition of the second generation stars naturally increases the tidal mass of the cluster and decreases $r_{\rm hm}$. The second process occurs because the added second generation is initially more compact compared to the distribution of the first generation at that time. As a result, the mixed stellar system {\tt g1+g2} have a smaller $r_{\rm hm}$ compared to the {\tt g1}-only stellar system. 

\begin{figure*}[ht]
\centering
\includegraphics[width=0.98\linewidth]{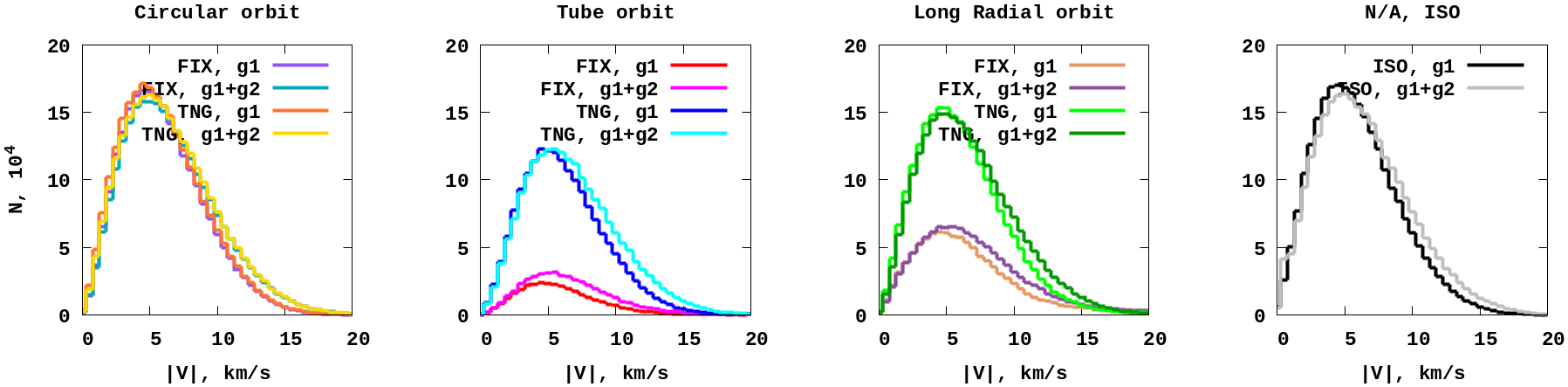}
\caption{The particles' 3D velocity distributions in the GCs for {\tt g1} and {\tt g1+g2} simulations at $T$ = -5 Gyr, i.e. after approximately one two-body relation time after inserting the {\tt g2} component. The distributions are similar in all cases as the {\tt g2} component represents a small fraction of the stars.}
\label{fig:vel}
\end{figure*}

Fig. \ref{fig:vel} presents the 3D velocity distribution of the GC stars for the {\tt g1}-only and the {\tt g1+g2} models at $T$ = -5 Gyr. In the case of TB and LR models in {\tt FIX} potential, we clearly see small values (for TB is $\sim$ 3 and for LR -- 6 km s$^{-1}$) compared with same model in TNG potential (for TB is $\sim$ 13 and for LR -- 15 km s$^{-1}$). This velocity behaviour is due to the fact that the clusters in the FIX potential have lost between $70-90\%$ of their initial mass after 3 Gyr of their evolution, see Fig.~\ref{fig:mass-loss}.

\begin{figure*}[ht]
\centering
\includegraphics[width=0.98\linewidth]{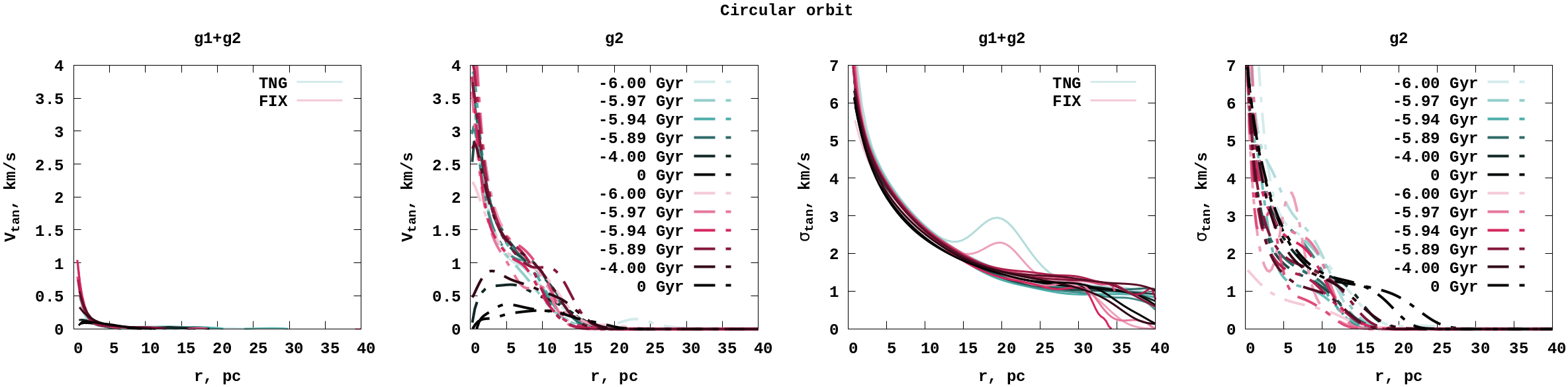}
\includegraphics[width=0.98\linewidth]{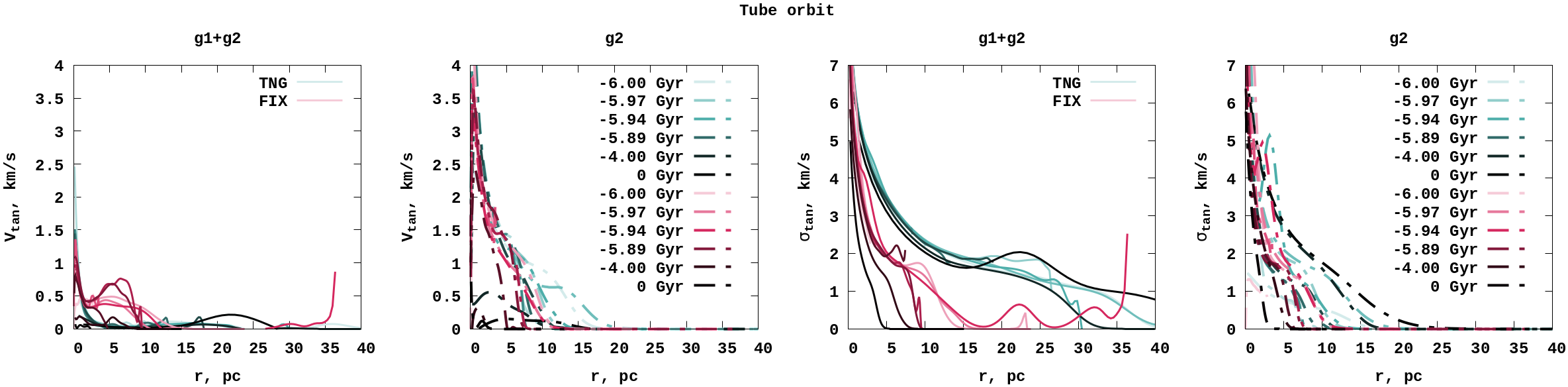}
\includegraphics[width=0.98\linewidth]{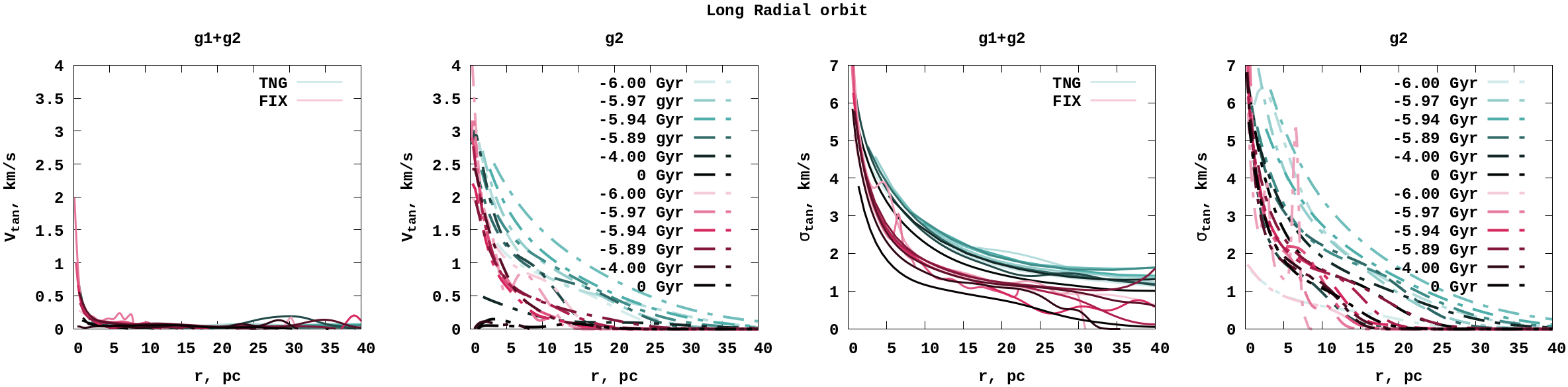}
\includegraphics[width=0.98\linewidth]{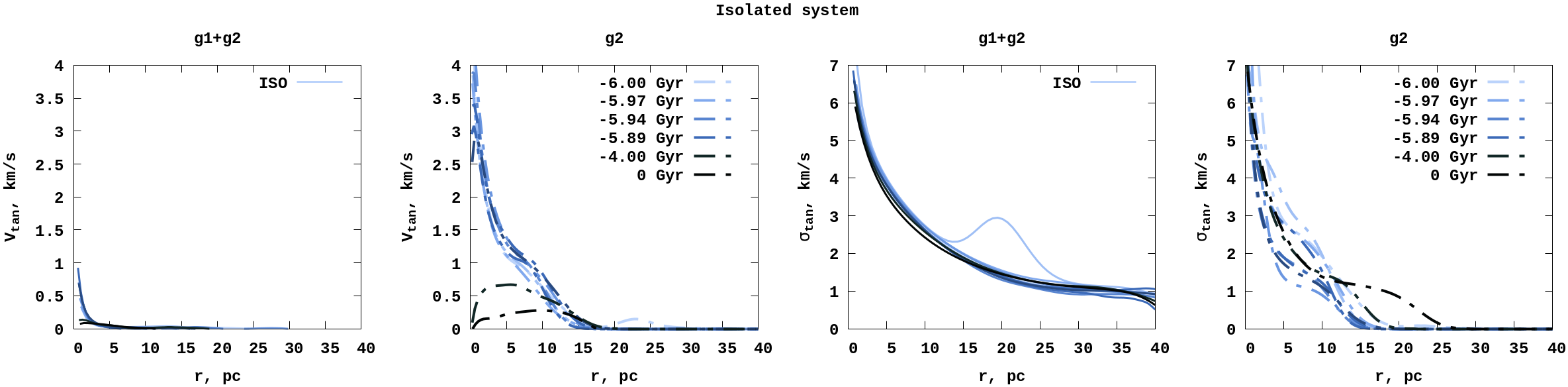}
\caption{Evolution of the mean tangential velocity (left two panels) and the tangential velocity dispersions (right two panels) for {\tt g1+g2} stars and for the {\tt g2} separately for the different external potentials and lookback times. Different rows of panels show different orbit families as labelled. Solid green, red, and blue curves show models in time variable {\tt TNG} potential, the {\tt FIX} potential, and the {\tt ISO} isolated case, respectively. Different curves show different moments of time for each potential as labelled.}
\label{fig:tangen-vel}
\end{figure*}

The two left panels in Fig.~\ref{fig:tangen-vel} show the evolution of the mean tangential (i.e. rotational) velocities for {\tt g1+g2} and {\tt g2} stellar populations. Solid green and red lines show models in the {\tt TNG} and {\tt FIX} potentials, respectively, and the blue curves show the isolated case, {\tt ISO}. The two right panels show the evolution of the tangential velocity dispersion.
Generally we conclude that irrespective of the orbit type and external potential, the mean tangential (i.e. rotational) velocity is always negligible when averaging over both {\tt g1+g2} stars, given that the {\tt g1} stars are initially isotropic and dominate the statistics. The mean initial tangetial velocity of the {\tt g2} stars decreases and becomes negligible by $\sim$ 2 Gyr for almost all potentials and orbit families. The exemption is the LR case where the tangential velocity decreases more slowly for the {\tt TNG} potential than for the {\tt FIX} potential.

The evolution of the tangential velocity dispersion (Fig.~\ref{fig:tangen-vel}, \textit{right panels}) shows a significantly less dynamical evolution compared to the tangential velocity itself. The tangential velocity dispersion changes significantly only for the TB orbit in the {\tt FIX} external potential due to the significant global mass loss in the system (see Fig.~\ref{fig:mass-loss}) 

\section{Internal kinematics of the second stellar generation}\label{sec:gen-g2}

In this section, we explore the internal kinematics of the {\tt g2} stars in globular clusters, focusing on how this population evolves under the combined effects of internal dynamics driven by the {\tt g1} stars and the external gravitational potential. A key question we aim to address is whether distinct kinematic features can be identified for the {\tt g2} population, shedding light on its formation and origin. By analysing the structural evolution and angular momentum of this initially disk-like population, we also examine how it interacts and mixes with the {\tt g1} stars over time.

\subsection{Structure and shape}\label{sybsec:vel}

\begin{figure*}[ht]
\centering
\includegraphics[width=0.98\linewidth]{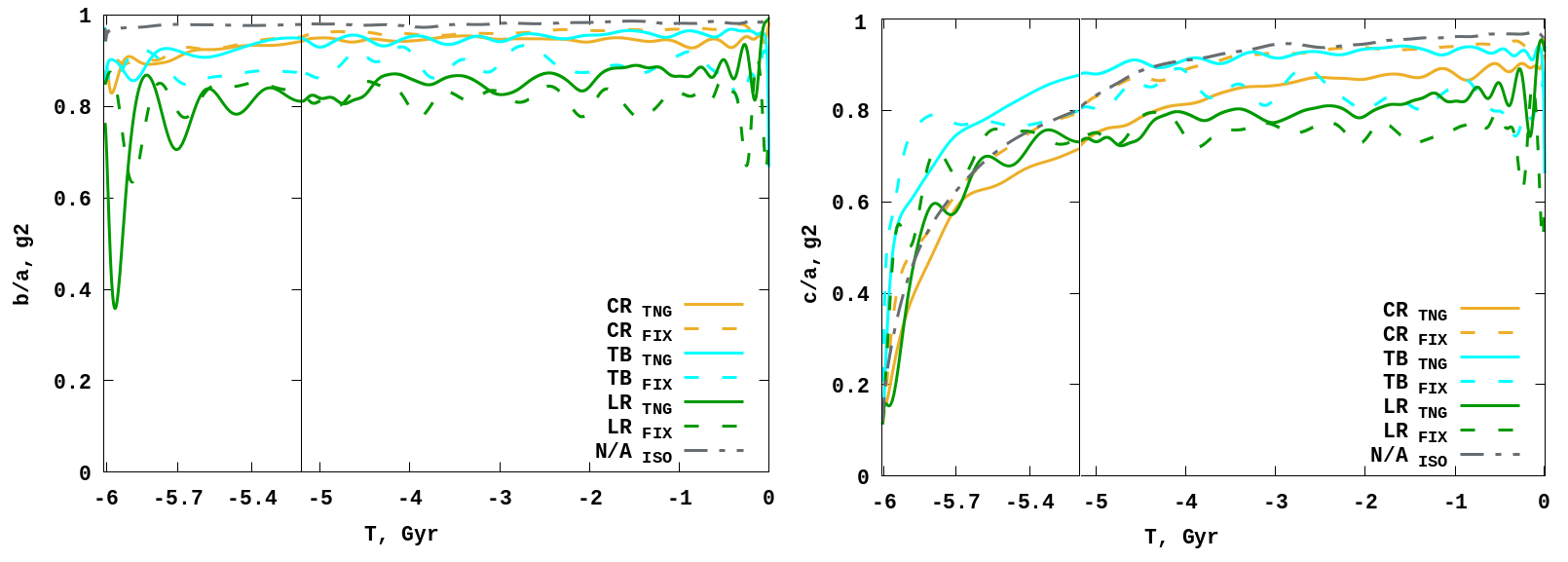}
\caption{Ratio of the moment of inertia matrix principal axes $b/a$ and $c/a$ for the second generation of stars - {\tt g2}. 
The vertical line at -5.1Gyr shows the two-body relaxation time (900 Myr). Rapid change is visible much earlier on a few times the 100 Myr VRR timescale (Eq.~\ref{eq:vrr}).
Solid lines represent TNG, dashed -- {\tt FIX} potentials, and dot-dashed -- isolated case.}
\label{fig:triax}
\end{figure*}

Recall that the {\tt g2} stars are initialized as a flattened axisymmetric disk. To study the evolution of the shape of the disk, we fit an ellipsoid with axes $a$, $b$, and $c$ using the matrix of moment of inertia (e.g. \citealt{Regaly2023}). For this analysis, we select the stars that are inside the tidal radius of the cluster. Fig.~\ref{fig:triax} shows the time evolution of the axial ratios $b/a$ (left panel) and $c/a$ (right panel). The system approximately maintains its axial symmetry, as $b/a$ is close to unity or to $c/a$. However, the initially disky structure evolves towards a spherical shape as $b/a$ and $c/a$ approach unity. This evolution is relatively rapid, the disk thickens rapidly during the first $\approx 300$ Myr, which is just an $\frac13$ fraction of the two-body relaxation time, it is $3\times$ the vector resonant relaxation time (Eqs.~\ref{eq:2-body}--\ref{eq:vrr}), as evident from the right panel of the figure. This behaviour is similar for all orbits and in both external potentials, which indicates that this process is mainly driven by the internal orbit-averaged dynamics of resonant relaxation rather than by close two-body encounters or the external forces (see Section~\ref{sybsec:iso-run}). We refer to Appendix~\ref{app:global} for a visualization of the internal morphology of the GCs (the two right panels of Figures~\ref{fig:cr-loc-tng}, \ref{fig:cr-loc-fixg}, \ref{fig:tb-loc-tng}, \ref{fig:tb-loc-fix}, \ref{fig:lr-loc-tng}, and \ref{fig:lr-loc-fix}).

\subsection{The evolution of angular momentum}\label{sybsec:mom}

While the second stellar generation tends to acquire a spherical shape in less than 500 Myr or $\frac12$ the relaxation time in physical space, here we show that it may preserve its rotational signature in angular momentum space for much longer, in some cases until present.

\begin{figure*}[ht]
\centering
\includegraphics[width=0.98\linewidth]{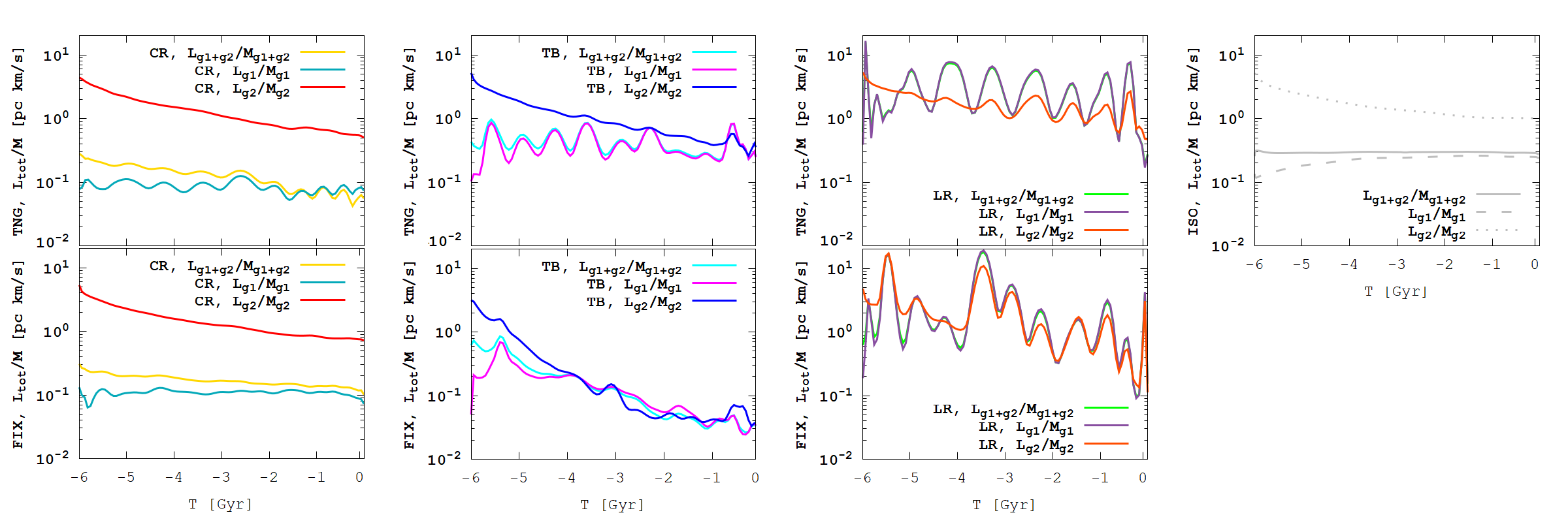}
\caption{The specific angular momentum of the {\tt g1} and {\tt g2} populations as a function of time in simulations with both subpopulations and the 
{\tt g1+g2} 
models. Different panels show different external potentials and different orbit types (first three coloumns: CR, TB, LR orbits; top row: TNG potential, bottom row: FIX potential, 4th column: isolated cluster). The specific angular momentum for the {\tt g2} component is calculated by dividing the total angular momentum magnitude of the {\tt g2} stars with the respective instantaneous total mass of {\tt g2} stars within the tidal radius at the same moment of time, and similarly for the {\tt g1} stars.}
\label{fig:mom-tot-spec}
\end{figure*}

Fig.~\ref{fig:mom-tot-spec} illustrates the evolution of the respective specific angular momenta of the two stellar populations $l = |\sum_i \vec{L}_i|/\sum_i m_i$ as a function of time for the stars which are bound to the cluster. The three curves show the result for the {\tt g1} and the {\tt g2} components in the two-component model and that for the {\tt g1}-only simulation, respectively. Different panels show different external potentials (FIX, TNG, ISO) and different orbits (CR, TB LR) as labelled. The specific angular momentum of the {\tt g2} component is by construction initially clearly larger than for the {\tt g1} component for the circular orbits and for the isolated models. In these cases, while the difference decreases due to two-body relaxation, but it remains very significant for 6 Gyr until present. However for the tube and long radial orbits, the external potential imparts large variations to the specific angular momentum for the {\tt g1} component and $l_{\rm g1}\ll l_{\rm g2}$ no longer holds after a few Gyr for the two TB orbits and immediately for the LR orbits. We conclude that clusters on smaller periapsis orbits than 5 kpc may loose their rotational signatures for the {\tt g2} component while those that are farther in the halo and/or on circular orbits may preserve memory of their initial rotation much beyond the half-mass two-body relaxation time, potentially until present.

\begin{figure*}[ht]
\centering
\includegraphics[width=0.98\linewidth]{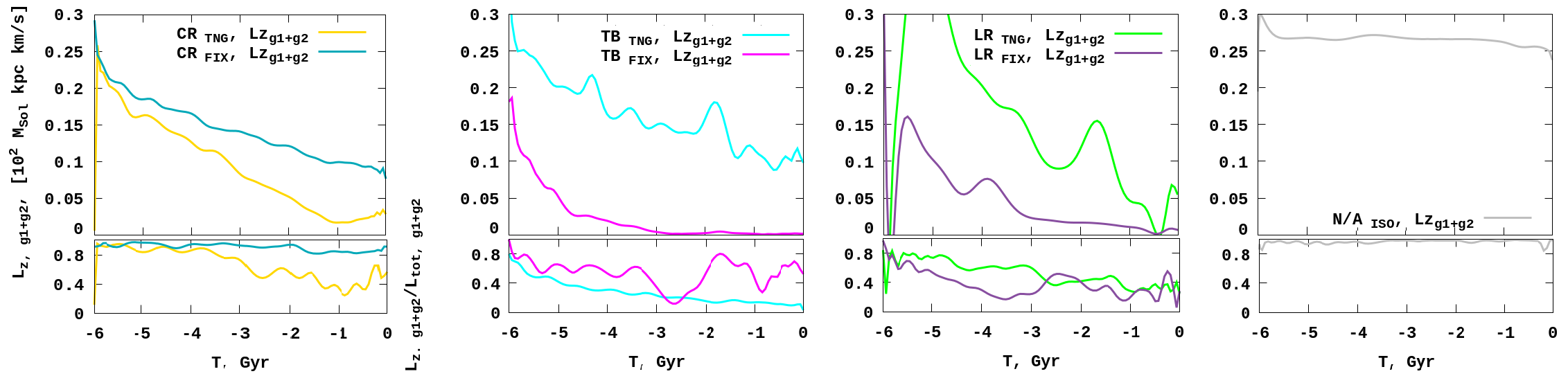}
\caption{ Evolution of the angular momentum $Z$ component for the combined stellar populations {\tt g1+g2}. \textit{Upper panels} show the evolution $L_{\rm z}$ component in physical units and \textit{bottom panels} -- $L_{\rm z} / L_{\rm tot}$ ratio. \textit{From left to right} -- circular, tube, long radial orbits and also the isolated case.}
\label{fig:mmom-tot}
\end{figure*}

\begin{figure*}[ht]
\centering
\includegraphics[width=0.98\linewidth]{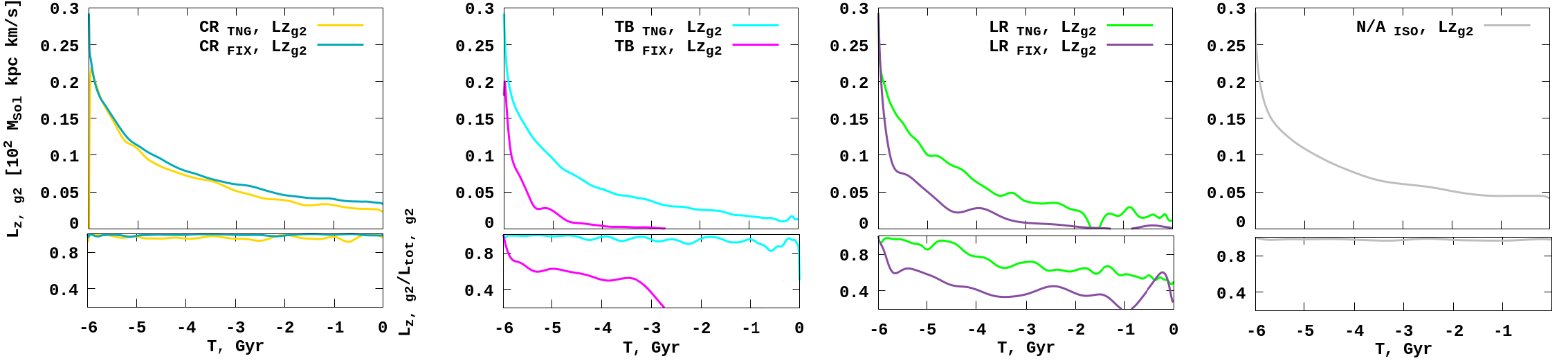}
\caption{Evolution of the angular momentum $Z$ component for the second stellar population {\tt g2}. \textit{Upper panels} represents the evolution $L_{\rm z}$ component in physical units and \textit{bottom panels} -- $L_{\rm z} / L_{\rm tot}$ ratio. \textit{From left to right} -- circular, tube, long radial orbits and also the isolated case.}
\label{fig:mom-g2}
\end{figure*}

Another interesting question is how much the direction of angular momentum may change due to the external potential. Fig.~\ref{fig:mmom-tot} and Fig.~\ref{fig:mom-g2} illustrates the evolution of the $z$-component of the total angular momentum for the combined stellar populations {\tt g1+g2} (Fig.~\ref{fig:mmom-tot}) and for the {\tt g2} component (Fig.~\ref{fig:mom-g2}) in both fixed and time varying external potentials for the stars within the tidal radius. The upper panels of the figure show the evolution of the $L_{\rm z}$ component in physical units, while the bottom part of the panels display the $L_{\rm z} / L_{\rm tot}$ ratio. From left to right, the panels depict the results for circular, tube, long radial orbits, and the isolated case. The {\tt g1+g2} curves show that the angular momentum direction varies greatly over time for not just the TB and LR orbits but also for the circular orbits in the TNG external potential, the direction is roughly conserved only for the circular case in the fixed potential and for the isolated cluster. The evolution for the {\tt g2} case (Fig.~\ref{fig:mom-g2}) is similarly variable as in the {\tt g1+g2} case, with the exception that the direction of the angular momentum happens to change much less not only for the ISO and CR cases but also for the TB orbit for the {\tt g2} component.

The total angular momentum is greatly affected by the change of the cluster mass (Fig.~\ref{fig:mass-loss}). Due to the faster mass loss in a fixed potential, the total angular momentum loss of the whole system occurs faster in a fixed potential than in the TNG potential (except for the circular orbit) as discussed in Sec.~\ref{sybsec:glo}. 
The dynamics of angular momentum loss also vary with the GC orbit type as already expected from the differences in the mass loss rate, Fig.~\ref{fig:mass-loss}. 

%

\subsection{Relevance to the observed clusters}\label{sybsec:obs}

In the previous subsection, we showed that the orbital type of GCs may determine whether the {\tt g2} stars retain their rotational imprint. Although we analysed only three orbits, the prediction is that if the {\tt g2} stars formed with initially higher angular momentum than the {\tt g1} stars, the rotational imprint will be retained for the clusters on circular orbits or with large pericentre distances, even if the fraction of {\tt g2} stars relative to {\tt g1} is initially small. Let us now examine data on multiple population clusters and their orbital parameters from the literature.

\begin{figure}
\centering
\includegraphics[width=0.98\linewidth]{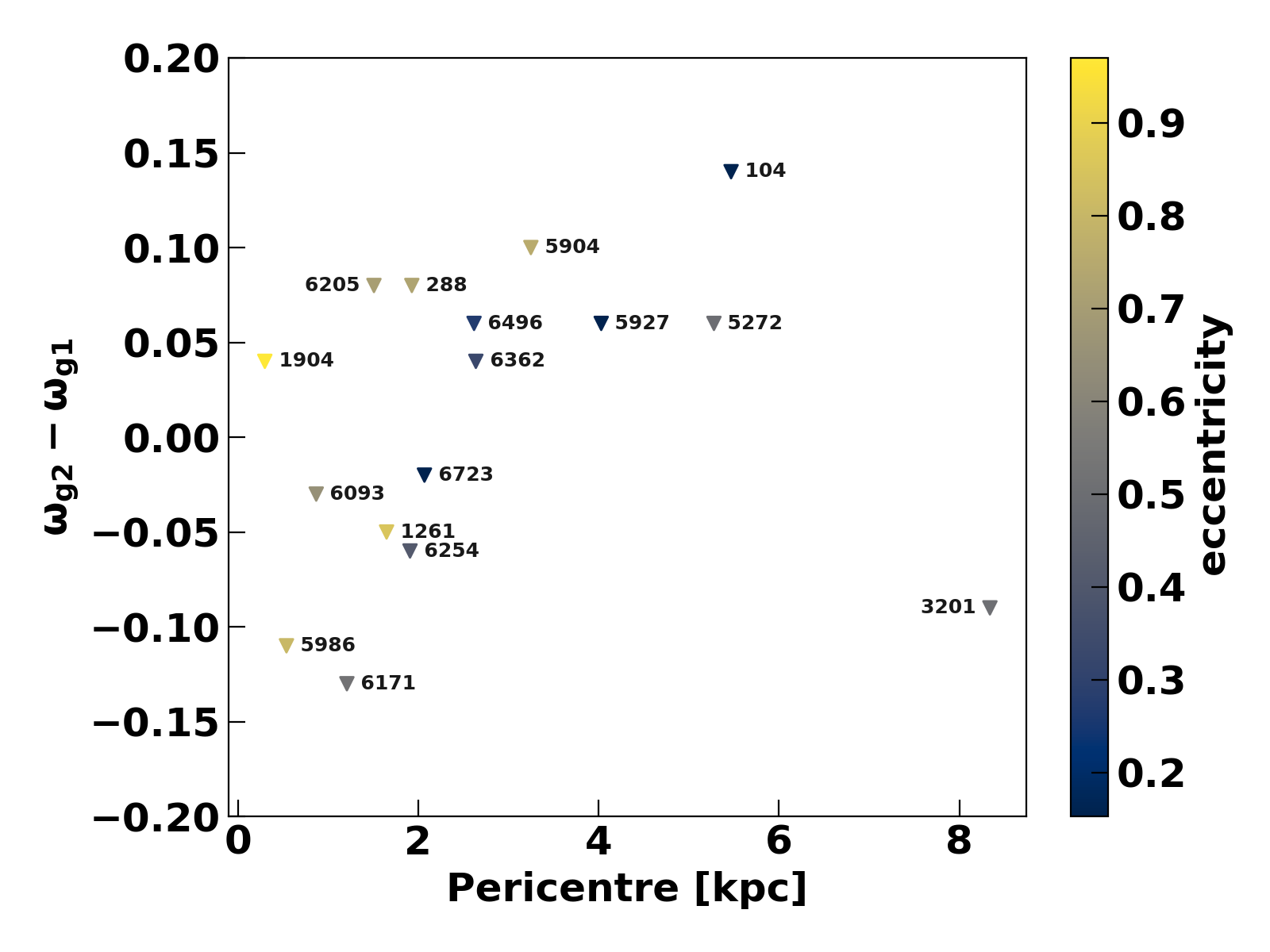}
\caption{Difference in rotation of {\tt g2} stars relative to {\tt g1} as a function of average pericentre distance over the past 500 Myr for a sample of clusters with multiple populations. Colour coding shows average eccentricities over the past 500 Myr. Positive values indicate that {\tt g2} stars rotate faster than {\tt g1} stars. Observational data on rotation taken from \citet{Dalessandro2024}, with orbital parameters from \citet{Baumgardt2021}. Corresponding cluster names from the NGC catalogue are shown next to the data points.}
\label{fig:observed}
\end{figure}

One of the most recent studies on kinematics of multiple stellar populations in GCs \citep{Dalessandro2024} presents evidence that in their sample {\tt g2} stars generally rotate faster than the {\tt g1} stars. Fig.~\ref{fig:observed} shows the difference in the net rotation for the {\tt g2} stars with respect to {\tt g1} as a function of average pericentre distance. Quantities $\omega_{g1}$ and $\omega_{g2}$ show the amount of rotation in the corresponding population (see Eqs.~10 and 2 in \citealt{Dalessandro2024} for the definition of $\omega$). The average pericentre and apocentre distances, from which we computed the average eccentricity, were taken from \citet{Baumgardt2021}. We ignored the error bars of both datasets. The figure illustrates that clusters with larger galactocentric pericentre distances and lower eccentricities tend to show higher rotation in {\tt g2} stars, consistent with our results.\footnote{One notable outlier is NGC 3201 which is beyond 8 kpc.}

On the other hand, \citet{Leitinger2024} analysed a larger sample of GCs and reported no significant evidence of higher rotation in {\tt g2} stars in general. The only exception in their sample is NGC 104 (47 Tuc), which is also present in the sample of \cite{Dalessandro2024}, where higher rotation in {\tt g2} stars was also observed. Interestingly, this cluster has a low eccentricity and a pericentre distance of more than 5 kpc.

\begin{figure}
\centering
\includegraphics[width=0.98\linewidth]{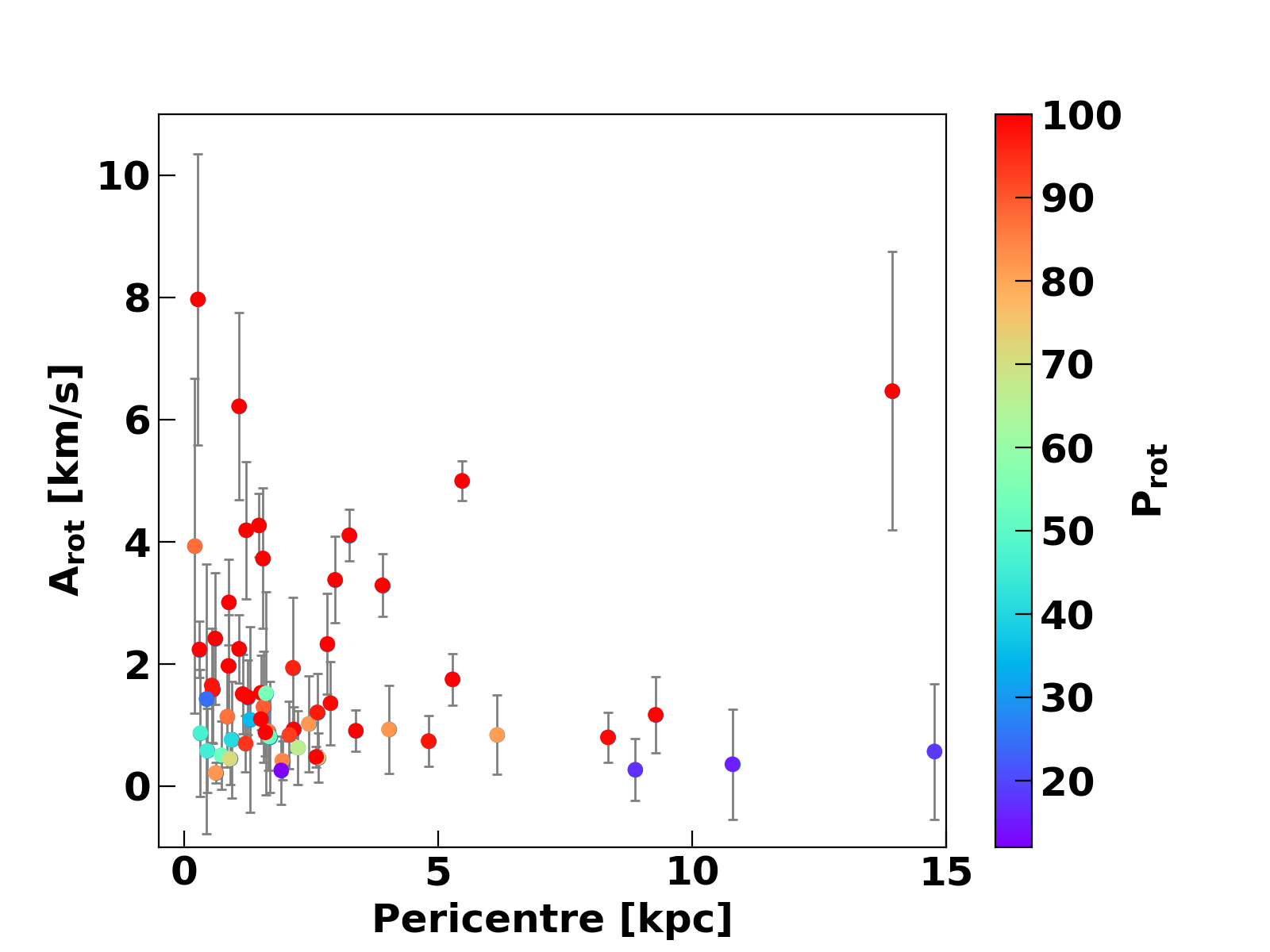}
\caption{Rotational velocity as a function of average pericentre distance over the past 500 Myr for a sample with measured rotational velocities. Colour coding shows the probability that the cluster is rotating. Rotational magnitudes (with their error bars) and probabilities are taken from \citet{Sollima2019}, with orbital parameters from \citet{Baumgardt2021}.}
\label{fig:observed-rot}
\end{figure}

Another prediction, independent of the second stellar generation, is that the external potential can spin up initially isotropic clusters for GCs on long radial orbits with high eccentricities and with small pericentre distances. Fig.~\ref{fig:observed-rot} shows the rotational velocities of clusters as a function of pericentre distance for a sample from \citet{Sollima2019} where the authors measured rotation and assigned probabilities of rotation. Although the figure does not show a strong trend, it suggests that clusters identified as rotating tend to exhibit higher rotational velocities at smaller pericentre distances. It is important to note that our prediction is valid only for initially isotropic clusters. Thus, clusters with high rotation but large pericentre distances may retain their intrinsic rotation rather than acquire it from the influence of the external potential.  

\section{Conclusions and discussion} \label{sec:conc}

We performed a set of direct $N$-body simulations of globular cluster with two stellar generations. We put these clusters on three types of orbits within the Galaxy (TB, LR and CR) which represent the reconstructed orbits of MW globular clusters. We evolved these clusters backward in time to a lookback time of 8 Gyr in the time varying potential of the TNG simulation for a Milky Way like galaxy. We then evolved the systems forward in time with stellar evolution, and introduced a disk of stars representing the second stellar generation after evolving the clusters for 2 Gyr. 

To understand the role of a time-varying Galactic potential, we compared these simulations with control runs that featured a static potential and also compared the results with the isolated cluster evolution. This allowed us to investigate the combined effects of internal evolution and external tidal forces on the structure and dynamics of the combined two stellar generations.

We found that generally the tidal stripping mass loss of the clusters, depends primarily on the orbit type, in particular- GCs on tube and radial orbits lead to higher mass loss in both static and time variable potentials. For the static potential, due to the higher mass of the Galaxy at early times, we observe a significantly higher mass loss of the GCs, independently of the presence or absence of the second stellar generation. This dependence is most prominent for the tube and radial orbits. This effect also exists for circular GC orbits but is less significant. 

The addition of the second generation to the star cluster (at least in the present form, i.e. $\sim$10\% of the current system mass at 2 Gyr, see in Table.~\ref{tab:mass-rat}) does not change significantly the mass loss behaviour of the system (see Fig.~\ref{fig:mass-loss}). The star clusters' main dynamical parameters, such as its tidal mass and half mass radius, also behave similarly with and without this second generation (see Fig.~\ref{fig:mtid-rhm-rtid}). 

The second generation stars supply the system with additional number of high mass remnants (BH and NS) which are concentrated in the central few pc region of the star cluster (see Figures in Appendix~\ref{app:global}). 

The initially disk-like second stellar population tends to become spherical rapidly well within the two-body relaxation timescale in all of our models (Fig.~\ref{fig:triax}), which is expected from vector resonant relaxation \citep{Meiron_Kocsis2019}. The principal axes of the clusters' geometry in physical space show that the clusters become almost spherical within $\sim$ 500 Myr (i.e. $5\times$ the VRR timescale and $\frac12$ the two-body relaxation timescale). Note that since the VRR time scales with star counts as $N^{1/2}$ while the two-body time scales with $N$, the mixing time will always be much less than the Hubble time even for the largest GCs well beyond those simulated in this work which may not yet be relaxed in terms of their energy distribution. In Eq.~\ref{eq:2-body} we showed that vector resonant relaxation is a factor 10 shorter for our system and a factor 35 for $10^6\msun$ clusters compared to 2-body relaxation. Thus, in terms of the clusters' principal axes in physical space, complete mixing is expected to be a very robust prediction. However, the simulations show that the total specific angular momentum preserves the rotational imprint of the second generation component for several Gyr, Fig.~\ref{fig:mom-tot-spec}. The rotational imprints for the second component cannot be distinguished from that of the first component for the long radial orbits with small periapsis distances because in this case the first component is spun up by the external potential. Thus, the external potential imparts a significant spin to the cluster in cases where the periapsis distance is only a few kpc, and the magnitude of the spin is expected to decrease with periapsis distance. At large distances the cluster's specific angular momentum carries memory of the initial conditions. 

Our results align with observations showing that clusters on circular or high-pericentre orbits tend to retain the rotational imprint of the second stellar generation ({\tt g2}), as seen in studies like \citet{Dalessandro2024}, where {\tt g2} stars are found to rotate faster than {\tt g1} stars in such clusters. In contrast, \citet{Leitinger2024} report no consistent trend of higher rotation in {\tt g2} stars across a broader sample, with exceptions in specific cases like NGC 104. For overall cluster rotation, our simulations predict that GCs on eccentric or low-pericentre orbits may acquire angular momentum from the Galactic potential. Starting from an initially isotropic distribution with nearly zero angular momentum, clusters on these orbits develop a net spin due to tidal effects. Observations by \citet{Sollima2019} support this, showing a tendency for clusters with smaller pericentre distances to exhibit higher rotational velocities, although this trend lacks statistical significance. To make more robust predictions about the acquisition of angular momentum from the external potential, further simulations with a larger sample of clusters on a range of orbits are needed, along with more detailed observational data to confirm potential correlations between rotation and orbital parameters in multiple population clusters.

The rotational imprints of {\tt g2} stars may be even more prominent in some Milky Way Gcs than in our simulations. The initial mass fraction of {\tt g2} stars in our models was set conservatively to $\sim$ 5\% of the original initial mass of {\tt g1} stars, a reasonable assumption given their formation from AGB ejecta \citep{D'Ercole+2008}. However, present-day clusters show higher fractions of second-generation stars \citep{Milone_Marino2022}, probably due to the preferential loss of {\tt g1} stars through tidal stripping. Furthermore, in our simulations, the {\tt g2} stars were set to be distributed initially on the same spatial scale as the {\tt g1} stars, which is consistent with some observed clusters, such as NGC 6362 \citep{Dalessandro2014}. However, many studies suggest that the {\tt g2} stars are more centrally concentrated \citep{Lardo2011, Bastian_Lardo2018}. 

On the other hand, we have introduced the first generation stellar population at a lookback time of 8 Gyr to avoid uncertainties in the external potential at early times, but the globular cluster formation may have peaked much earlier at -12 Gyr \citep[e.g.][]{ElBardy_2019}. Furthermore, we introduced the second stellar generation at 2 Gyr after the first generation to allow {\tt g1} stars to relax. However, observations suggest that second-generation stars may have formed much earlier, likely within 100 Myr of the first generation \citep{Marino2012}. An earlier formation and a smaller age gap could allow for more mixing to take place between the stellar populations during their evolution to the present.

Taking these considerations into account would improve the physical realism of our models. Additionally, more models are needed to describe clusters of different masses, concentrations, and radial locations within the Galaxy. Even more importantly, models may also be improved by incorporating the formation and evolution of hard binaries, possibly using regularised codes such as \textsc{NBODY6++GPU} \citep{Wang2015} or \textsc{PETAR} \citep{Wang2020}. Given that we have demonstrated in this work that the time-varying potential plays an important role in the mass-loss evolution of star clusters, it may also be important to incorporate more realistic time-varying potentials from higher-resolution cosmological simulations such as TNG-50 \citep{Nelson2019}, as these data become publicly available. Future models with these improvements will more accurately describe the dynamics of multiple stellar populations in GCs, thereby enhancing our understanding of their formation history and dynamical evolution.







\begin{acknowledgements}


The work of TP and MI was funded by the Science Committee of the Ministry of Education and Science of the Republic of Kazakhstan (Grant~No.~AP22787256). 

PB and MI thank the support from the special program of the Polish Academy of Sciences and the U.S. National Academy of Sciences under the Long-term program to support Ukrainian research teams grant No.~PAN.BFB.S.BWZ.329.022.2023.

PB acknowledges the support by the National Science Foundation of China under grant NSFC~No.~12473017.

This work was supported by the UK Science and Technology Facilities Council Grant Number ST/W000903/1 (to B.K.).

\end{acknowledgements}

\bibliographystyle{aa}  
\bibliography{gc-disk}   

\begin{appendix}
\section{Visualisation of the GCs' orbits}\label{app:det-orb}
We present orbital evolution for the GCs with tube and long radial orbits in {\tt \#411321} TNG external and {\tt FIX} potentials. The orbital evolution is presented in three planes, ($X-Y$, $X-Z$, and $R-Z$, where $R$ is the planar Galactocentric radius). The total time of integration is a 8 Gyr lookback time and is shown by the coloured line.

\begin{figure*}[ht]
\centering
\includegraphics[width=0.90\linewidth]{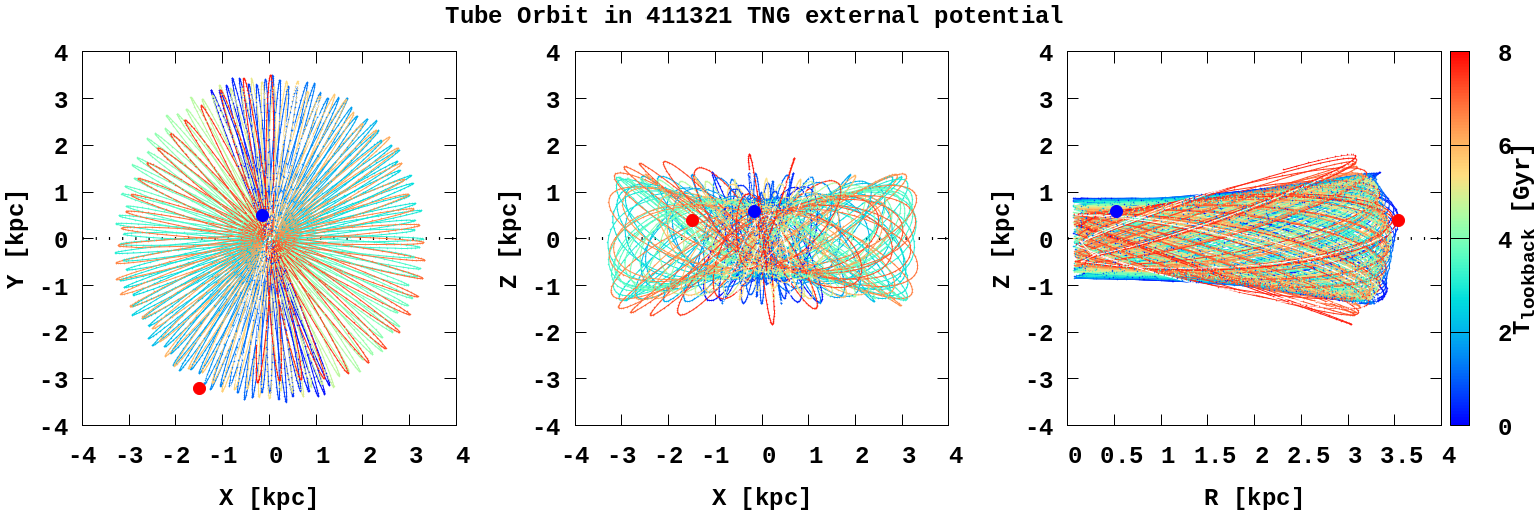}
\includegraphics[width=0.90\linewidth]{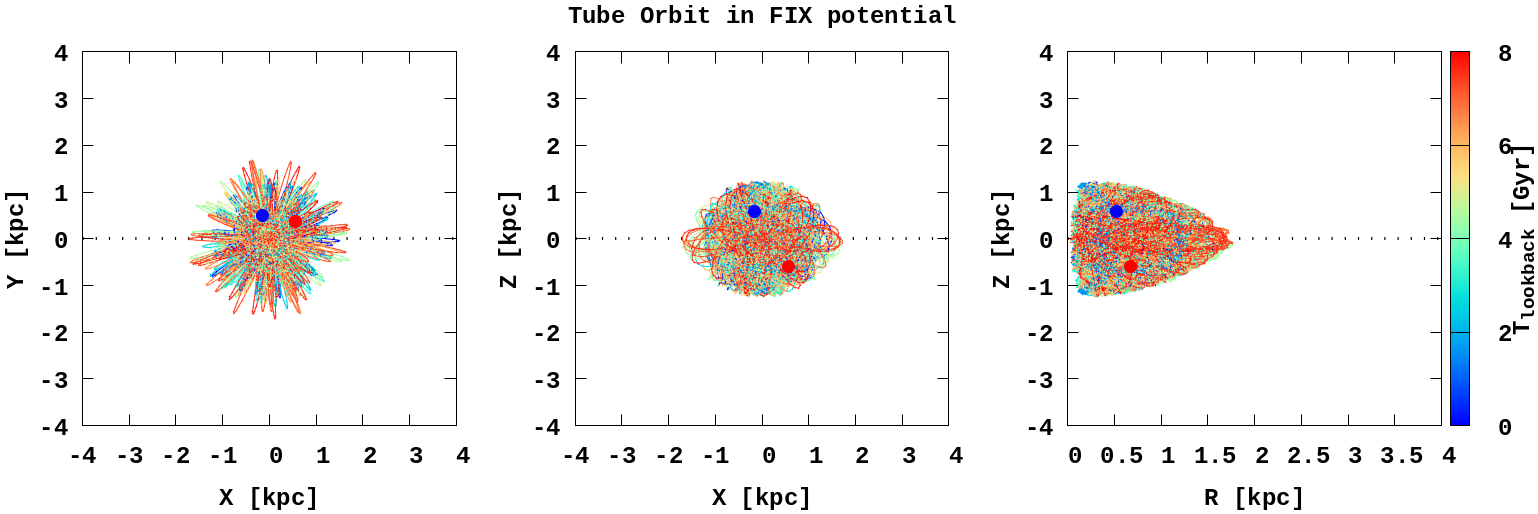}
\caption{
The simulated GC orbit for the GC on a tube orbit (NGC 6401)  in the {\tt 411321 TNG} and {\tt FIX} potentials.}
\label{fig:tube-orb}
\end{figure*}

\begin{figure*}[ht]
\centering
\includegraphics[width=0.90\linewidth]{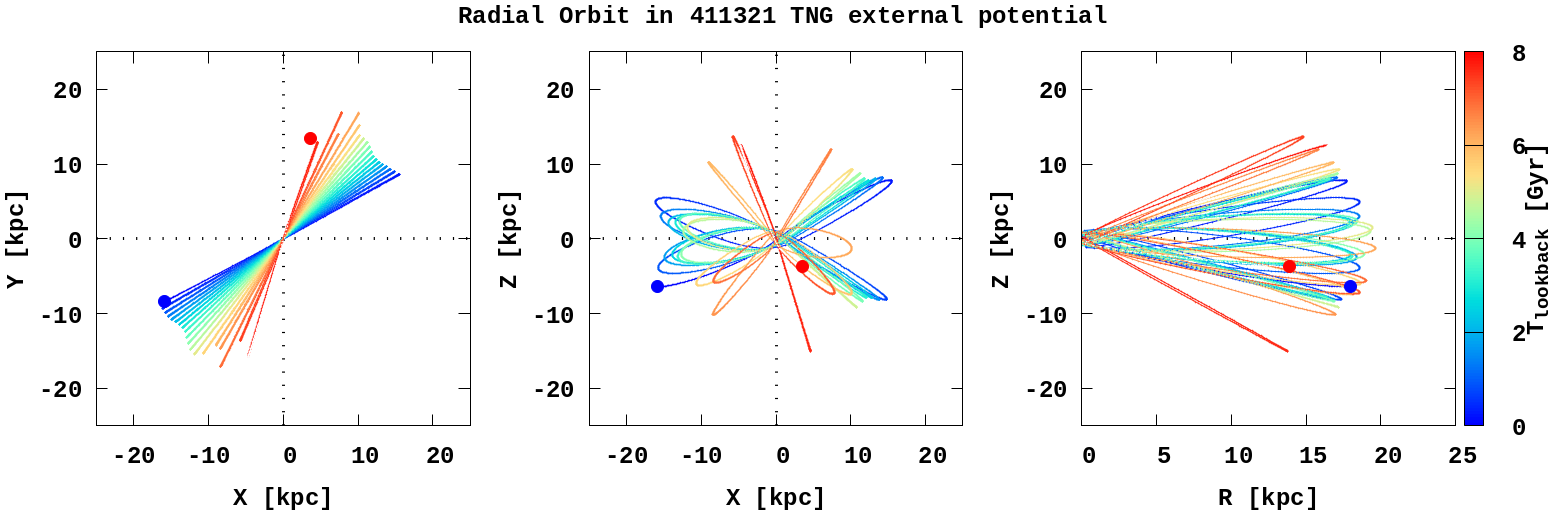}
\includegraphics[width=0.90\linewidth]{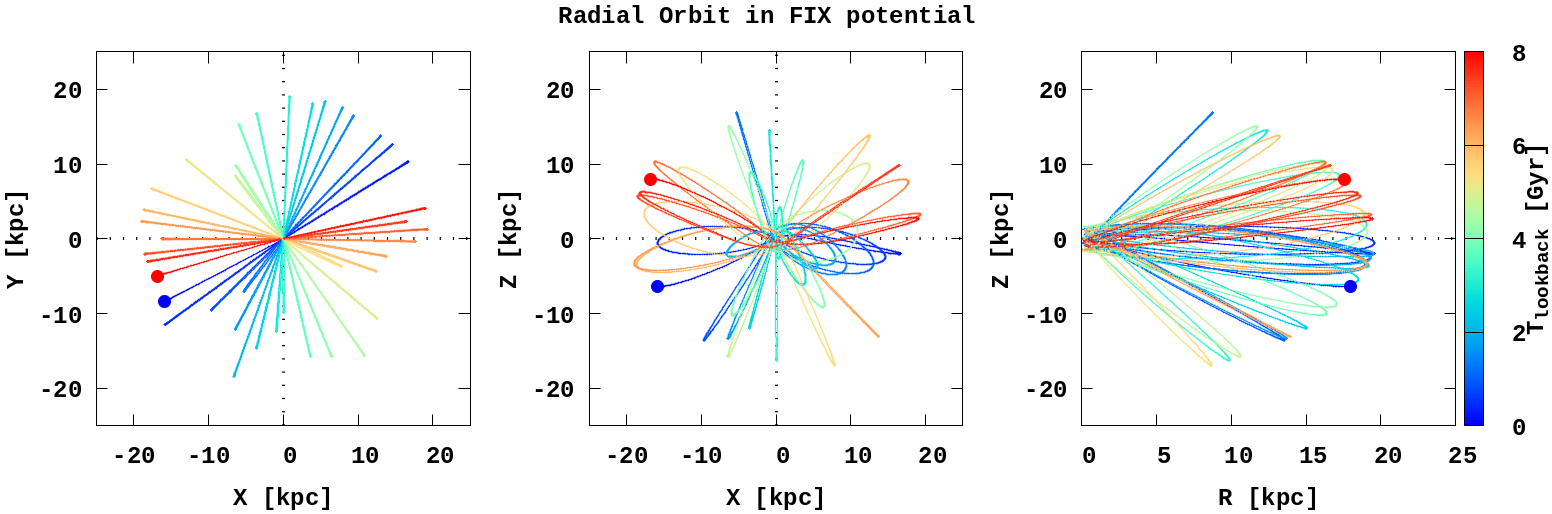}
\caption{The simulated GC orbit for the GC on long radial (LR) orbit (NGC 1904) in the {\tt 411321 TNG} and {\tt FIX} potentials.}
\label{fig:lr-orb}
\end{figure*}

\section{Large and local scale kinematics} \label{app:global}

In Figures below we present the volume density distribution of the GCs for both generations. The cluster tidal radius (or ``Jacobi'', or ``King'' radius) was calculated based on the numerical iteration of the $M_{\rm tid}$ and $r_{\rm tid}$ values in the equation:
\be
r_{\rm tid} = \left[ \frac{G \cdot M_{\rm tid}}{4\Omega^2-\kappa^2} \right]^{1/3} 
\ee
where $G$, $M_{\rm tid}$, $\Omega$, and $\kappa$ are respectively the gravitational constant, the cluster tidal mass, the circular, and the epicyclic frequencies of a near-circular orbit in Galaxy at the GC current position. For more detail, the reader can follow the papers \citet{King1962} and \citet{Ernst2011}. 

\begin{figure*}[ht]
\centering
\includegraphics[width=0.33\linewidth]{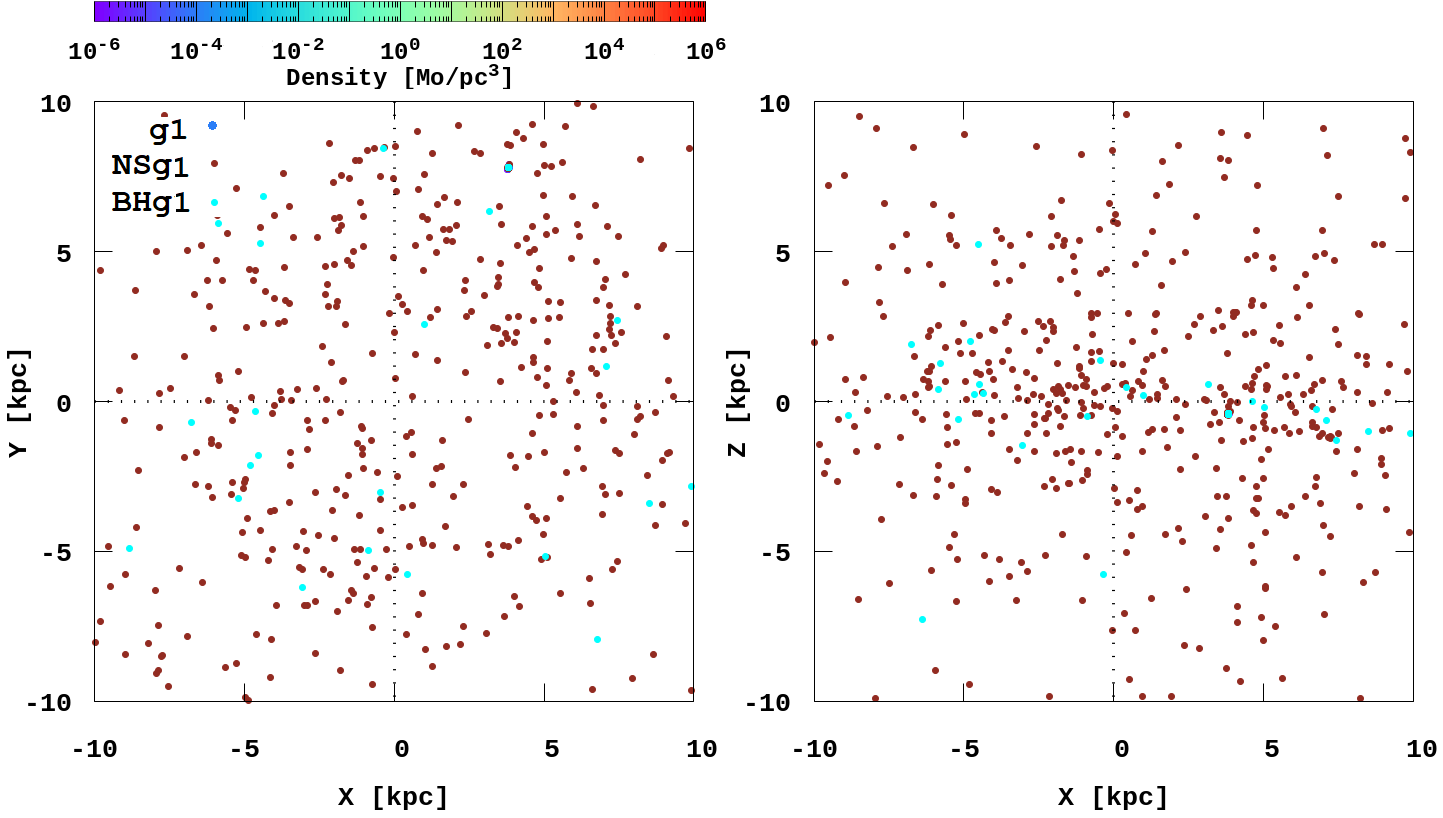}
\includegraphics[width=0.33\linewidth]{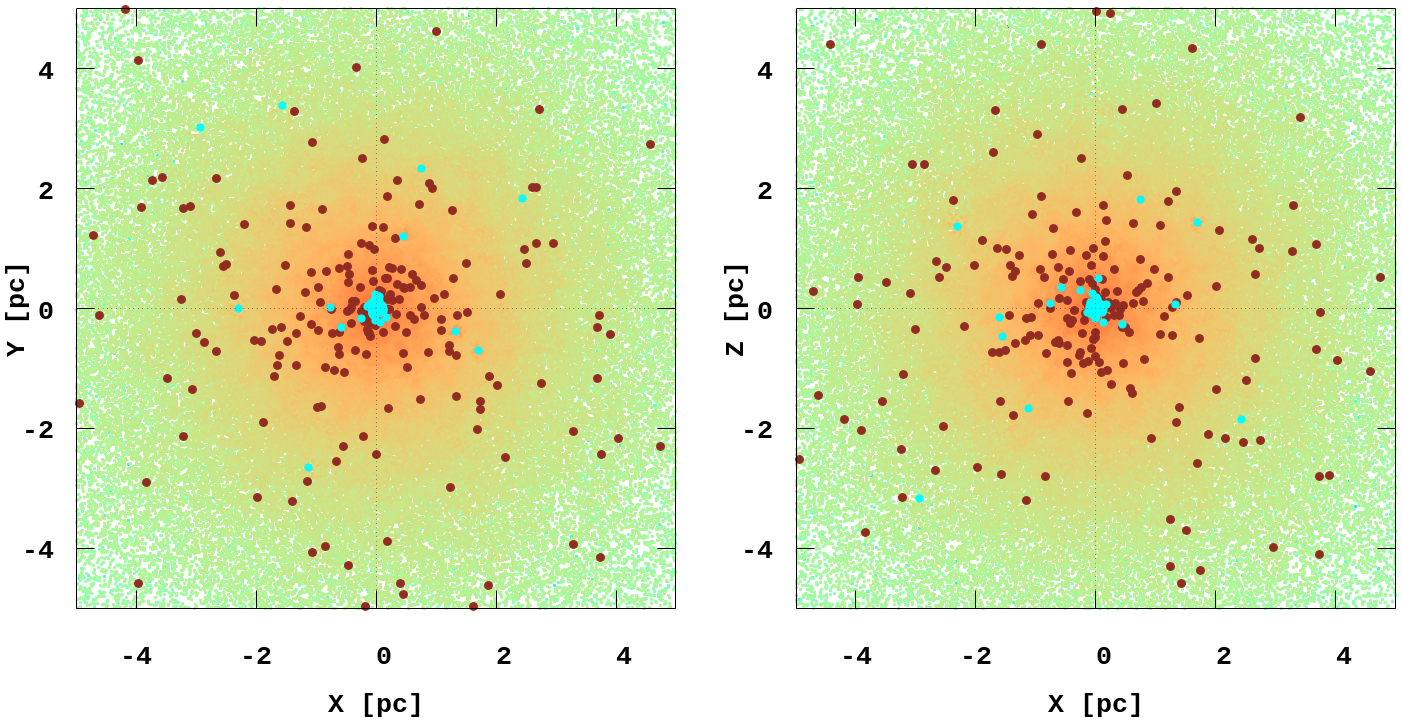}
\includegraphics[width=0.33\linewidth]{pic/cr-tng-den-loc-gc-001640.png}
\includegraphics[width=0.33\linewidth]{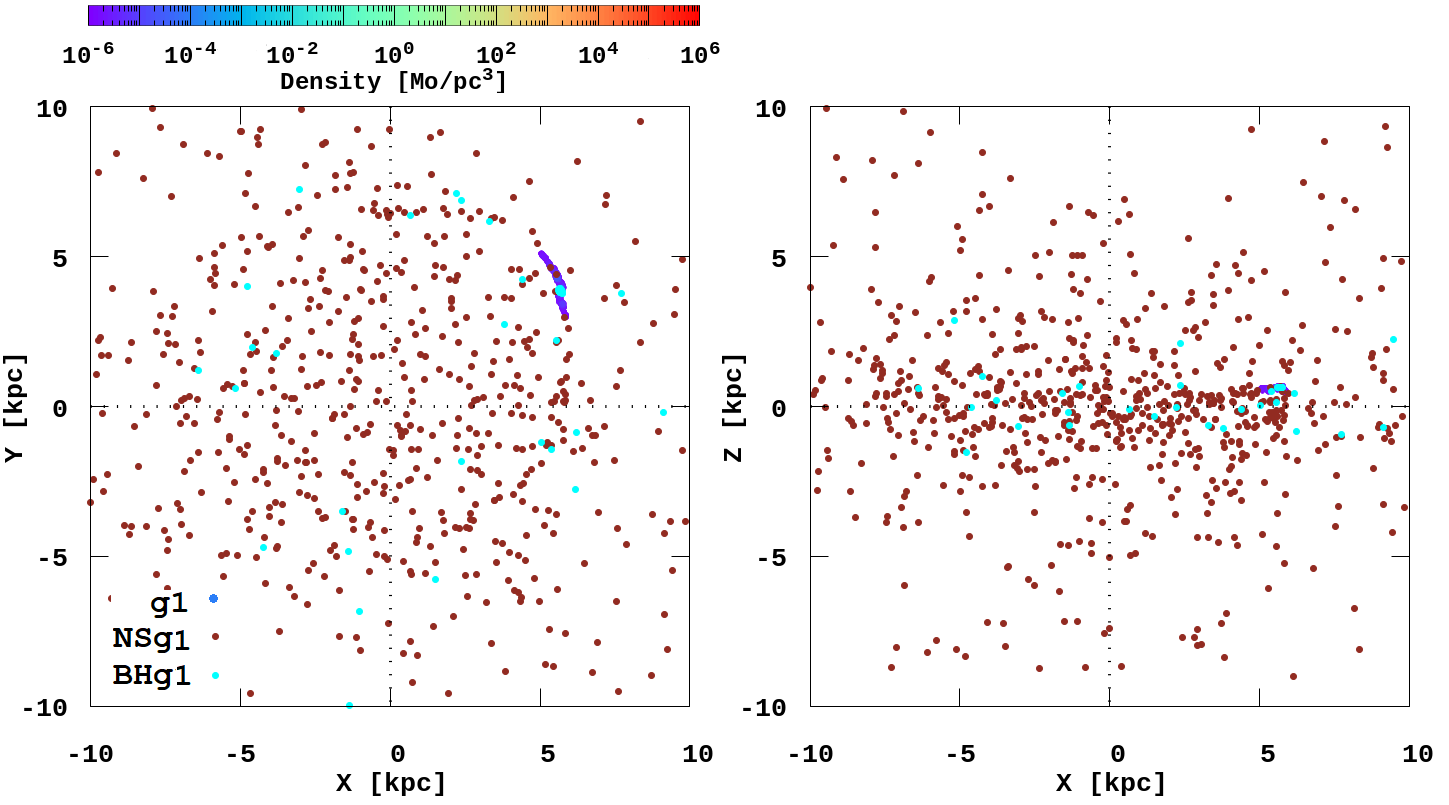}
\includegraphics[width=0.33\linewidth]{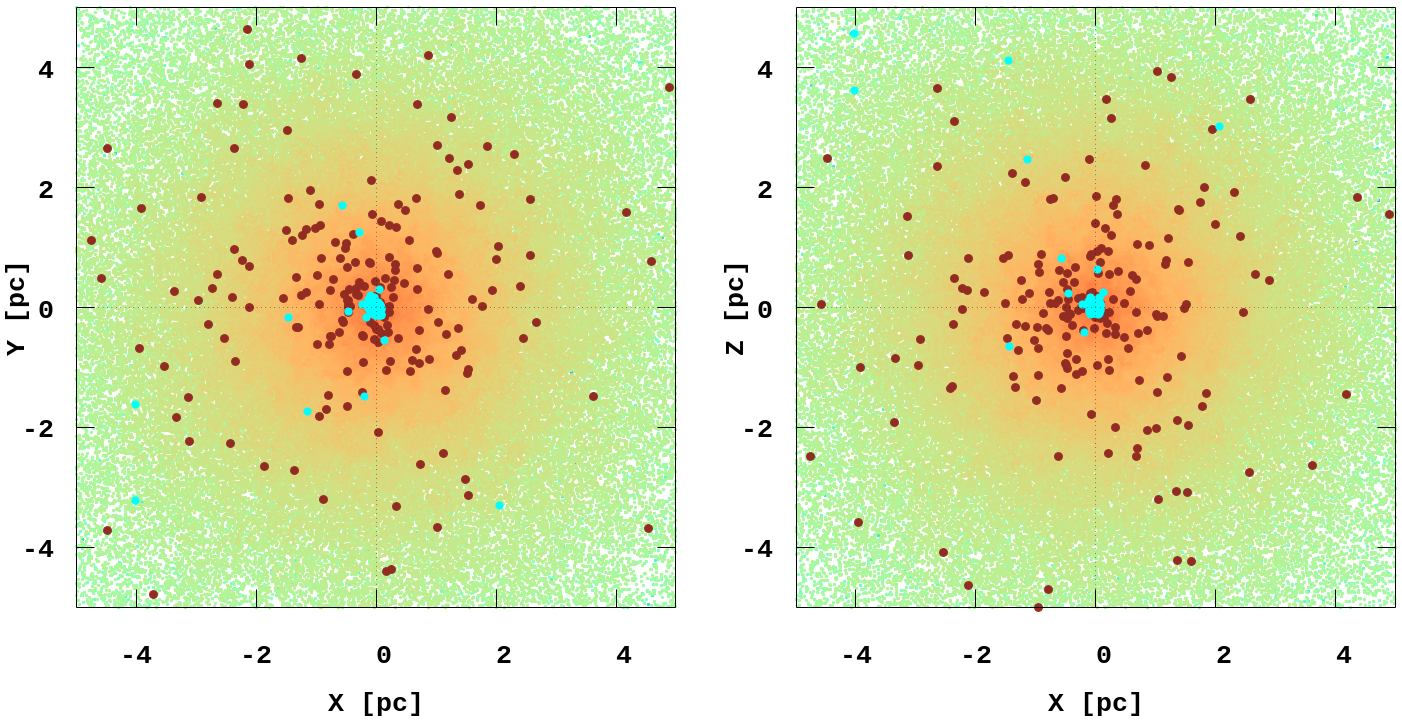}
\includegraphics[width=0.33\linewidth]{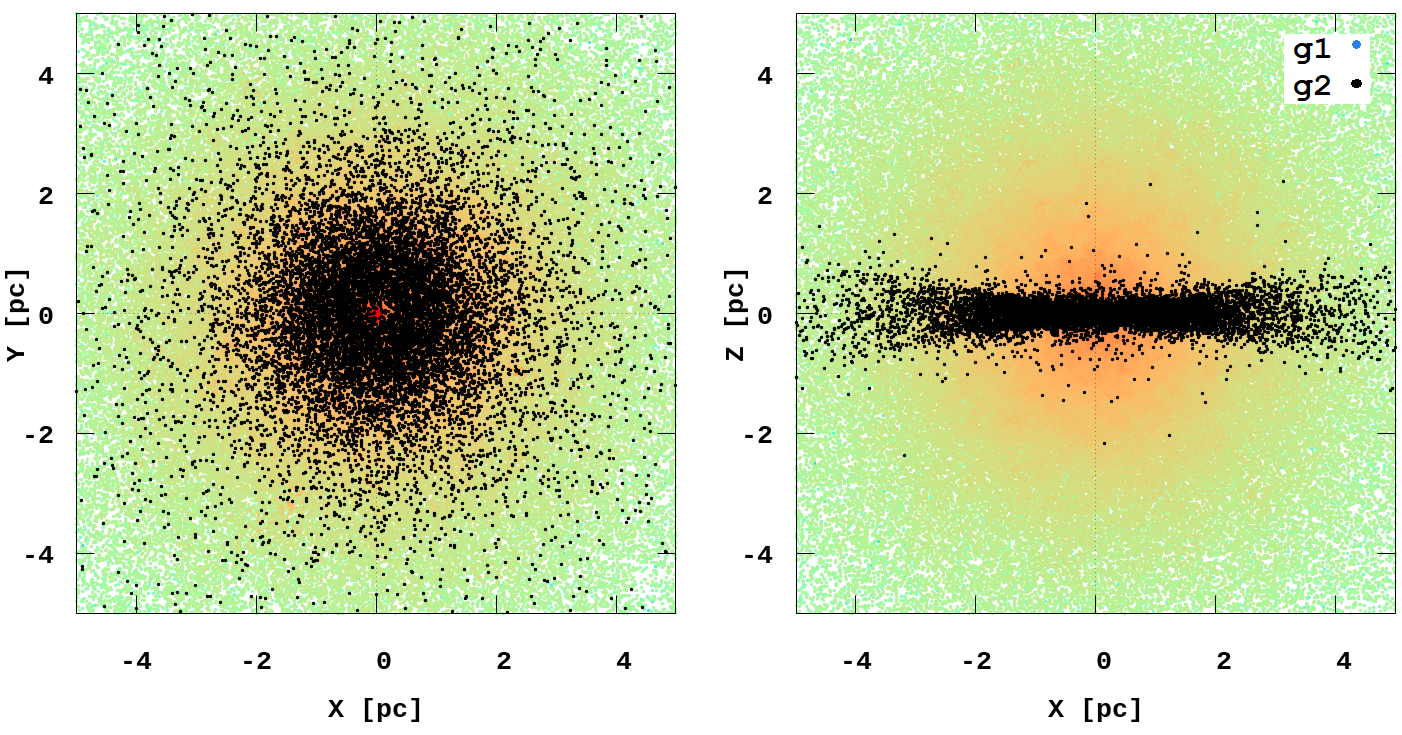}
\caption{GC density distribution at -6.0 Gyr of the dynamical evolution for circular type of orbit. In two \textit{left panels} the GC presented in global Galactocentric coordinates system. Cyan and brown dots -- BH and NS in GC. 
Density of the central GC's core (5$\times$5 pc) is presented in two \textit{middle  panels}. In the two \textit{right panels} black dots represent the spatial distributions of the second generation of stars {\tt g2} in GC. \textit{Upper} and \textit{lower panels} show the density distributions within the GC in the {\tt TNG} and {\tt FIX} potentials, respectively.}
\label{fig:cr-init}
\end{figure*}

\begin{figure*}[ht]
\centering
\includegraphics[width=0.33\linewidth]{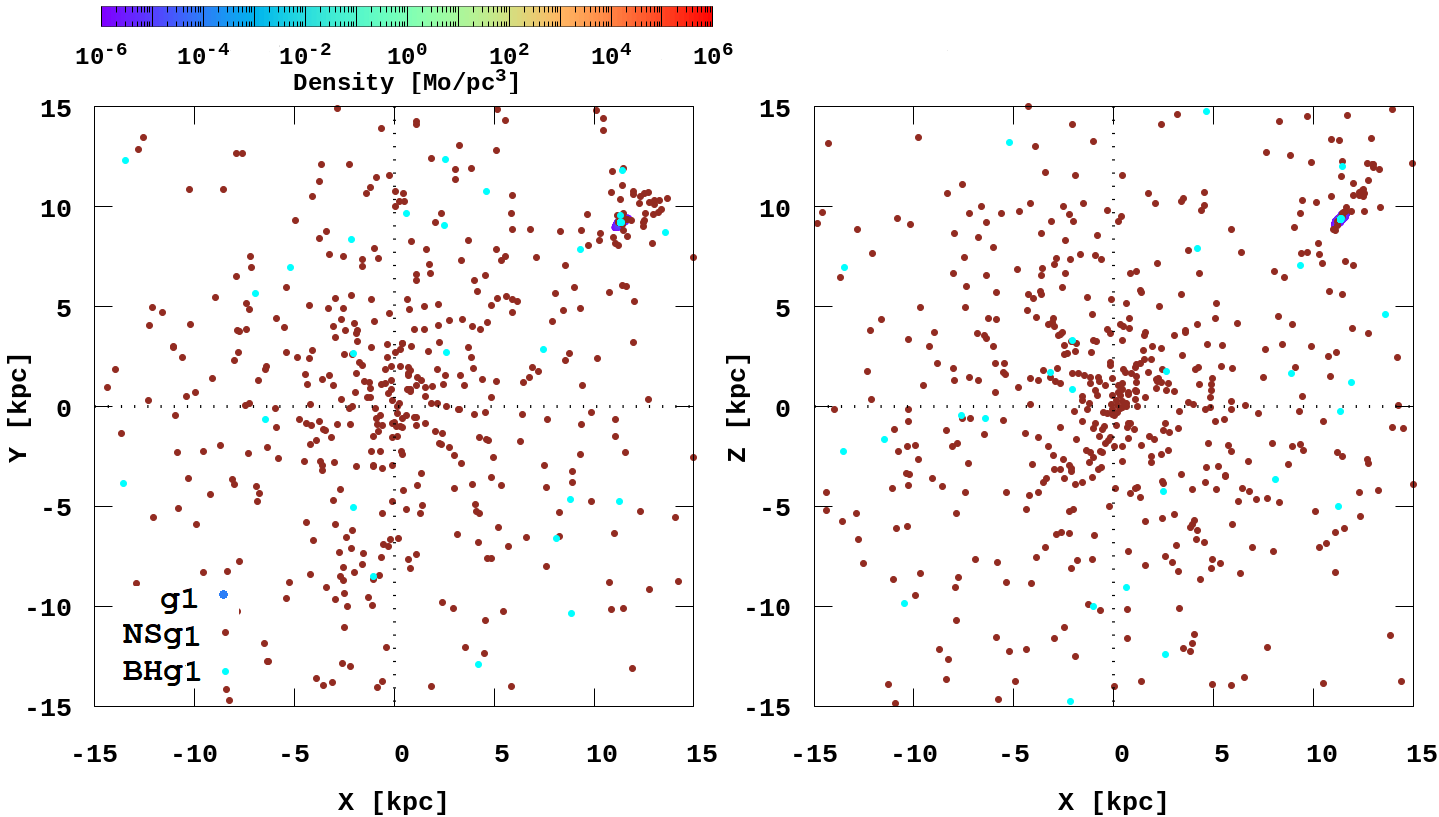}
\includegraphics[width=0.33\linewidth]{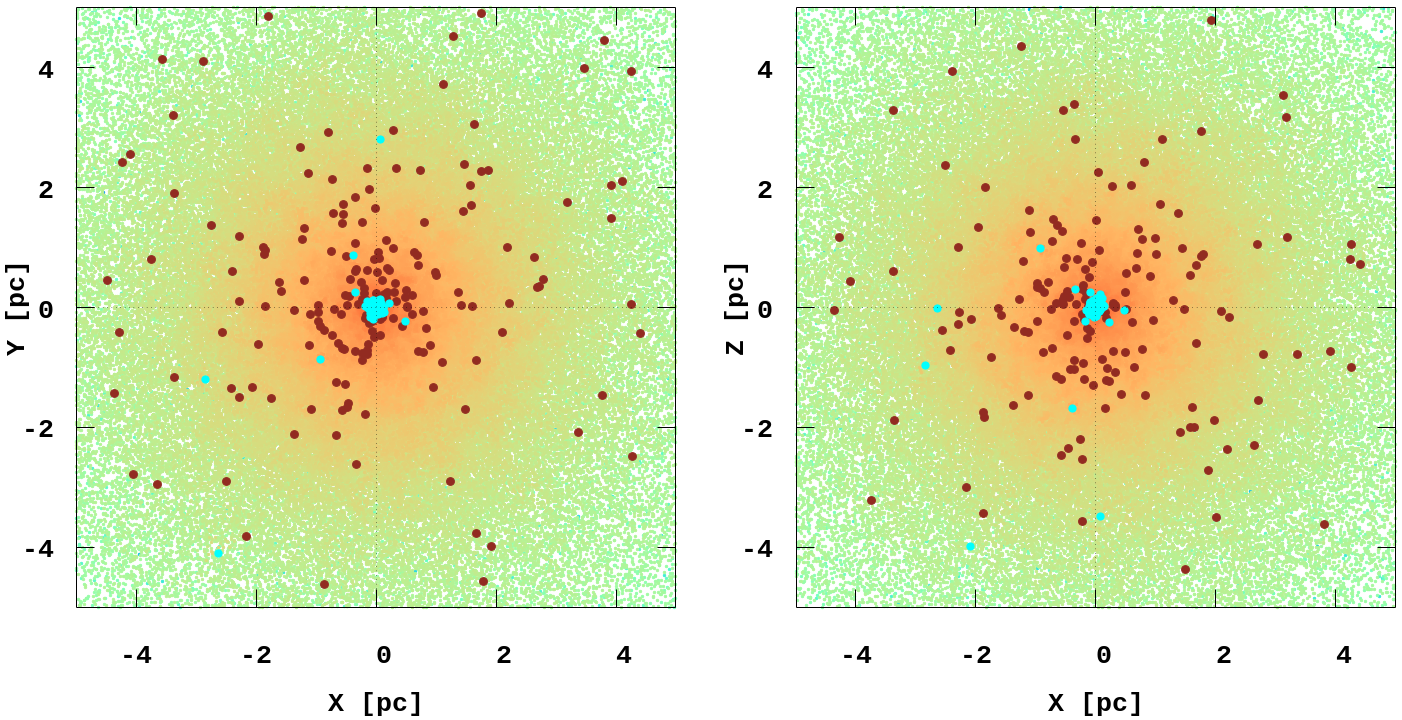}
\includegraphics[width=0.33\linewidth]{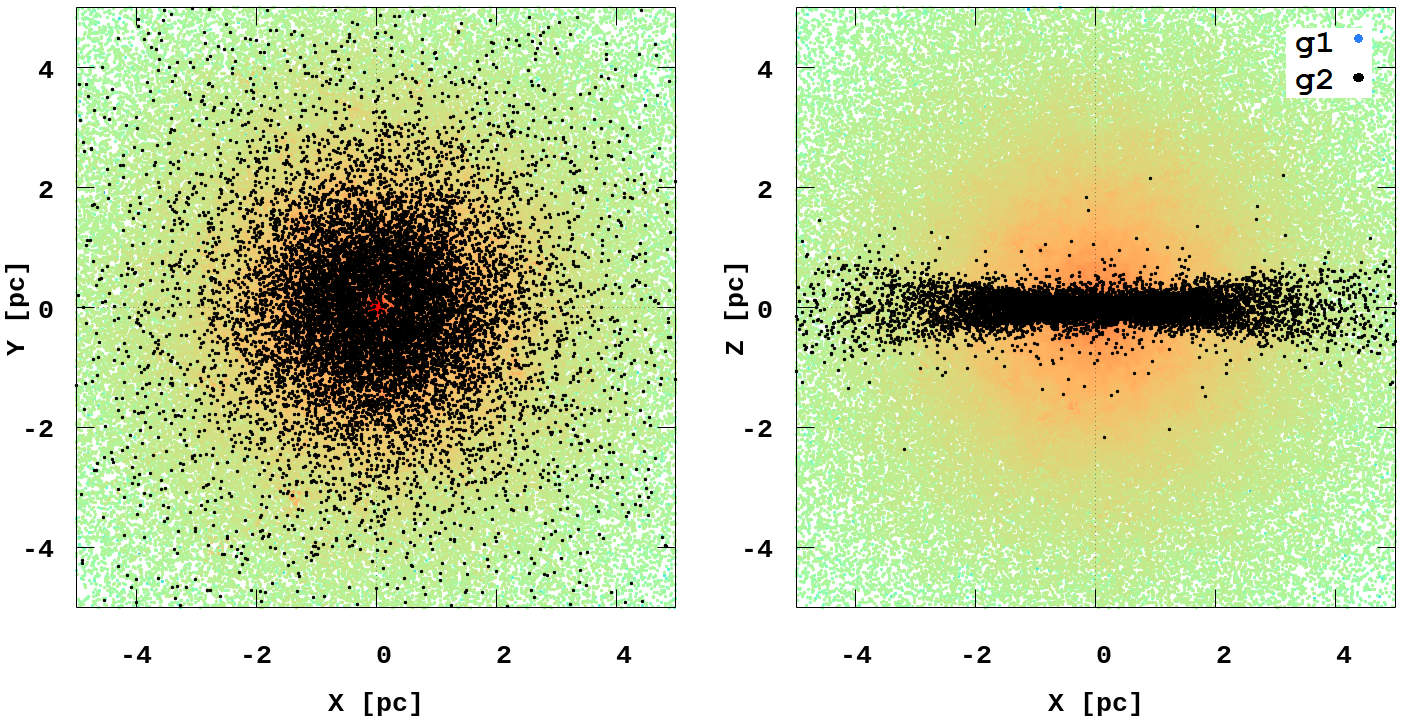}
\includegraphics[width=0.33\linewidth]{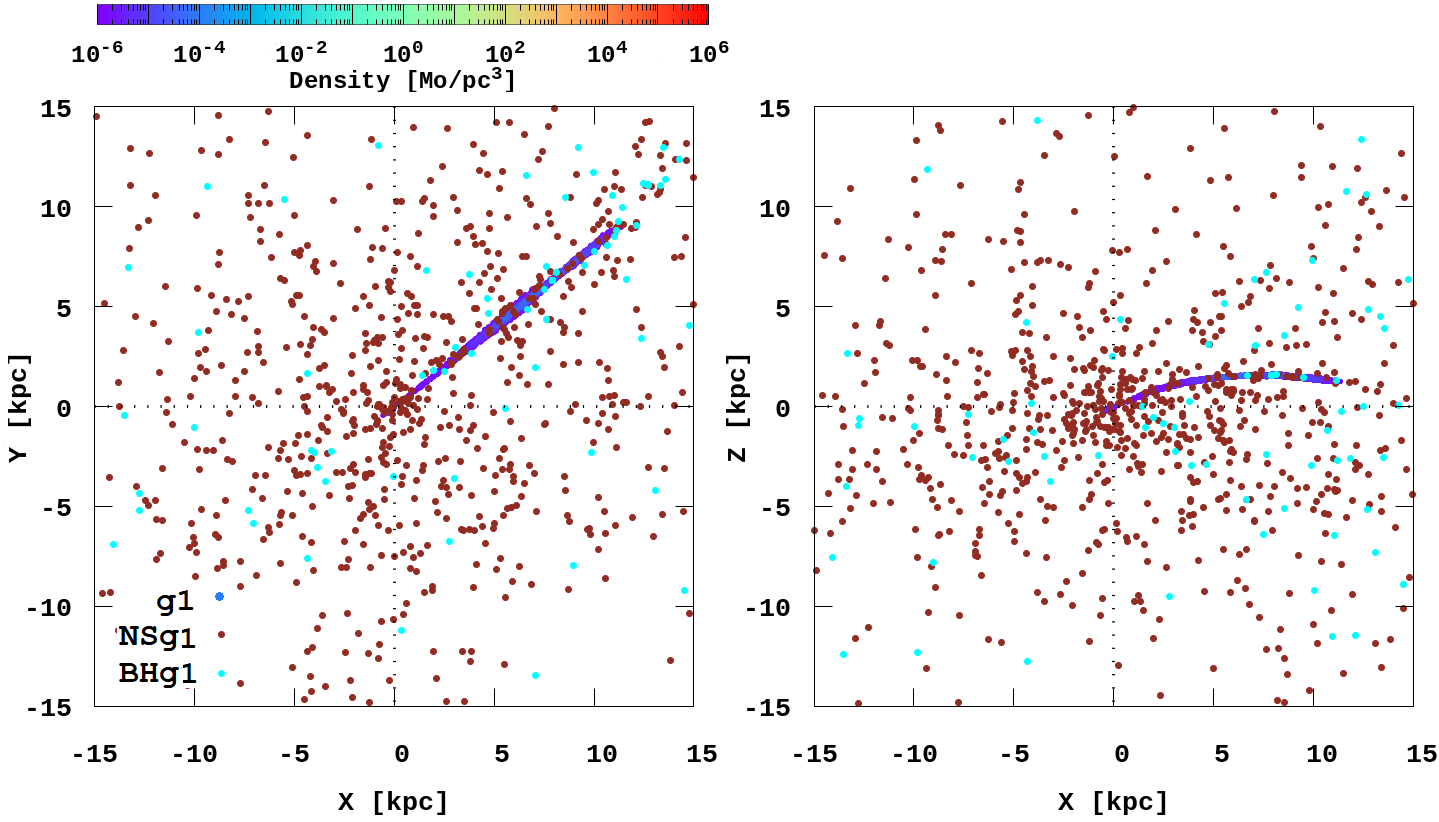}
\includegraphics[width=0.33\linewidth]{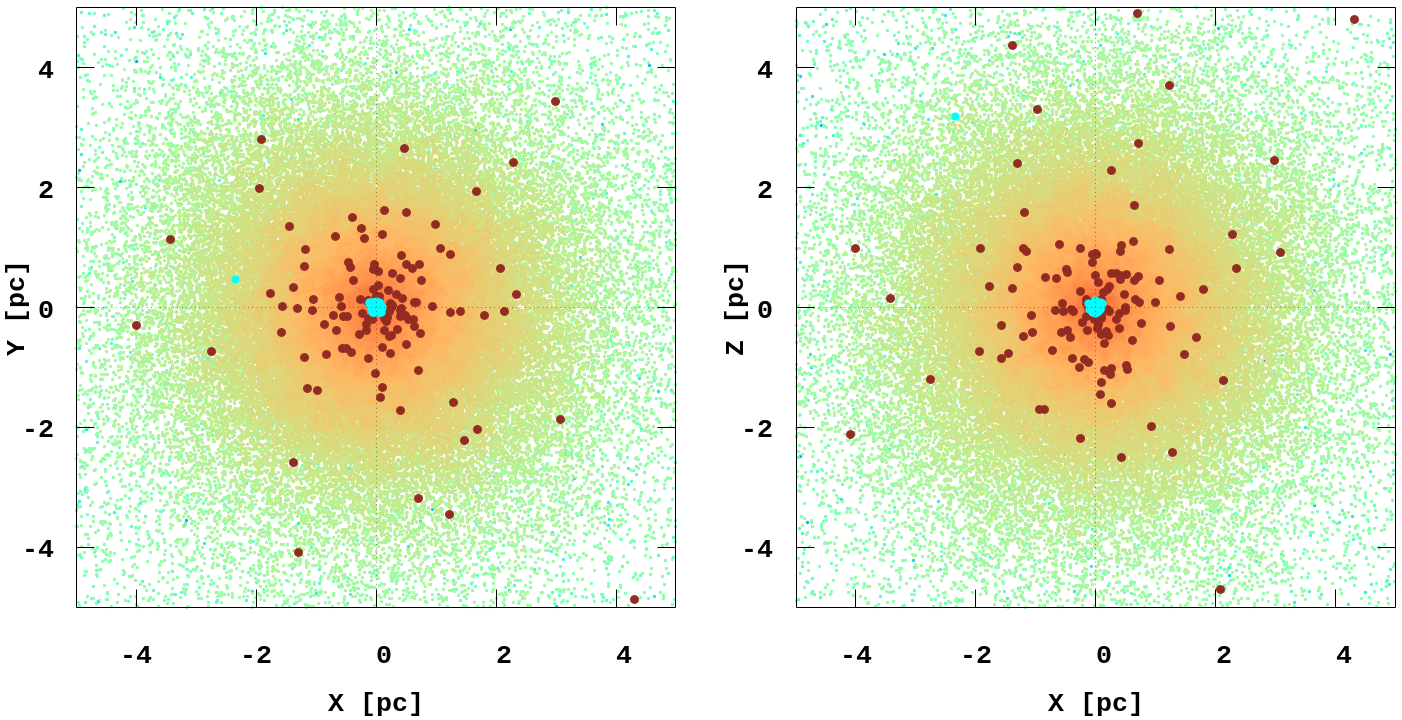}
\includegraphics[width=0.33\linewidth]{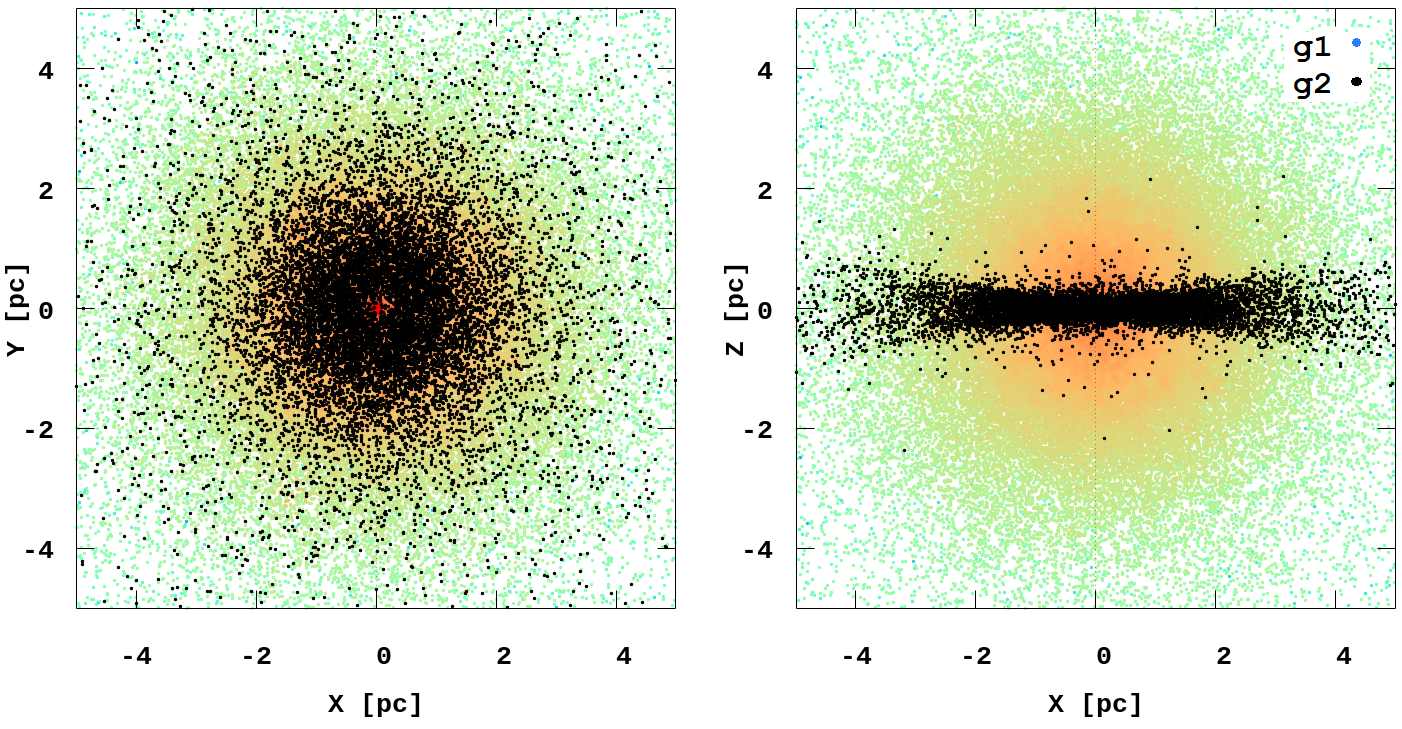}
\caption{Same as Fig. \ref{fig:cr-init}, but for long radial orbit.}
\label{fig:lr-init}
\end{figure*}

In Fig. \ref{fig:cr-init} and \ref{fig:lr-init} we present the GC density distribution with  different box resolution at 2.0 Gyr of the dynamical evolution in for the CR and LR orbits in external potentials {\tt 411321} and {\tt FIX}. On the {\tt g1} generation picture frames we clearly see the wide distribution of BH and NS stellar remnants over a large volume of cluster Galactic orbit. For the BH's we also clearly observe the well separated central population settled due to the dynamical mass segregation and zero or small kick velocities due to the fallback scheme in our stellar evolution routines. 

\begin{figure*}[ht]
\centering
\includegraphics[width=0.45\linewidth]{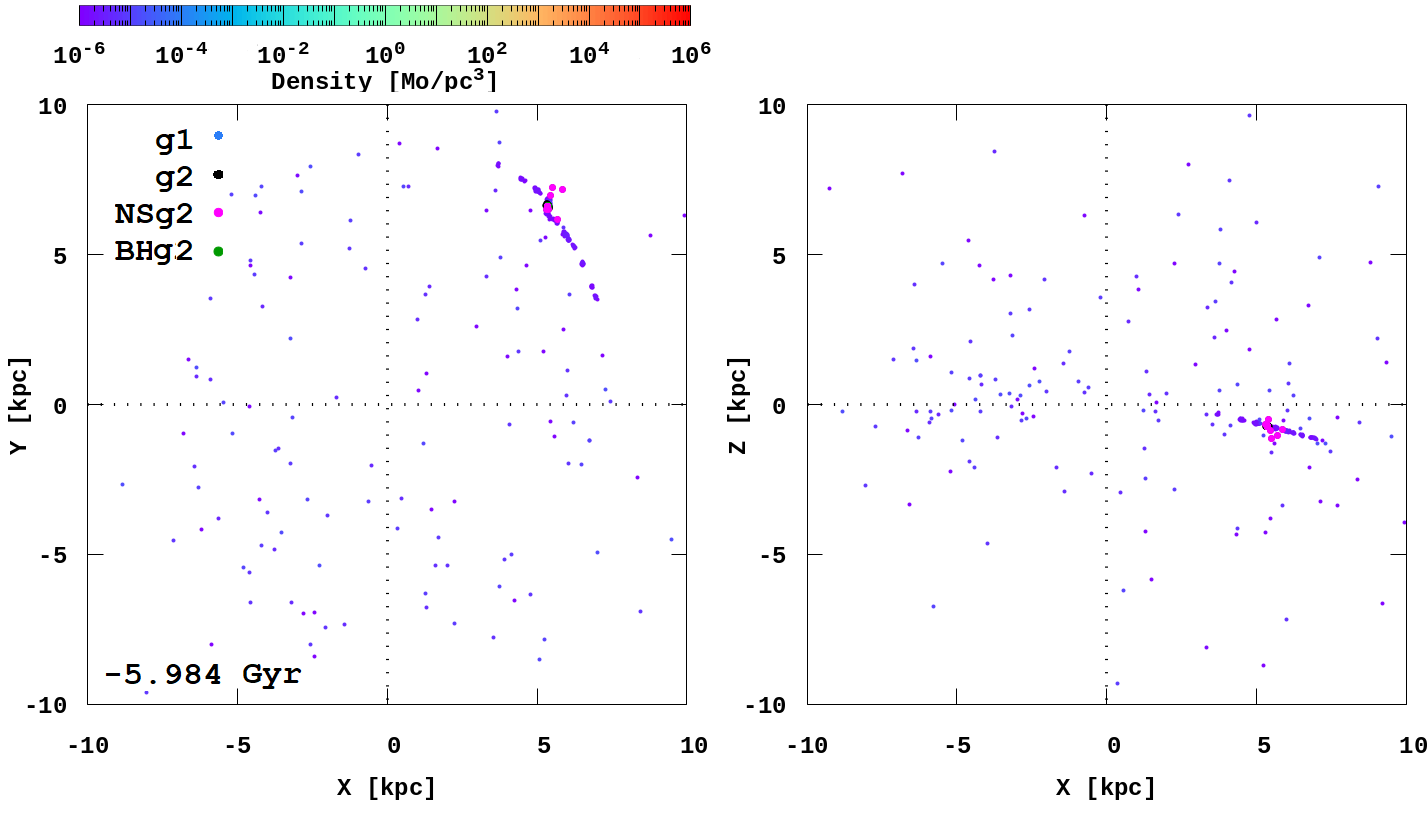}
\includegraphics[width=0.45\linewidth]{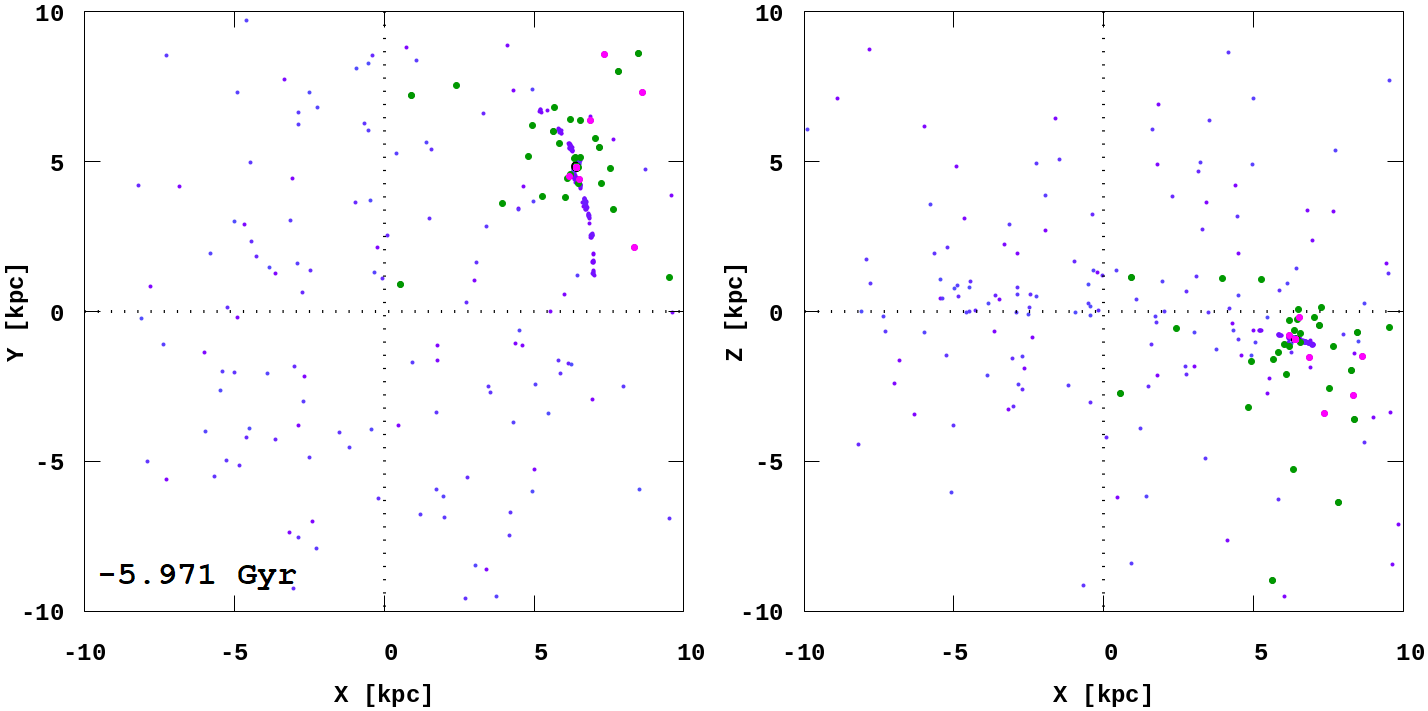}
\includegraphics[width=0.45\linewidth]{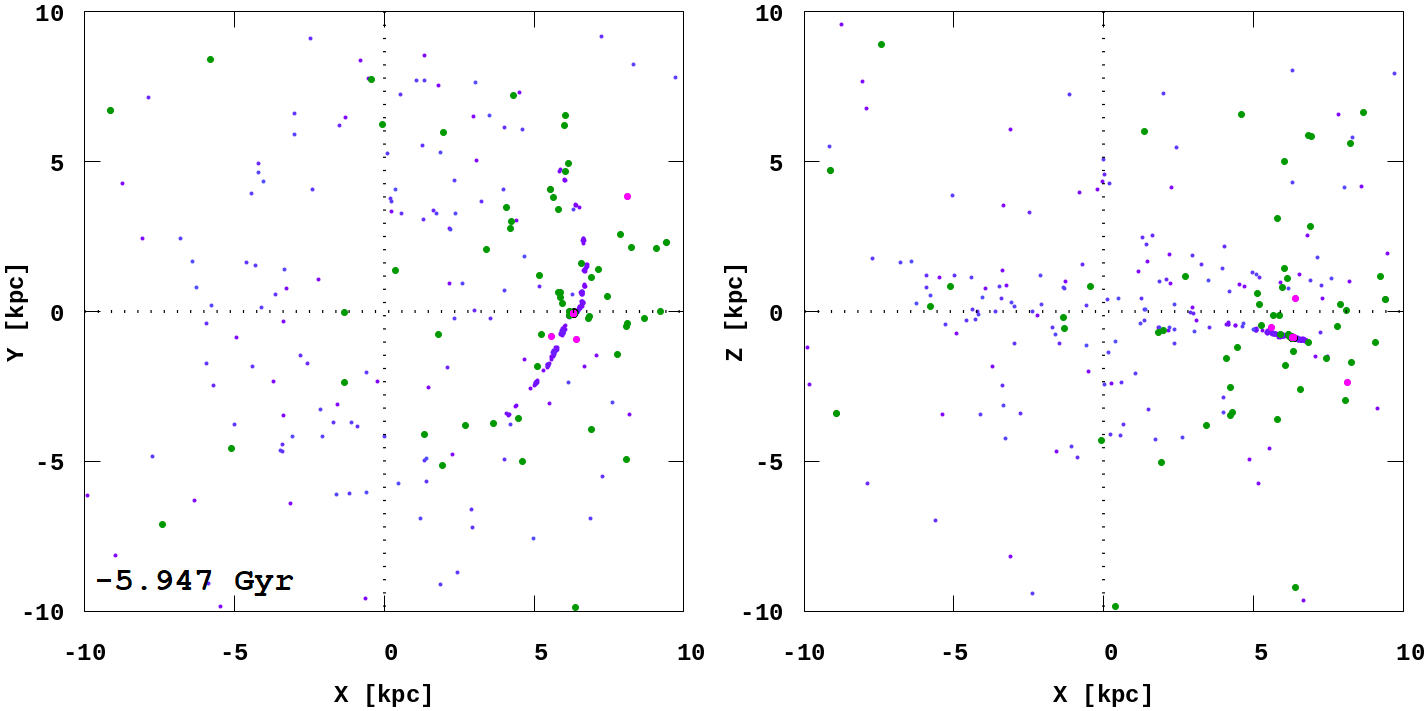}
\includegraphics[width=0.45\linewidth]{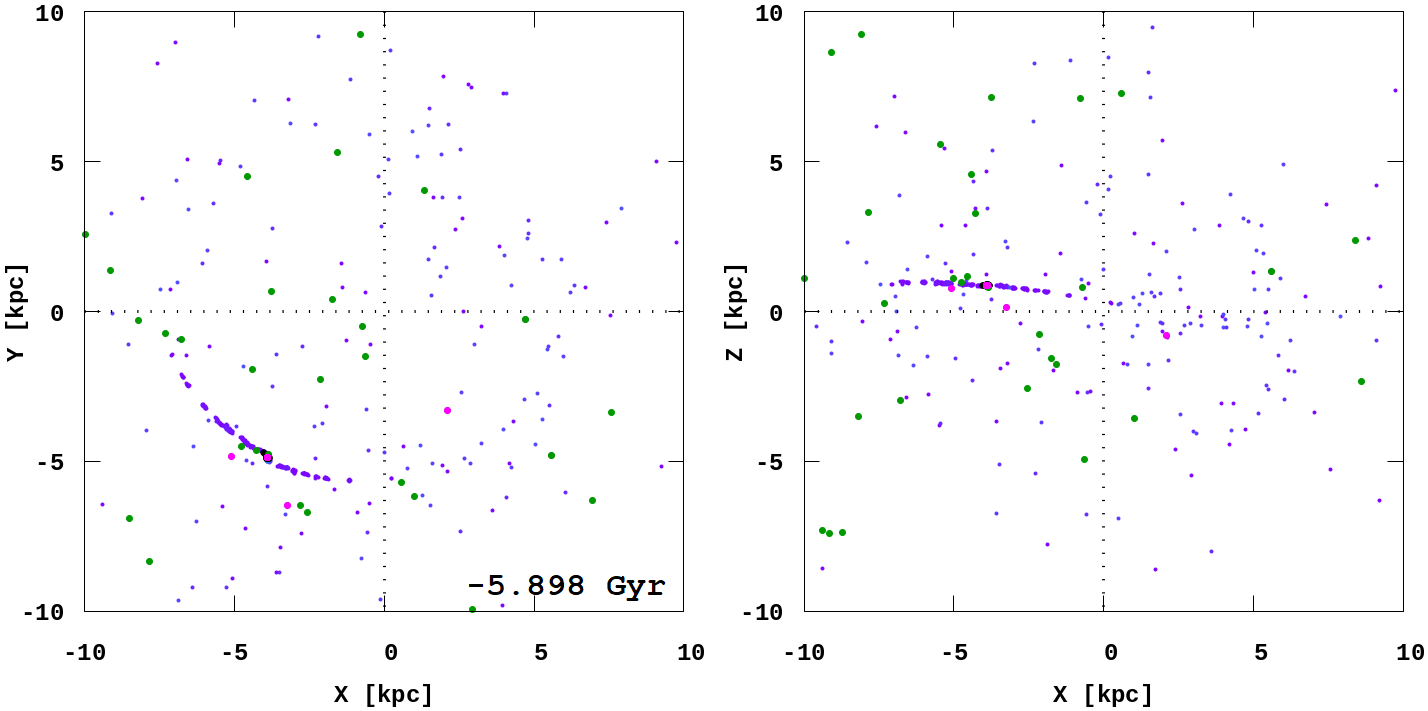}
\includegraphics[width=0.45\linewidth]{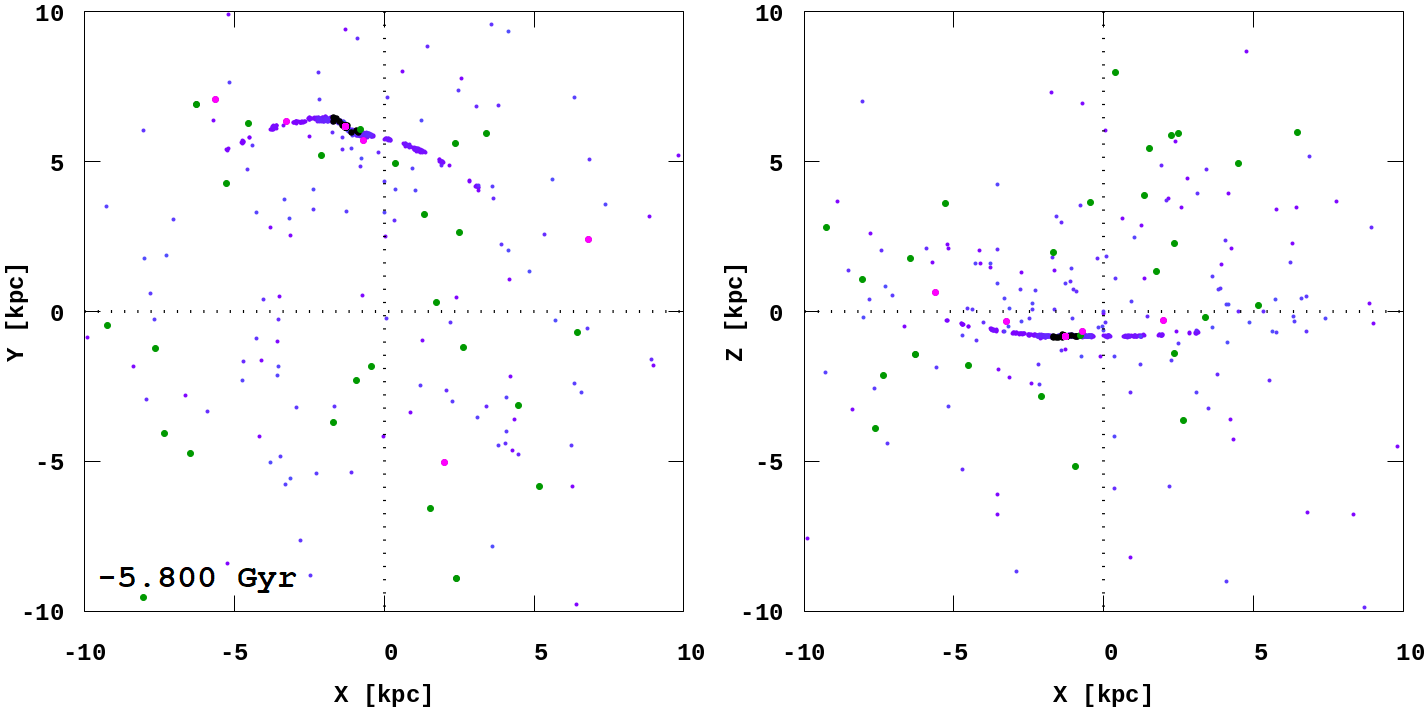}
\includegraphics[width=0.45\linewidth]{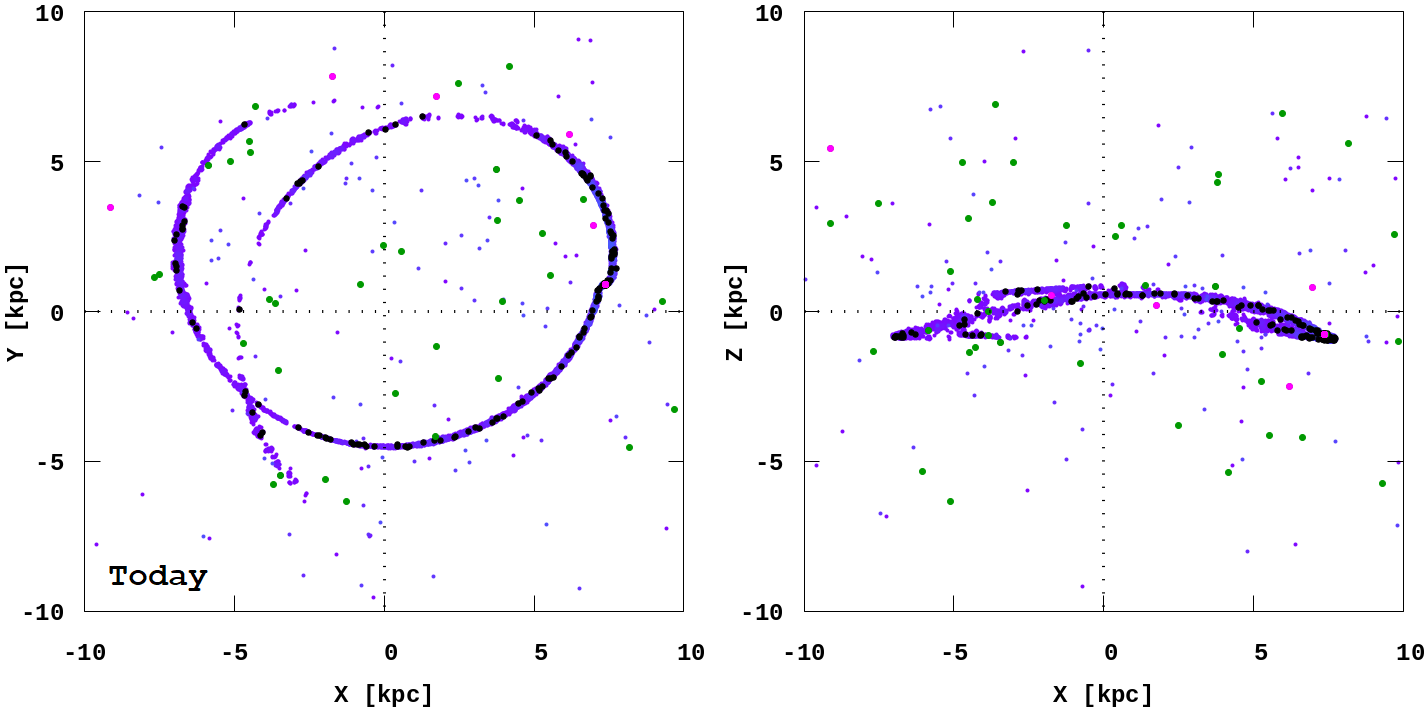}
\caption{The distribution of GC density (palette colour) for the circular type of orbit at different moments in time in TNG potential for both generations of stars {\tt g1+g2}. Black dots represent the distribution of the second generation of stars {\tt g2}, magenta dots -- BH, dark green dots -- NS only from {\tt g2}. The orbital and stellar evolution is presented in two planes -- ($X$, $Y$) and ($X$, $Z$) for lookback time $T = -5.984\,$ Gyr, -5.971 Gyr, -5.947 Gyr, -5.898 Gyr, -5.800 Gyr, and today, \textit{from left to right}.}
\label{fig:cr-glo-tng}
\end{figure*}

\begin{figure*}[ht]
\centering
\includegraphics[width=0.45\linewidth]{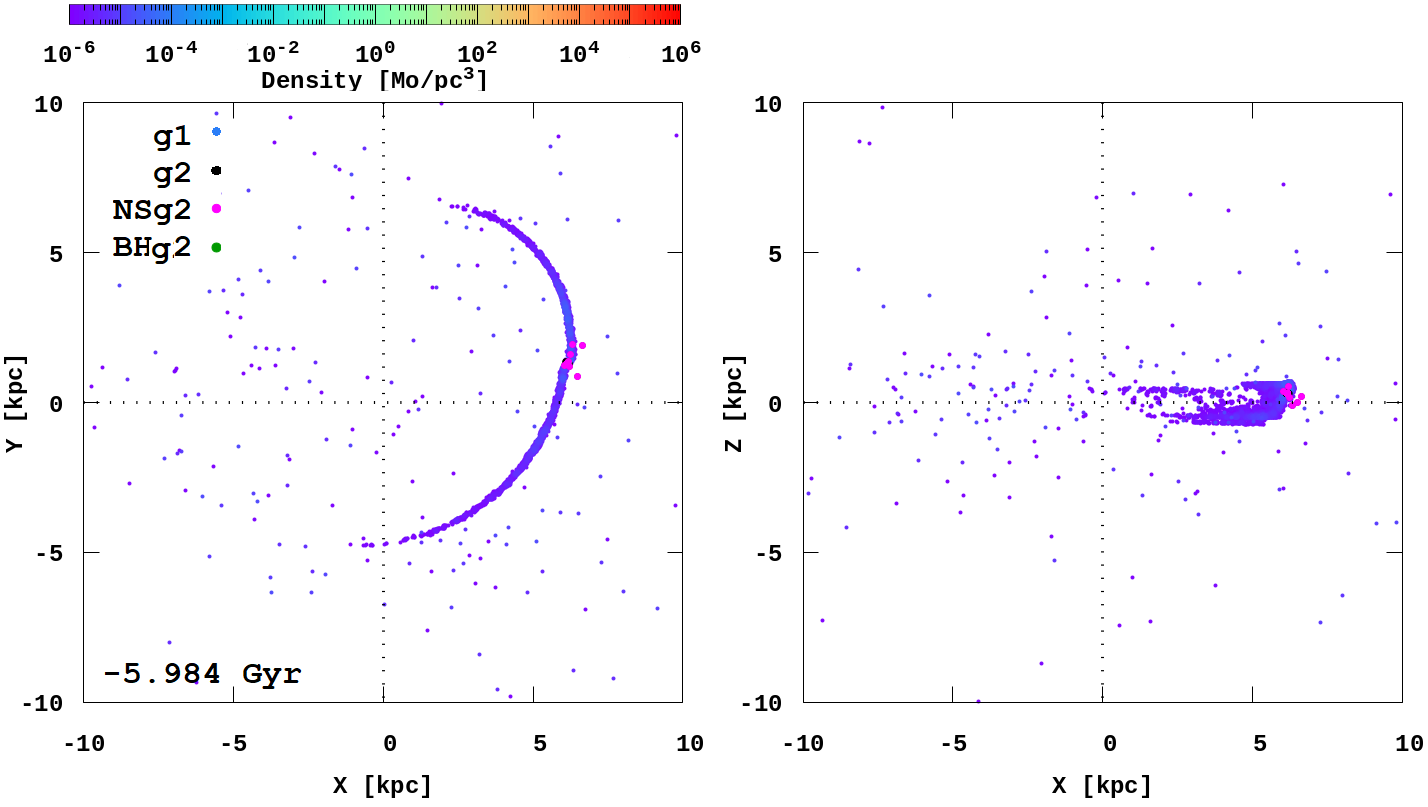}
\includegraphics[width=0.45\linewidth]{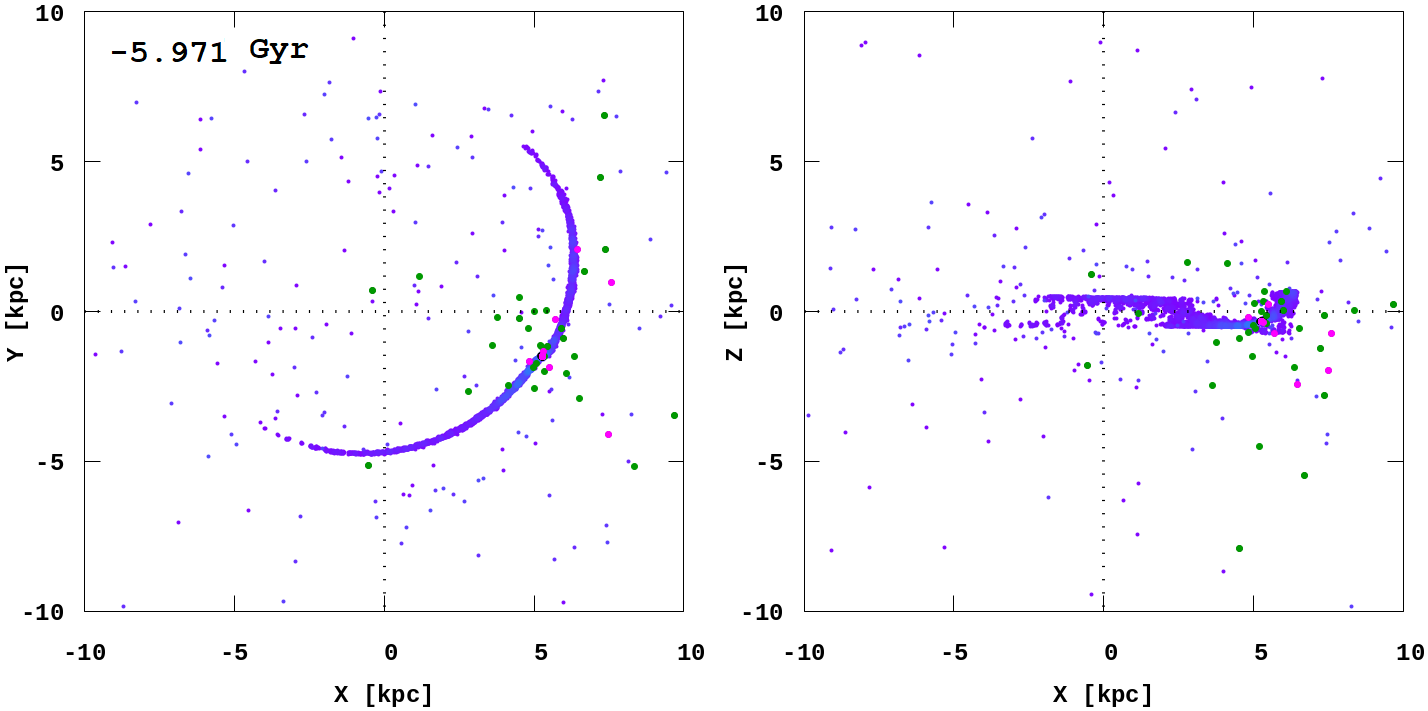}
\includegraphics[width=0.45\linewidth]{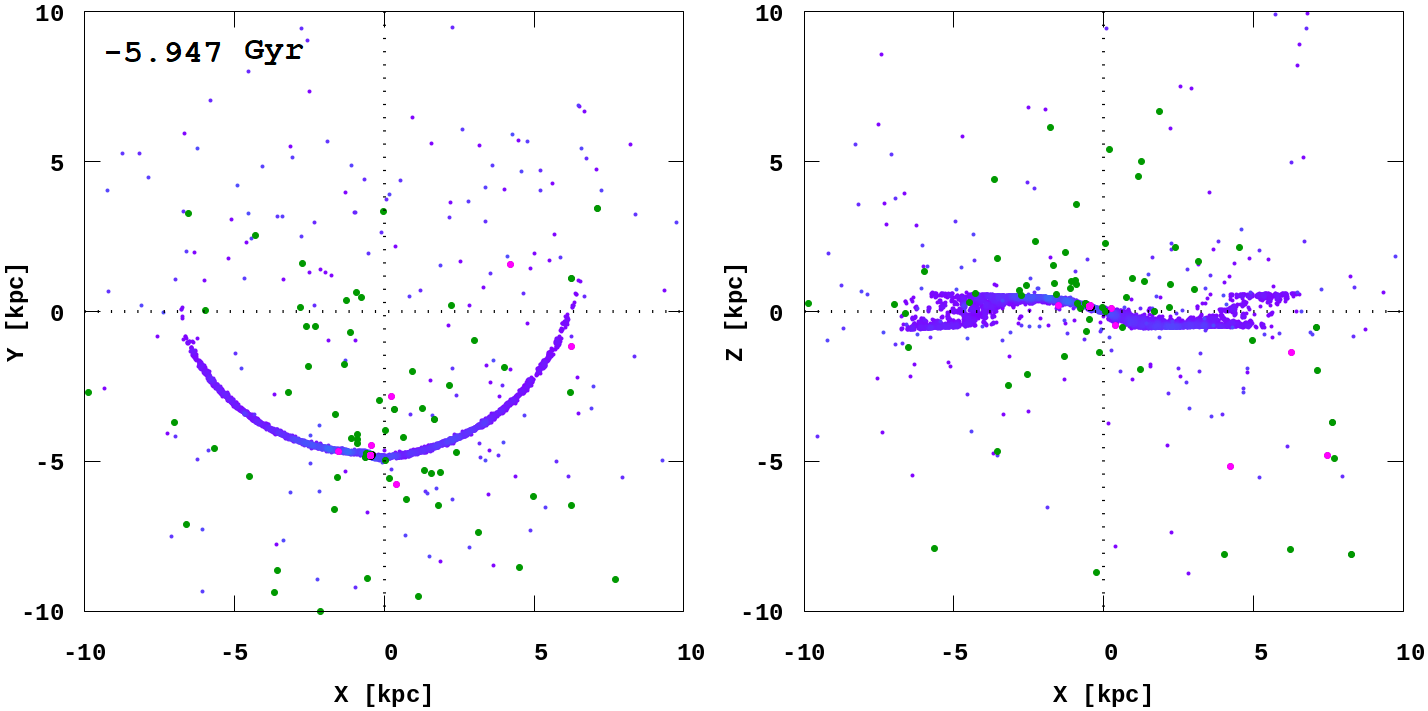}
\includegraphics[width=0.45\linewidth]{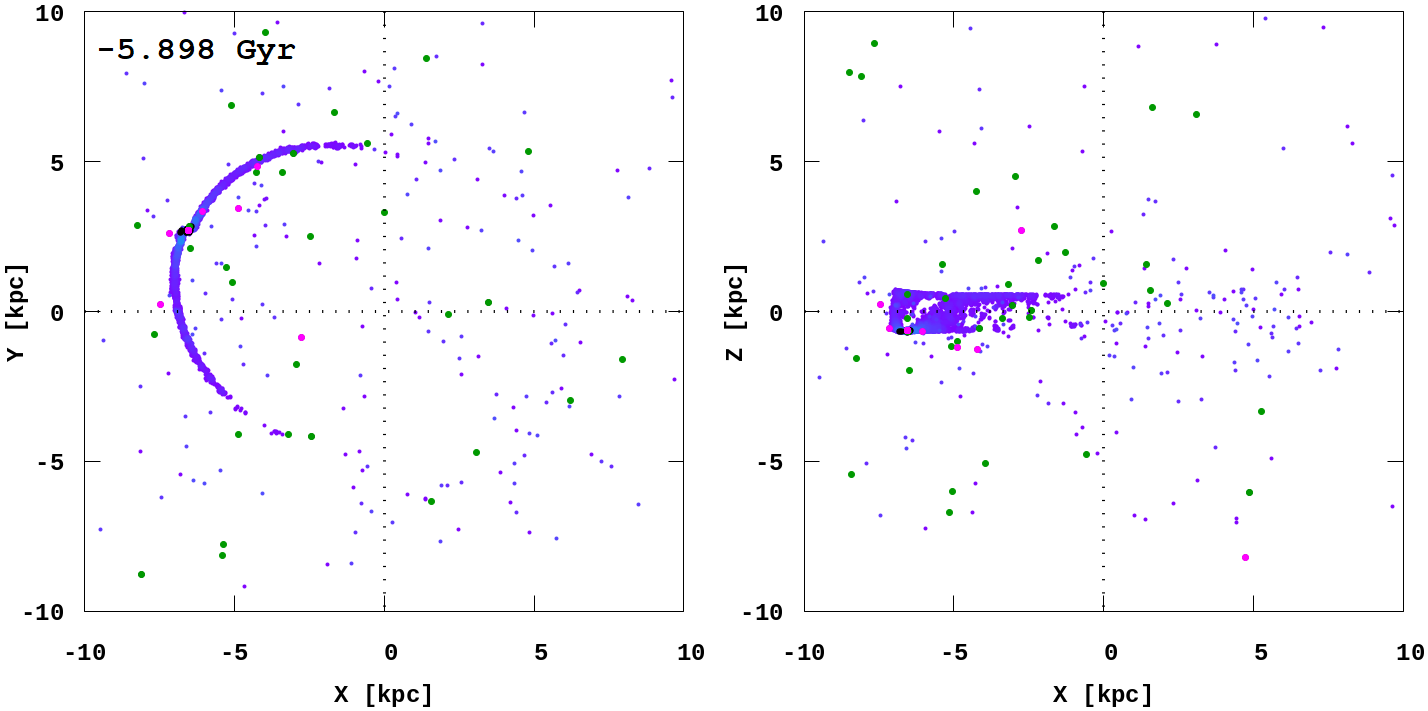}
\includegraphics[width=0.45\linewidth]{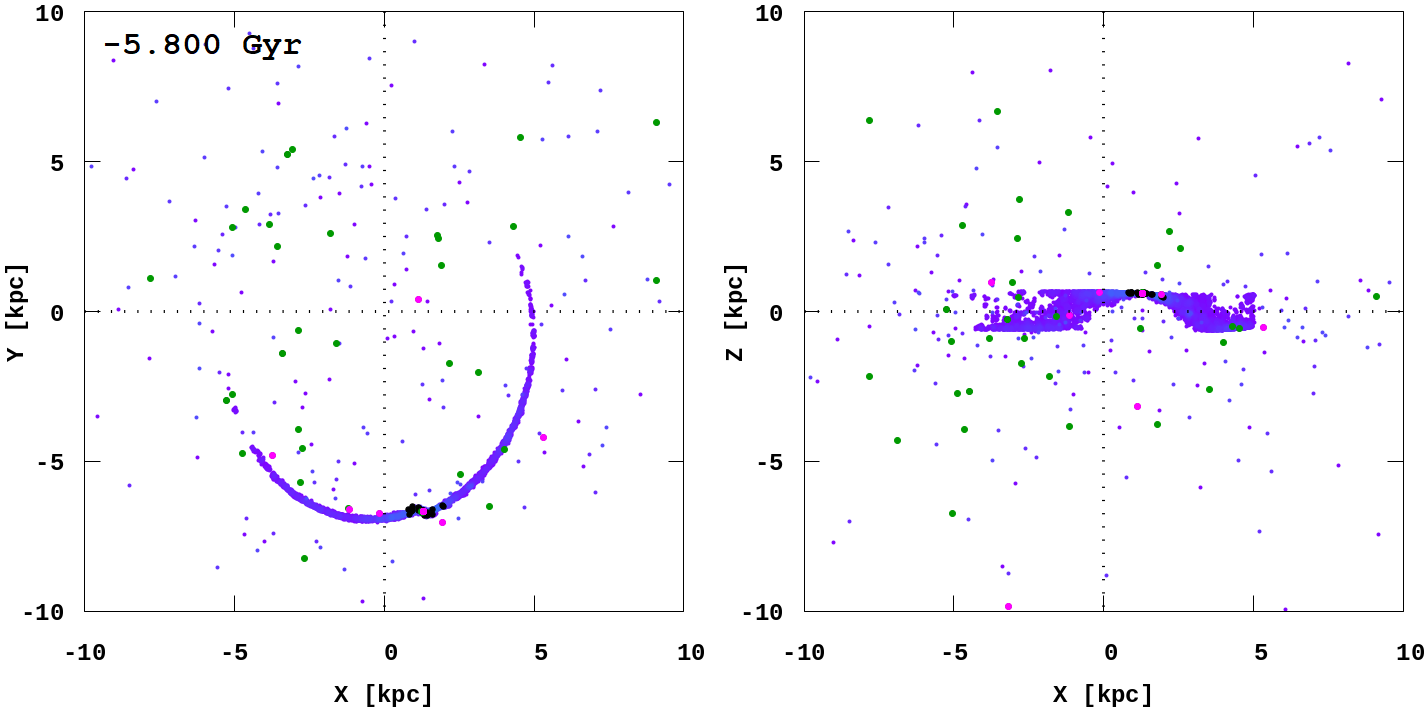}
\includegraphics[width=0.45\linewidth]{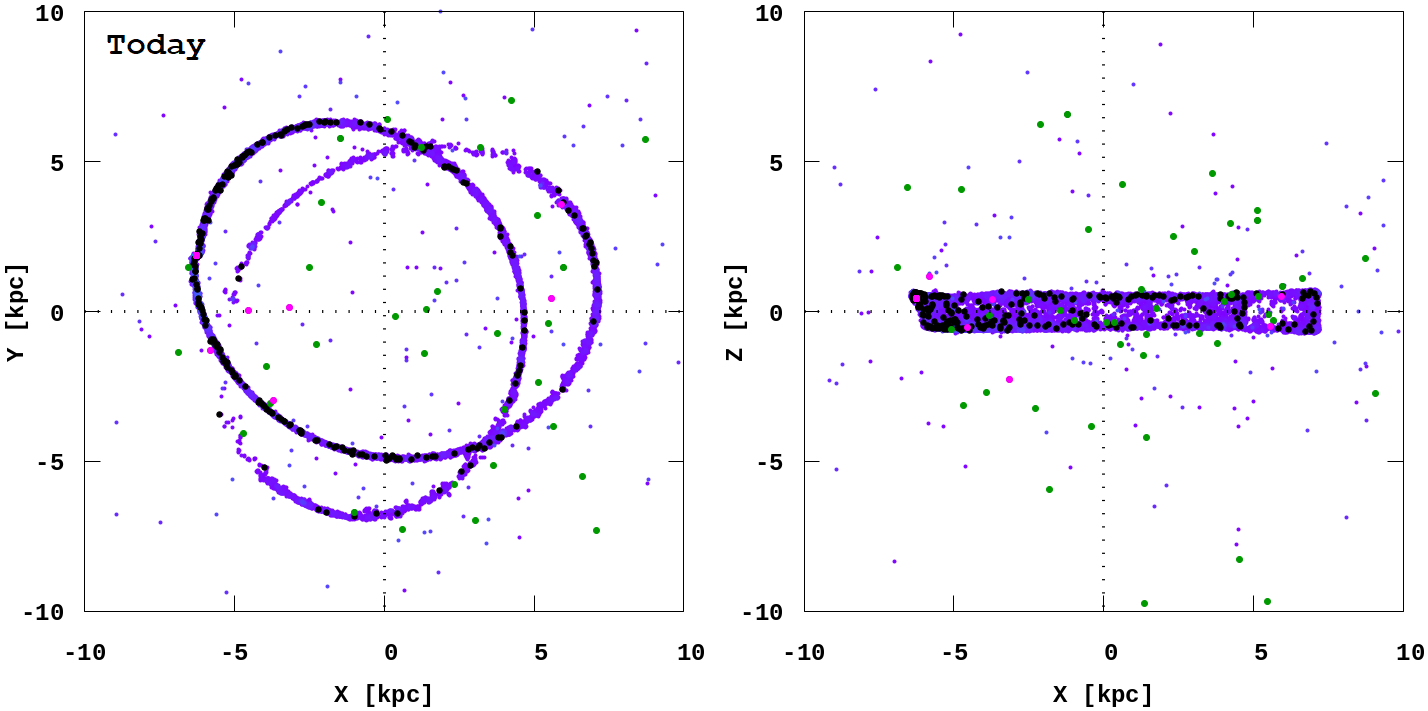}
\caption{Same as in Fig. \ref{fig:cr-glo-tng}, but in FIX potential for circular type of orbit.}
\label{fig:cr-glo-fix}
\end{figure*}

\begin{figure*}[ht]
\centering
\includegraphics[width=0.45\linewidth]{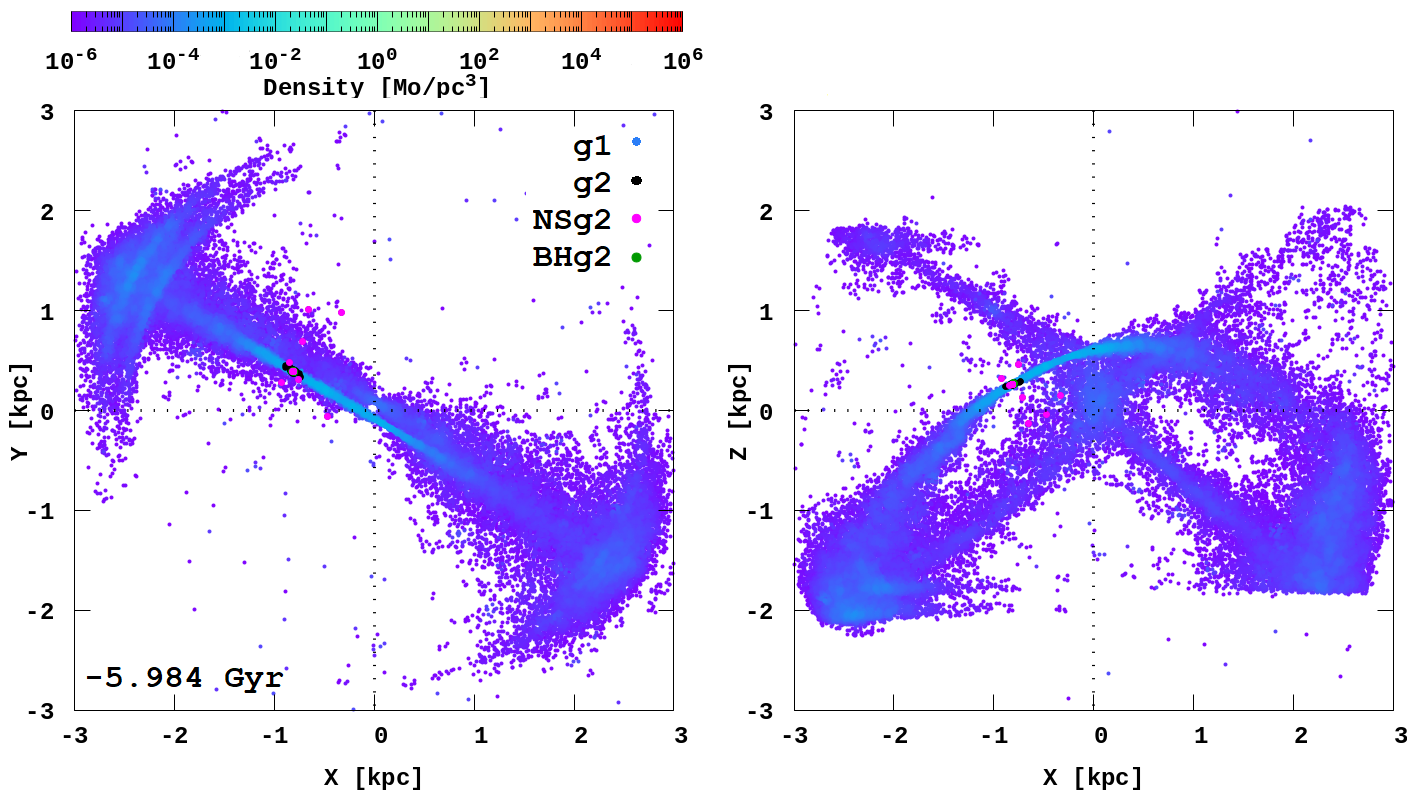}
\includegraphics[width=0.45\linewidth]{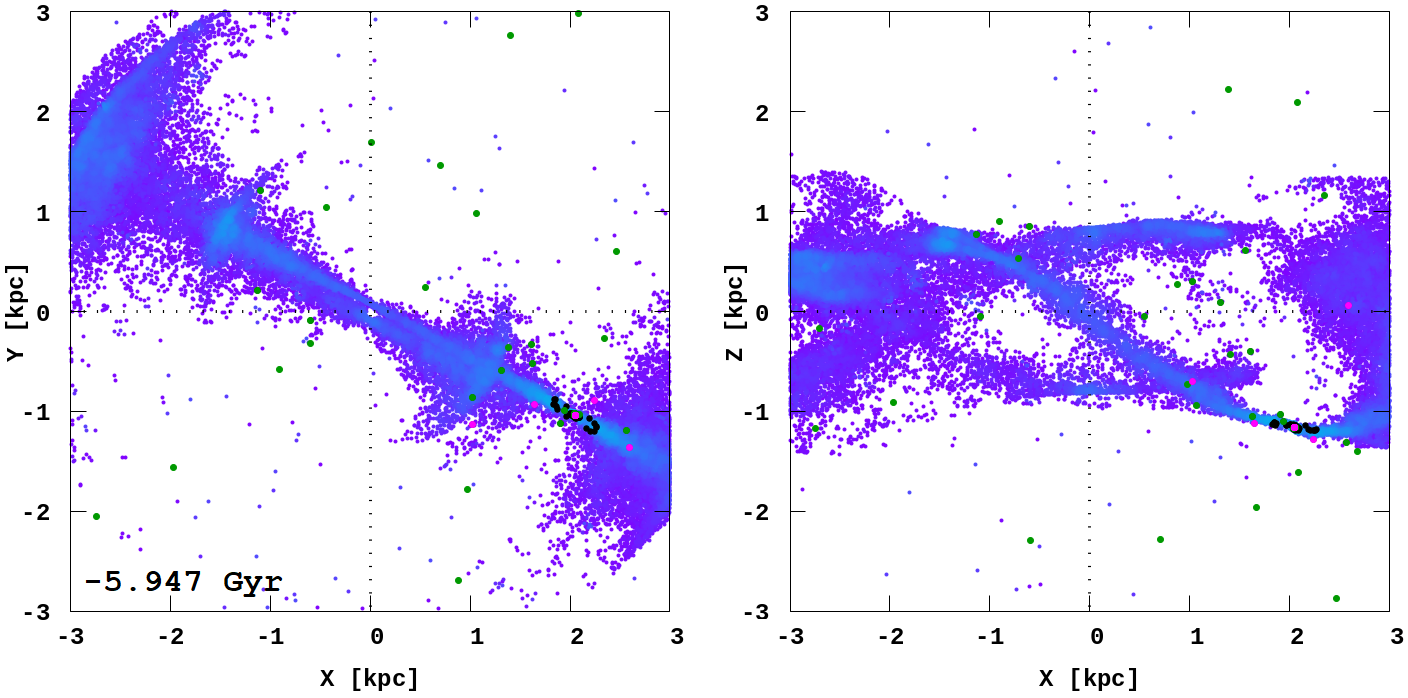}
\includegraphics[width=0.45\linewidth]{pic/tb-tng_glo-pos-dis_001680.png}
\includegraphics[width=0.45\linewidth]{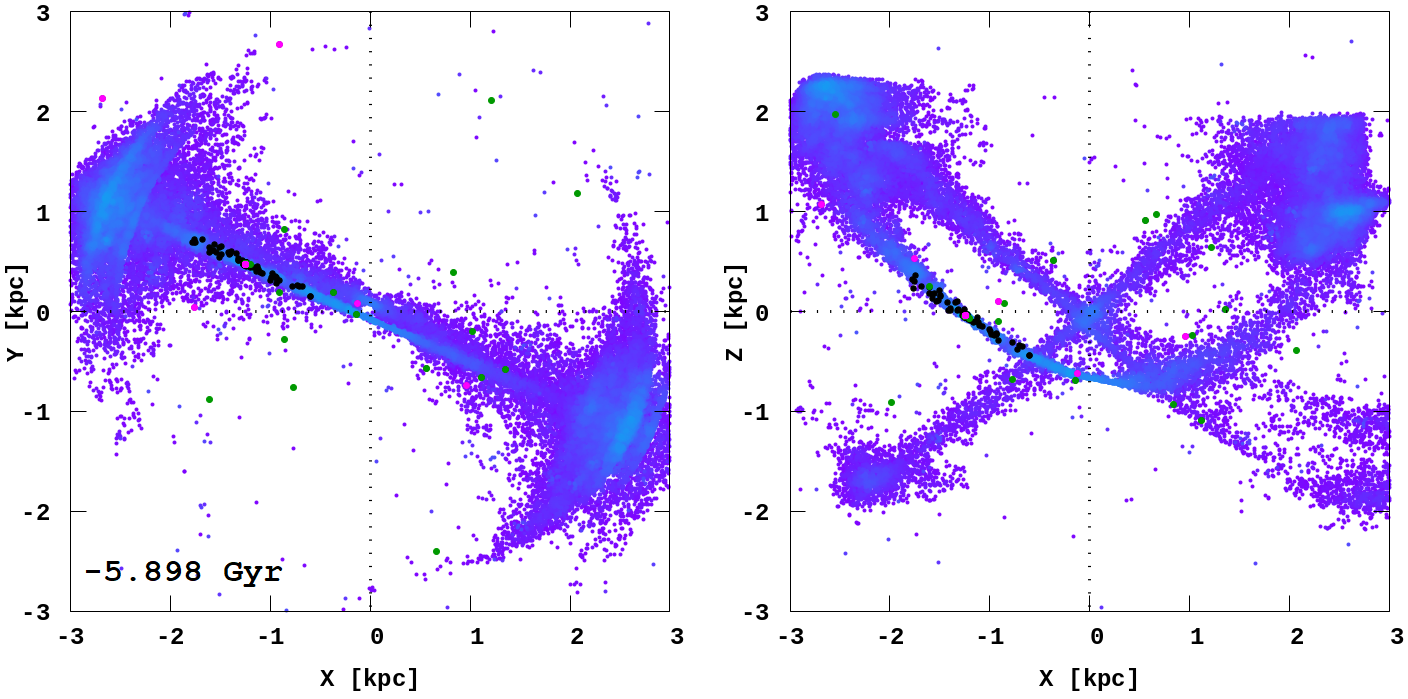}
\includegraphics[width=0.45\linewidth]{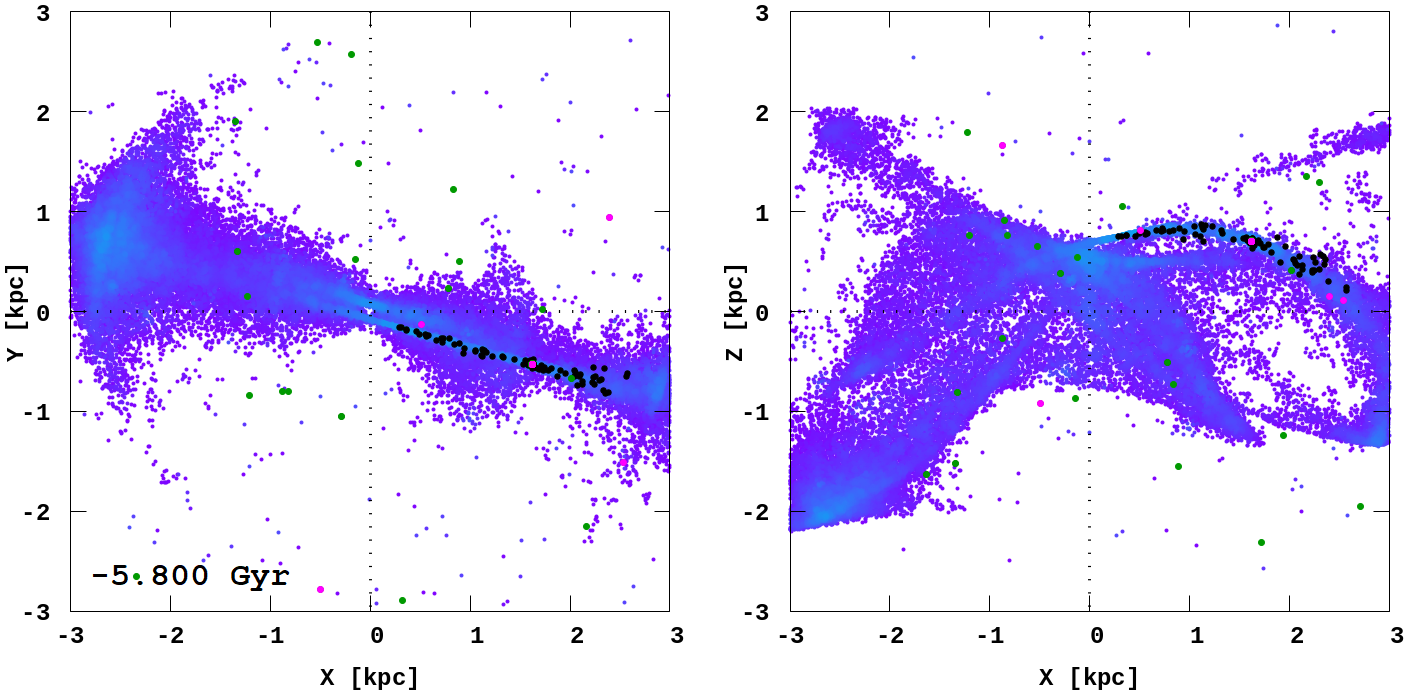}
\includegraphics[width=0.45\linewidth]{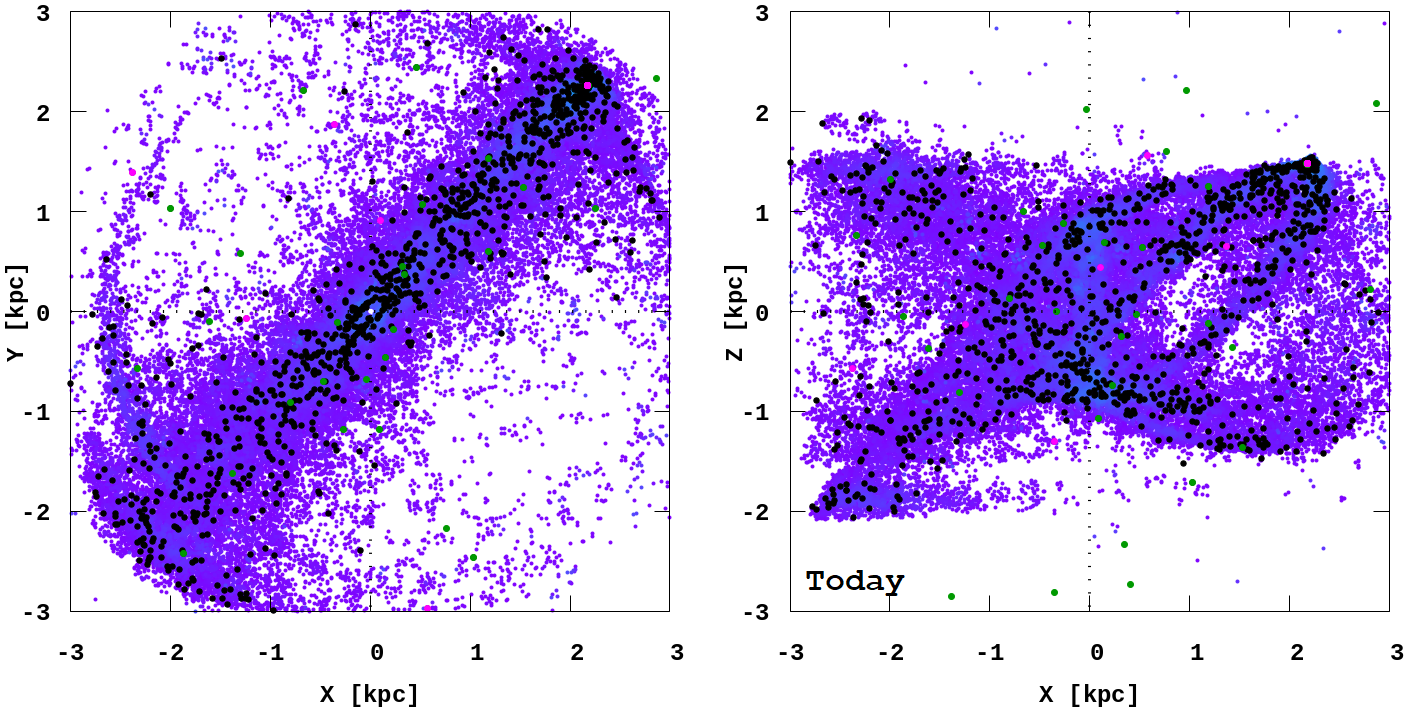}
\caption{Same as in Fig. \ref{fig:cr-glo-tng}, but in TNG potential for tube type of orbit.}
\label{fig:tb-glo-tng}
\end{figure*}

\begin{figure*}[ht]
\centering
\includegraphics[width=0.45\linewidth]{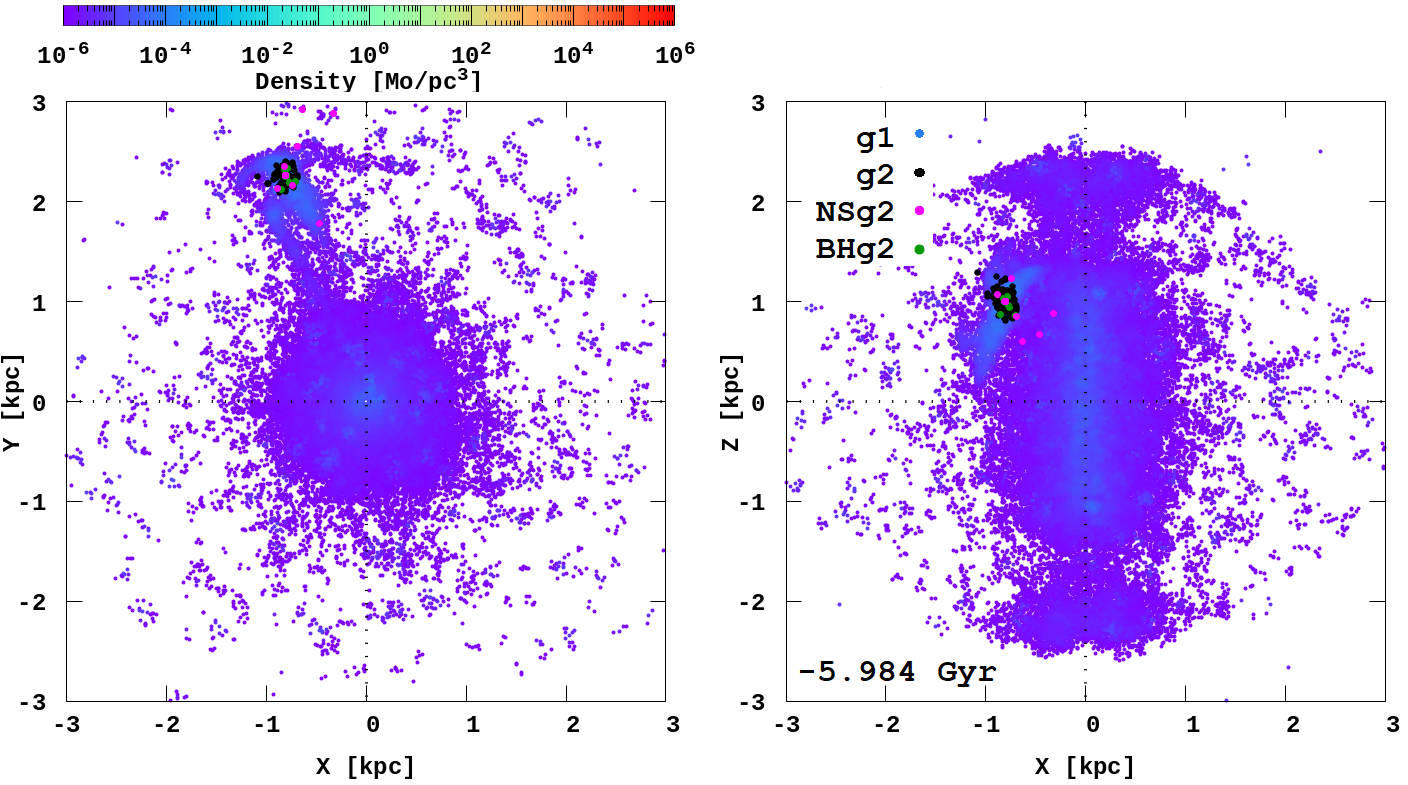}
\includegraphics[width=0.45\linewidth]{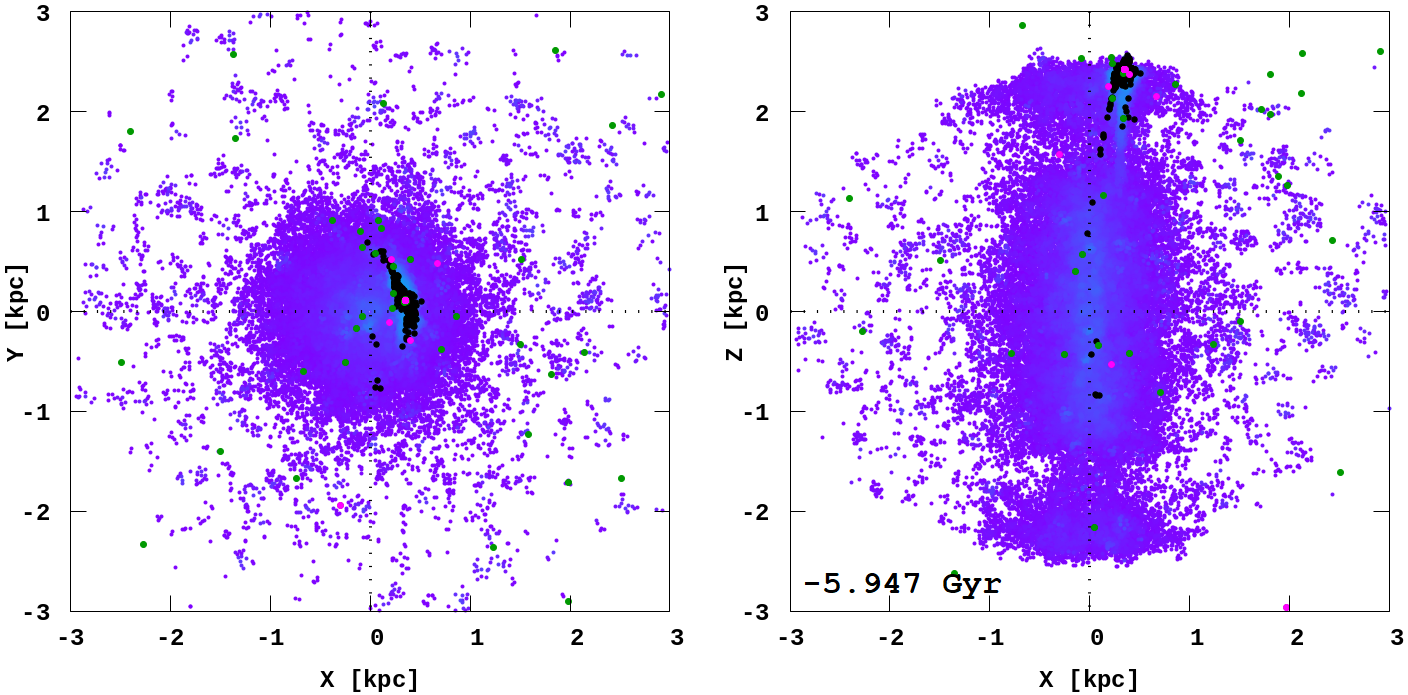}
\includegraphics[width=0.45\linewidth]{pic/tb-fix_glo-pos-dis_001680.png}
\includegraphics[width=0.45\linewidth]{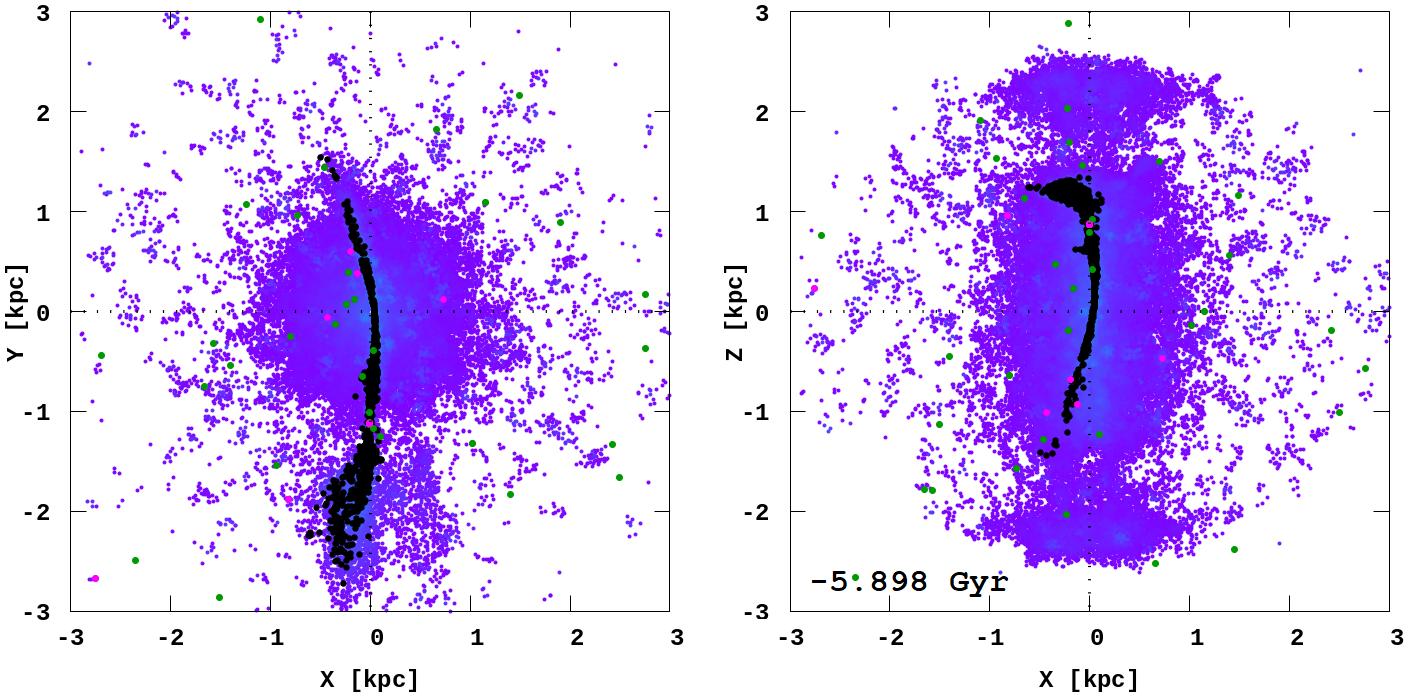}
\includegraphics[width=0.45\linewidth]{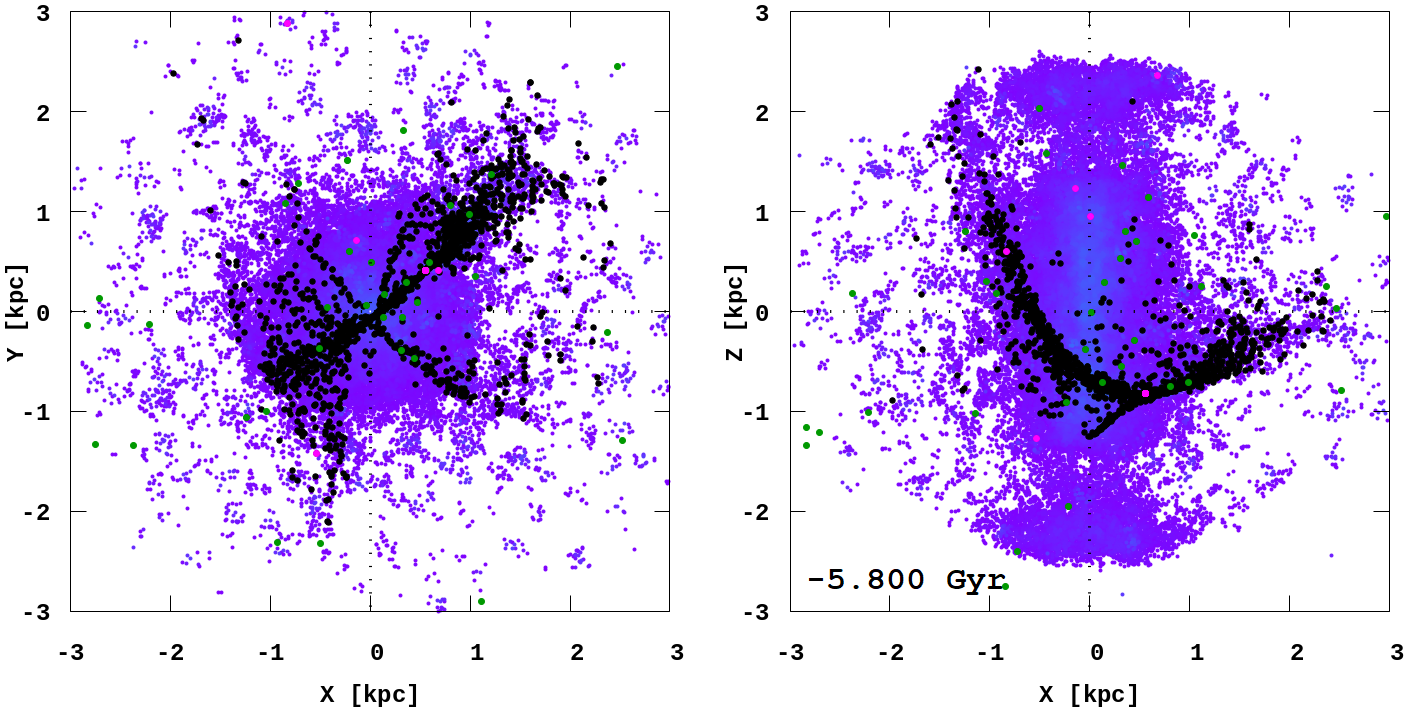}
\includegraphics[width=0.45\linewidth]{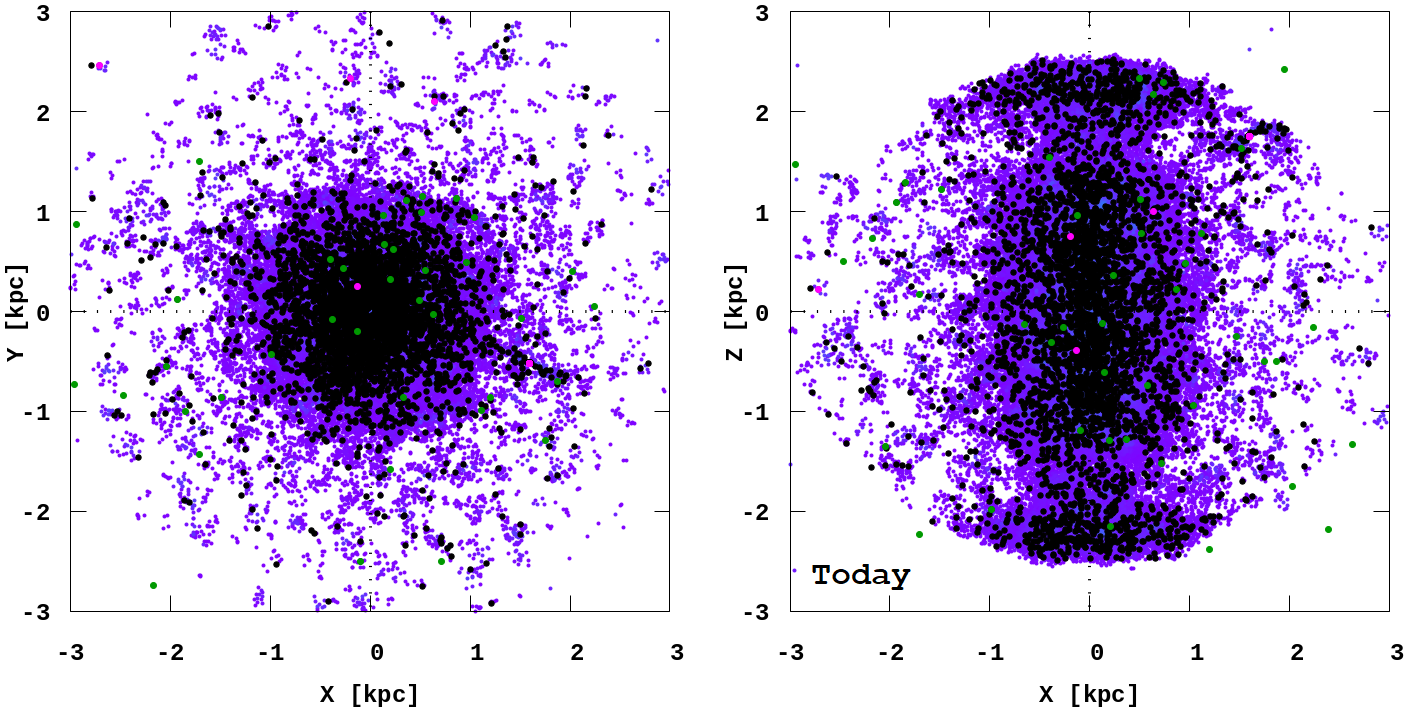}
\caption{Same as in Fig. \ref{fig:cr-glo-tng}, but in FIX potential for tube type of orbit.}
\label{fig:tb-glo-fix}
\end{figure*}

\begin{figure*}[ht]
\centering
\includegraphics[width=0.45\linewidth]{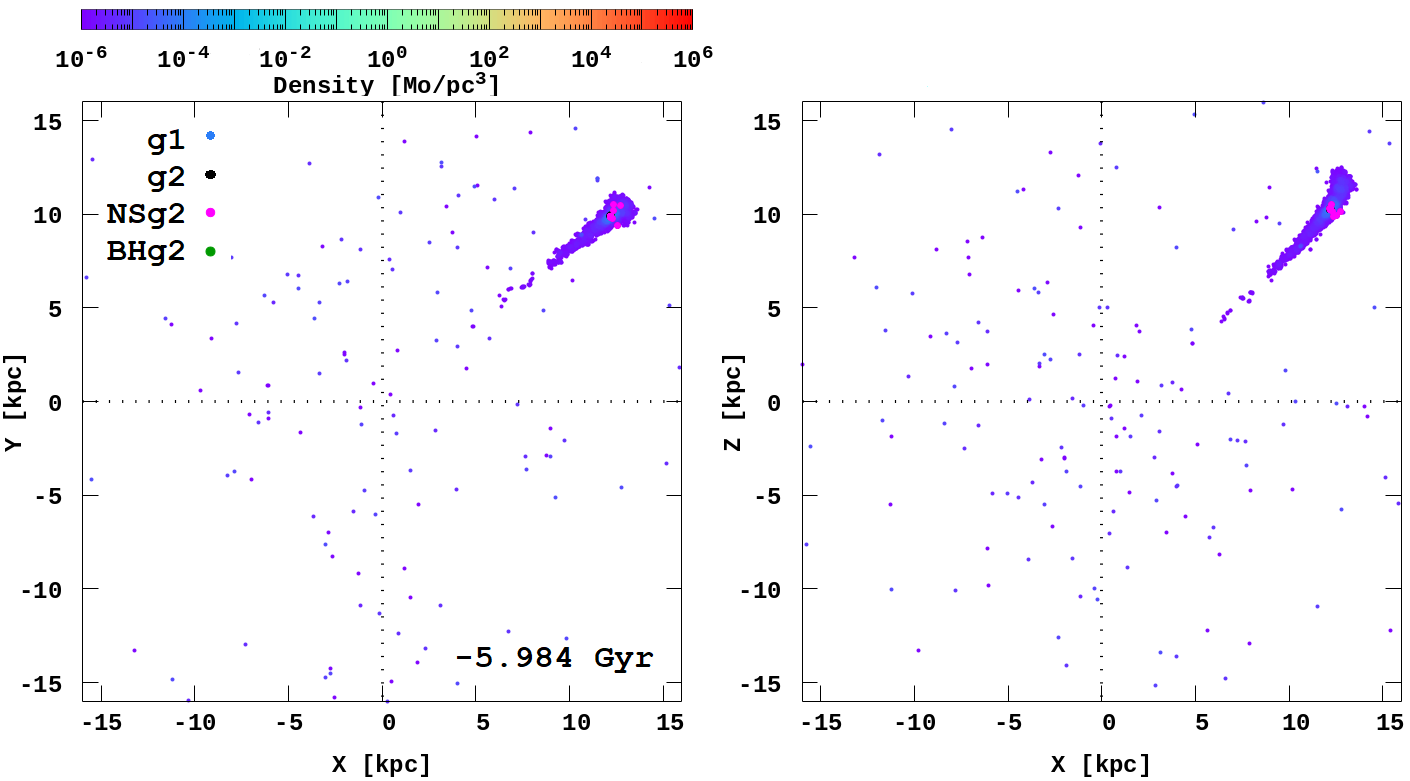}
\includegraphics[width=0.45\linewidth]{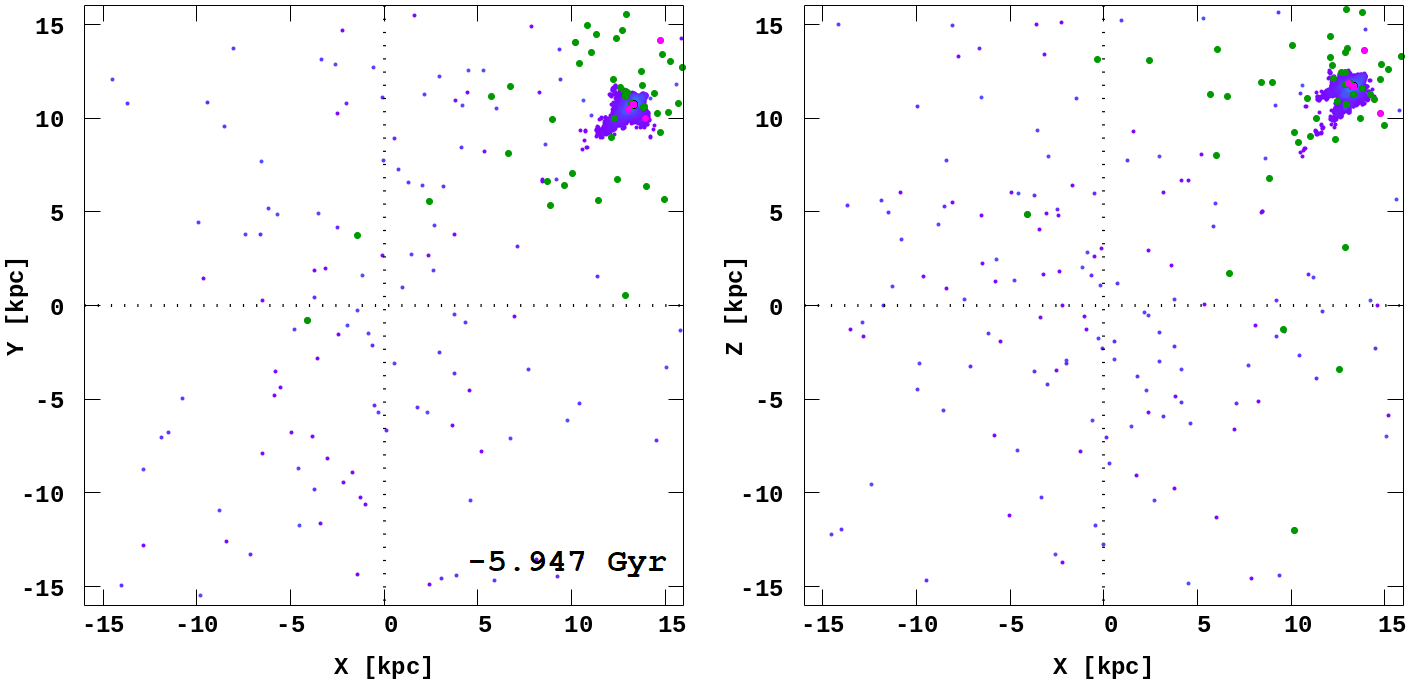}
\includegraphics[width=0.45\linewidth]{pic/lr-tng_glo-pos-dis_001680.png}
\includegraphics[width=0.45\linewidth]{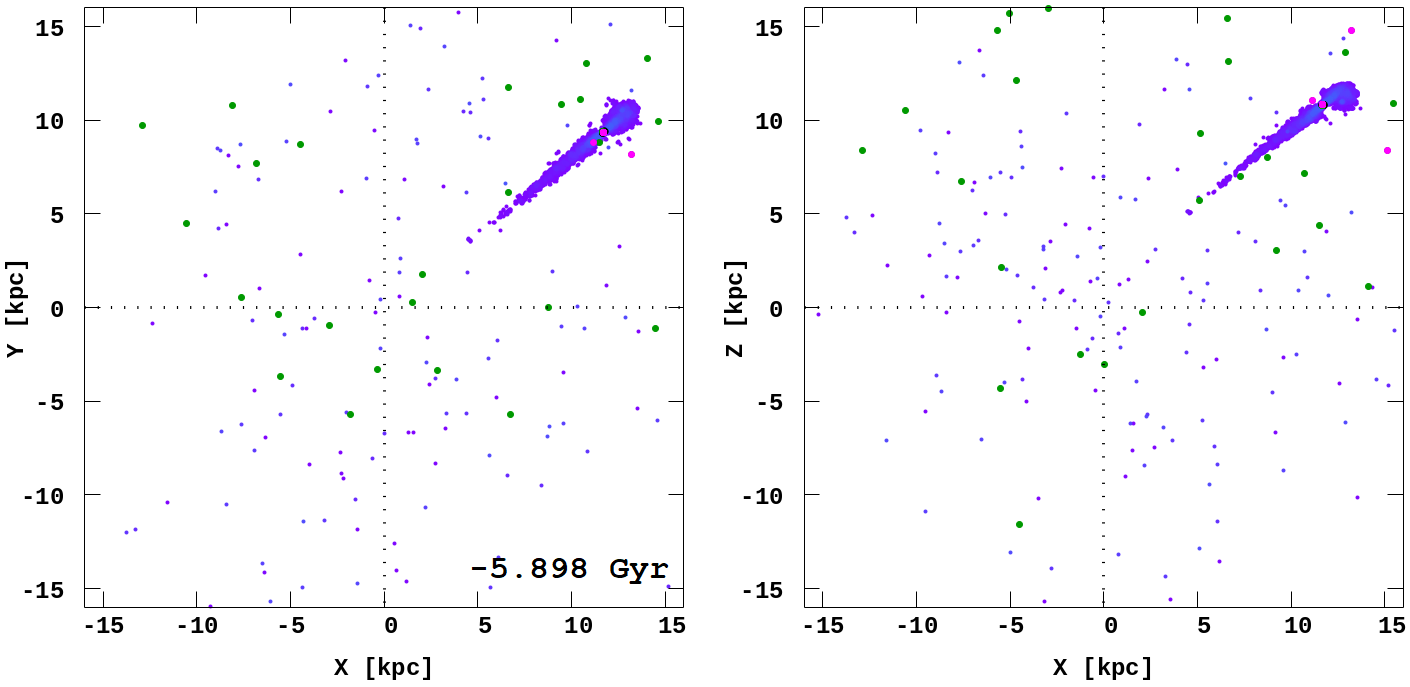}
\includegraphics[width=0.45\linewidth]{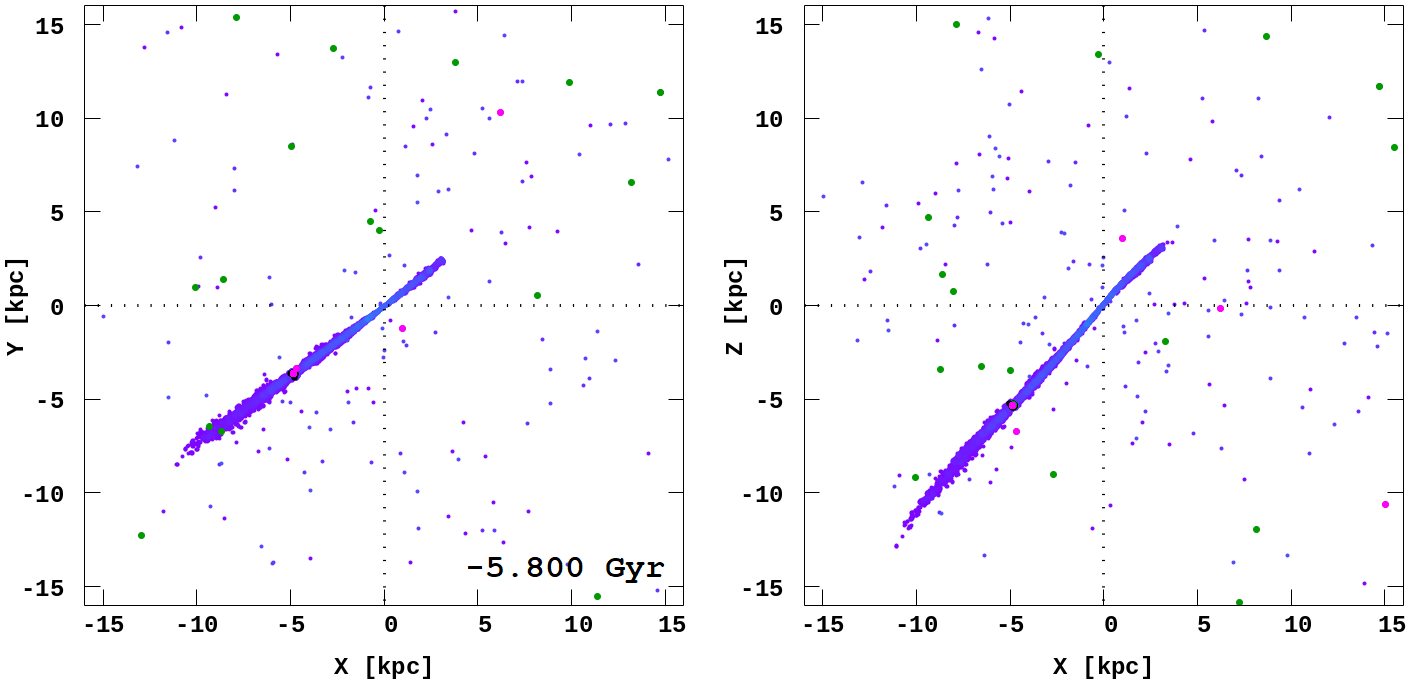}
\includegraphics[width=0.45\linewidth]{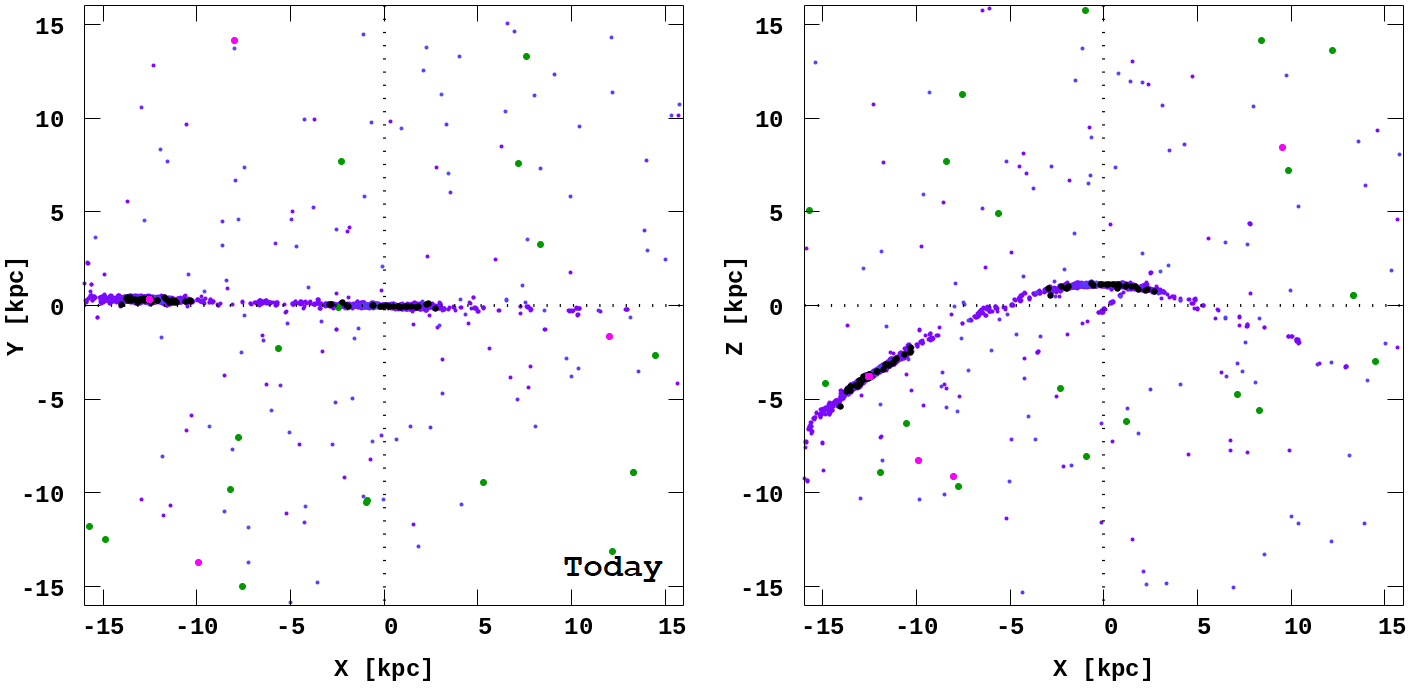}
\caption{Same as in Fig. \ref{fig:cr-glo-tng}, but in TNG potential for long radial type of orbit.}
\label{fig:lr-glo-tng}
\end{figure*}

\begin{figure*}[ht]
\centering
\includegraphics[width=0.45\linewidth]{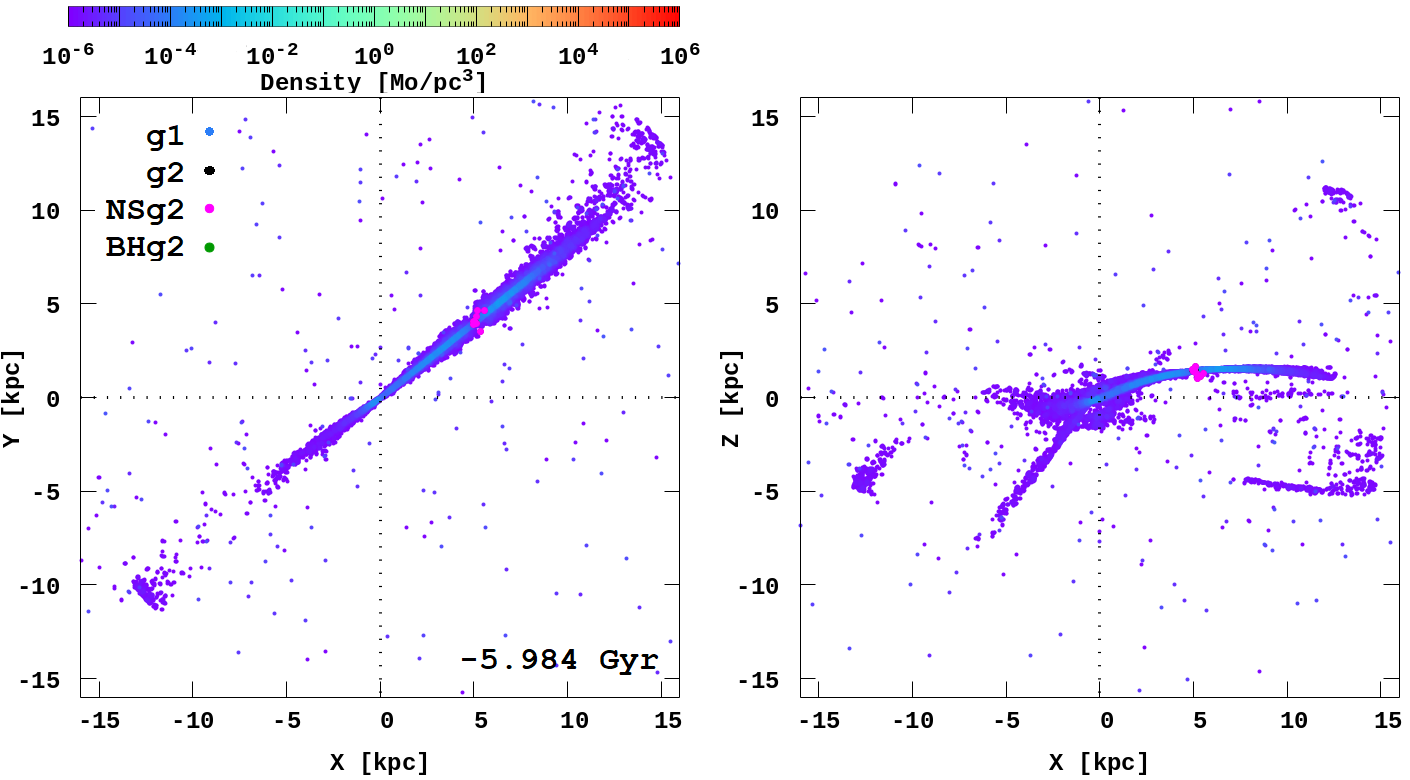}
\includegraphics[width=0.45\linewidth]{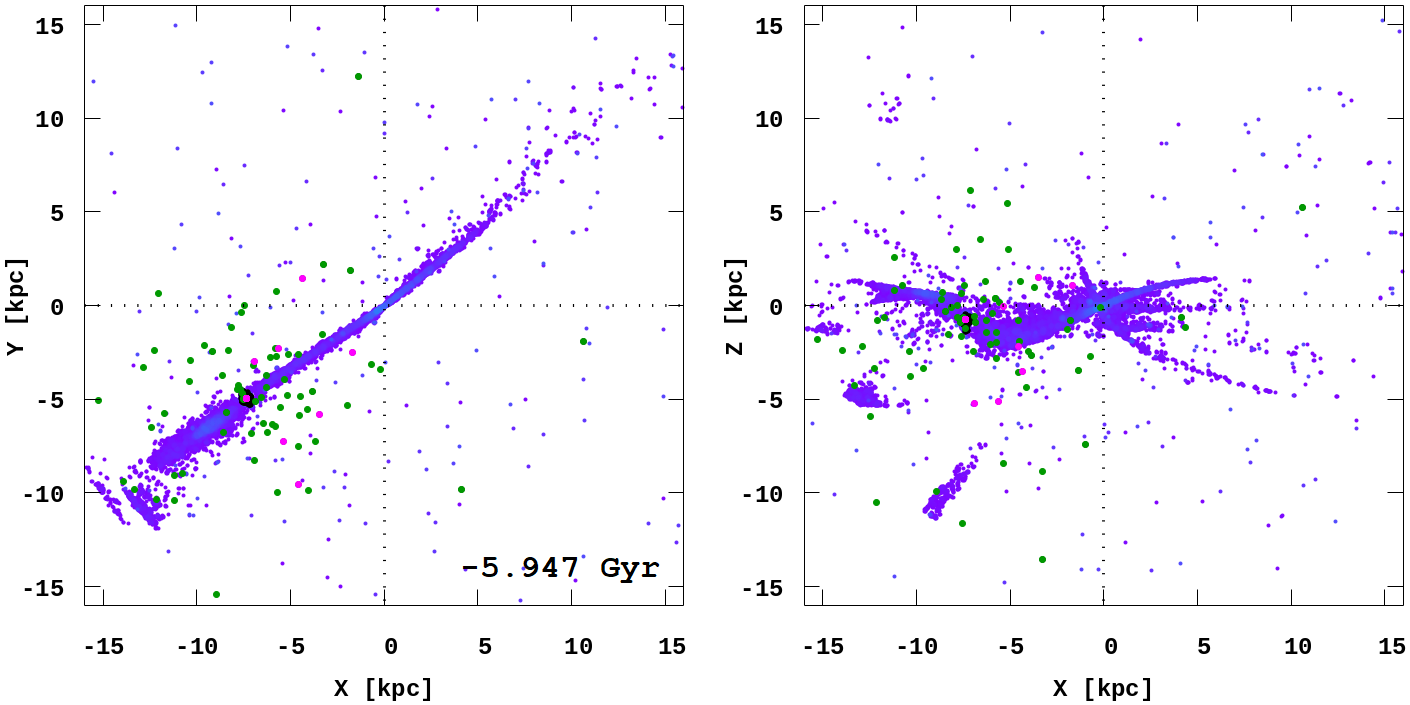}
\includegraphics[width=0.45\linewidth]{pic/lr-fix_glo-pos-dis_001680.png}
\includegraphics[width=0.45\linewidth]{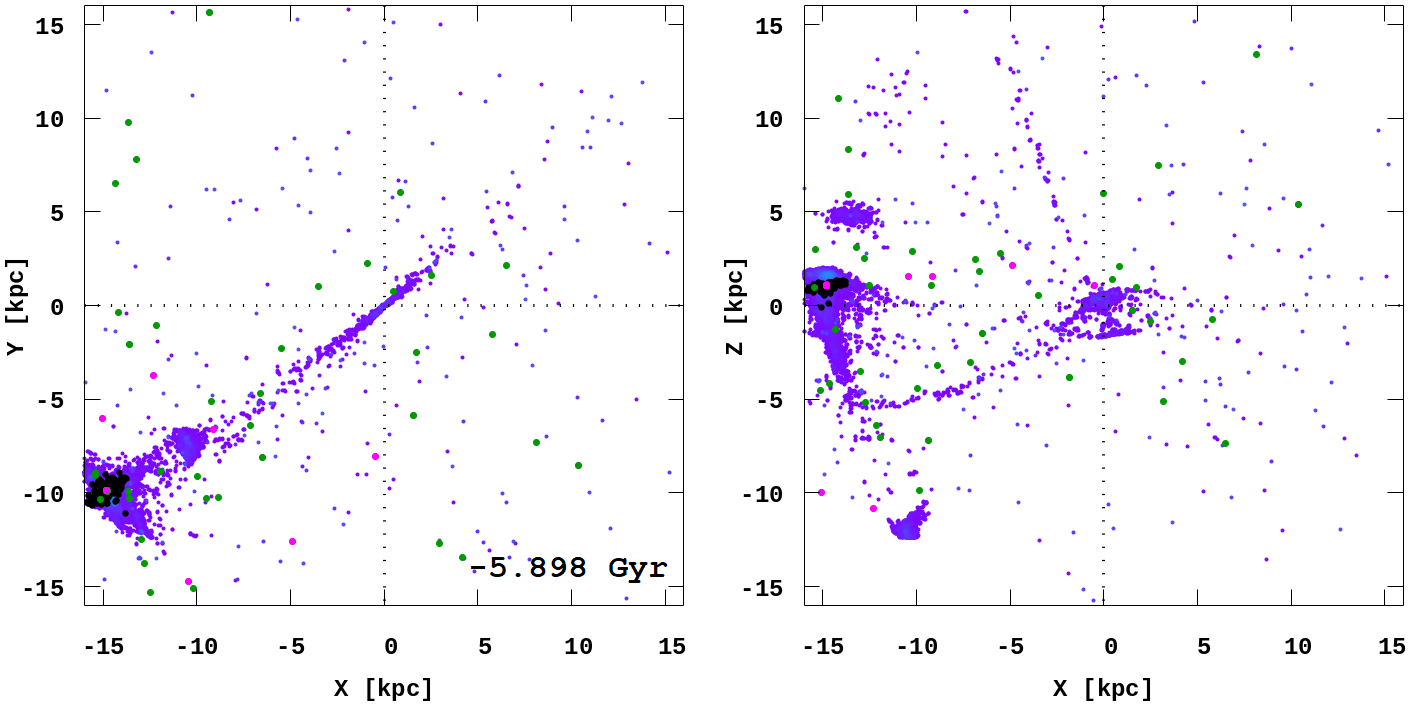}
\includegraphics[width=0.45\linewidth]{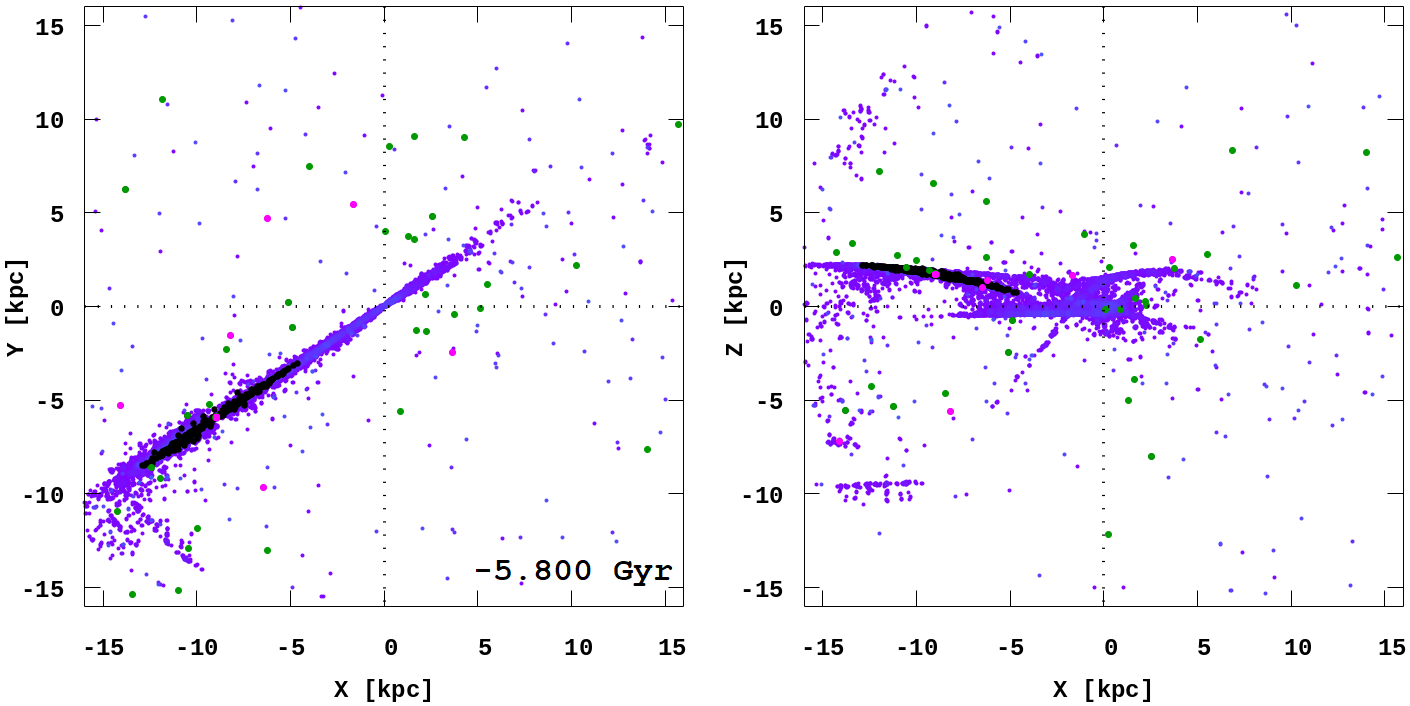}
\includegraphics[width=0.45\linewidth]{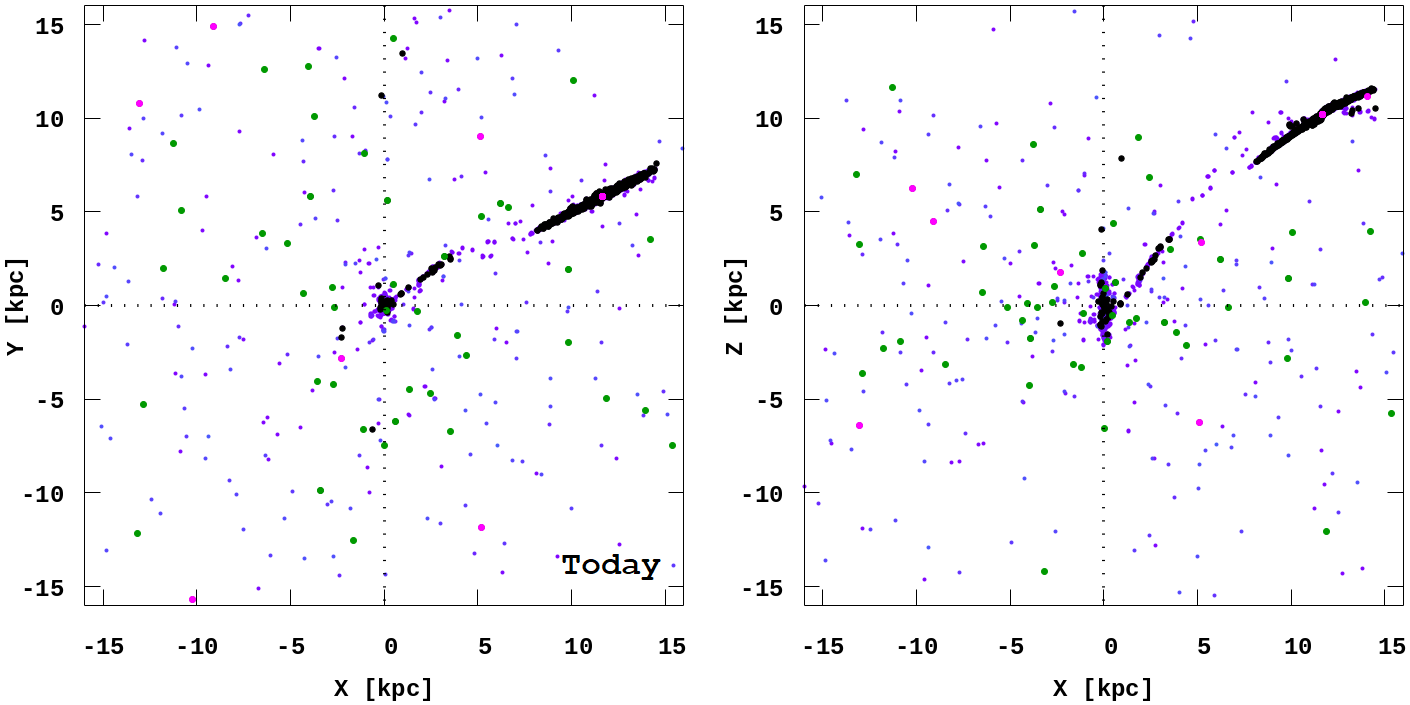}
\caption{Same as in Fig. \ref{fig:cr-glo-tng}, but in FIX potential for long radial type of orbit.}
\label{fig:lr-glo-fix}
\end{figure*}

In Figures \ref{fig:cr-glo-tng}, \ref{fig:cr-glo-fix}, \ref{fig:tb-glo-tng}, \ref{fig:tb-glo-fix}, \ref{fig:lr-glo-tng} and \ref{fig:lr-glo-fix} we present the distribution of GC density (palette colour) in global Galactocentric view. In these visualisations we also present the {\tt g2} distribution with the BH and NS remnants only for the {\tt g2} generation. The orbital and stellar evolution is present for selected snapshots at $T$ = -5.984, -5.971, -5.947, -5.898, -5.800 and zero lookback time. From these figures we can make a clear conclusion about the today complete mixing of {\tt g1} and {\tt g2} generations for TB orbit in both potentials and the significant {\tt g2} stars distribution along the orbits in a case of CR and LR. 

\begin{figure*}[ht]
\centering
\includegraphics[width=0.92\linewidth]{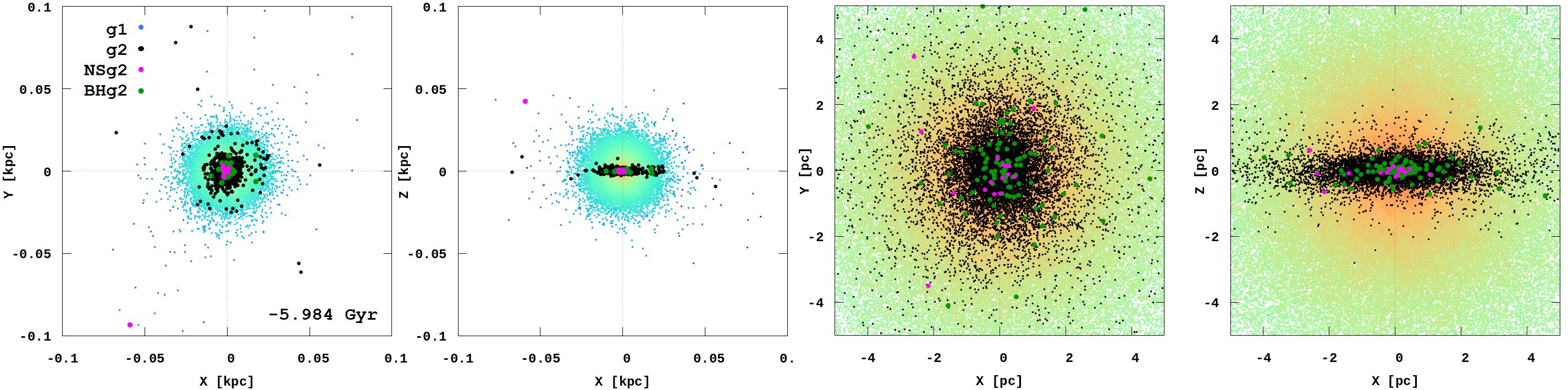}
\includegraphics[width=0.92\linewidth]{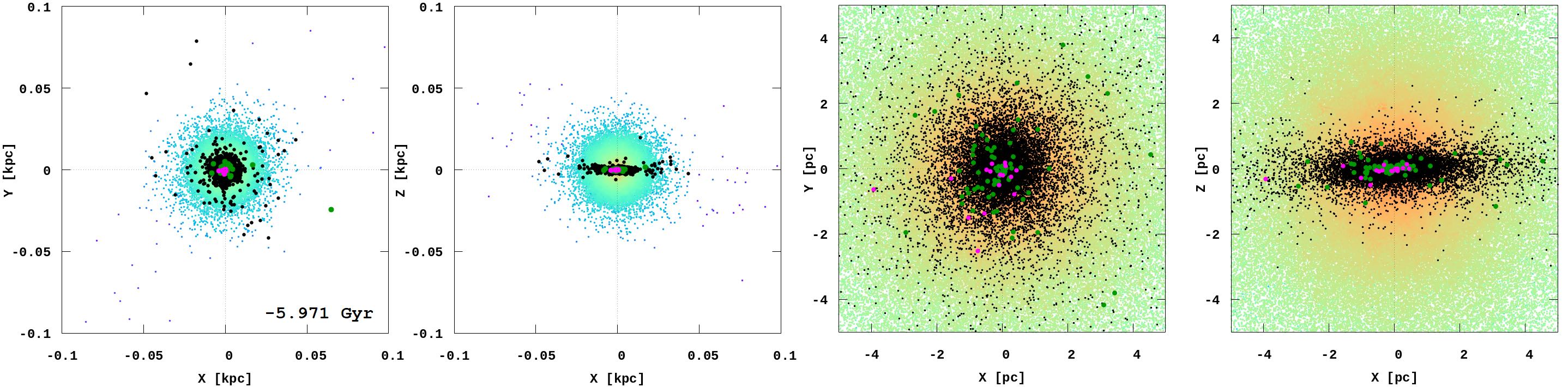}
\includegraphics[width=0.92\linewidth]{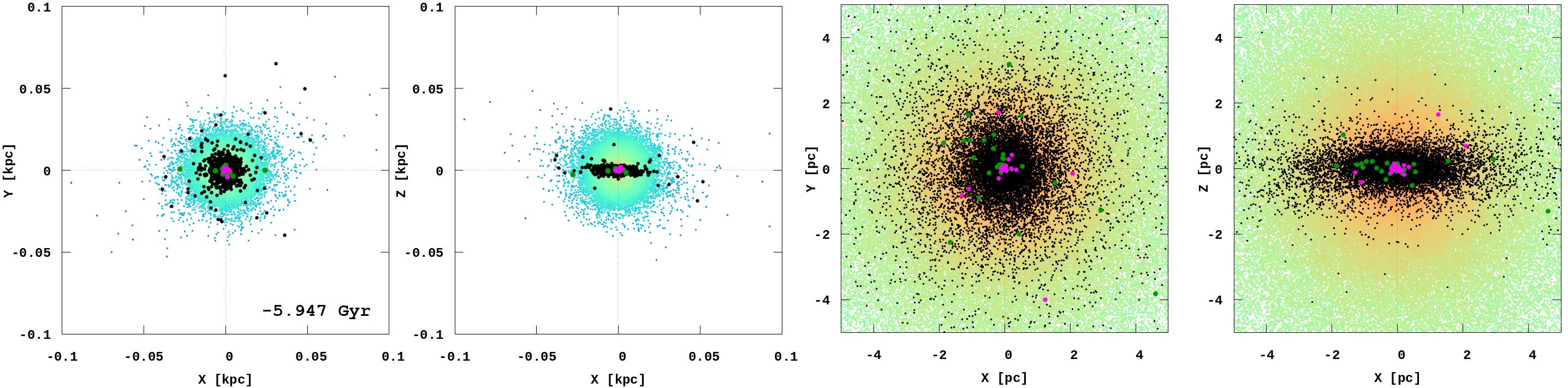}
\includegraphics[width=0.92\linewidth]{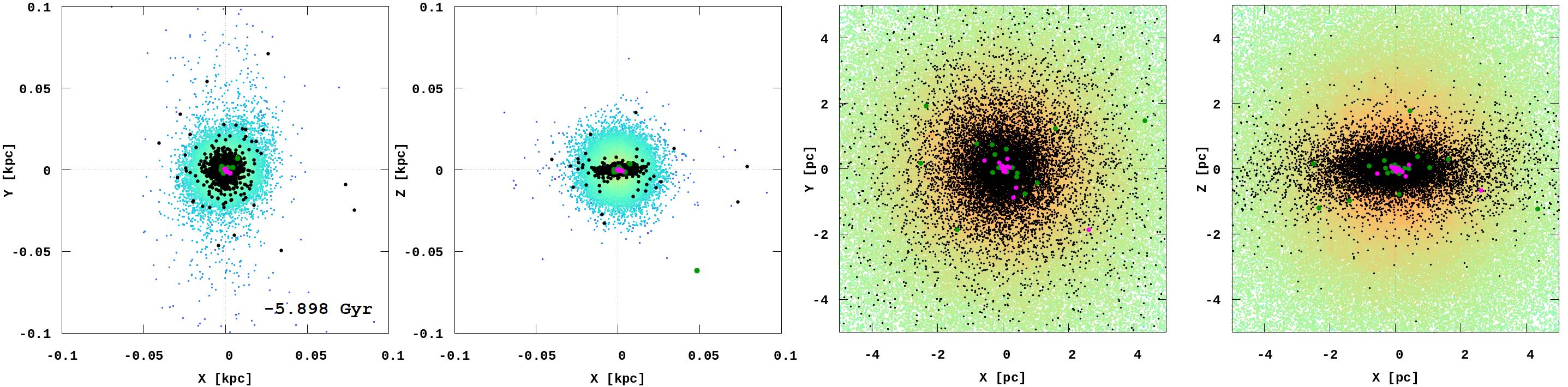}
\includegraphics[width=0.92\linewidth]{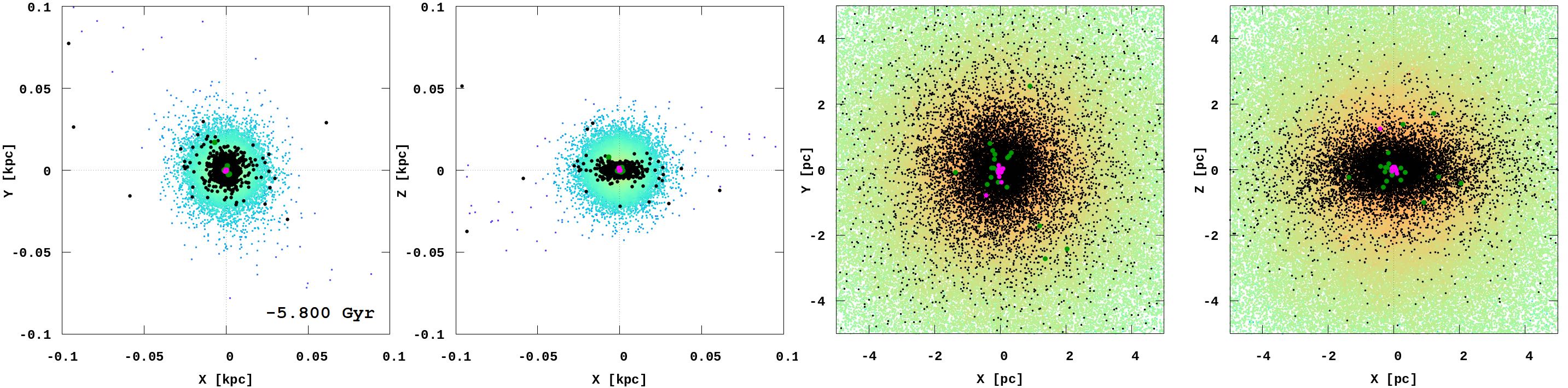}
\includegraphics[width=0.92\linewidth]{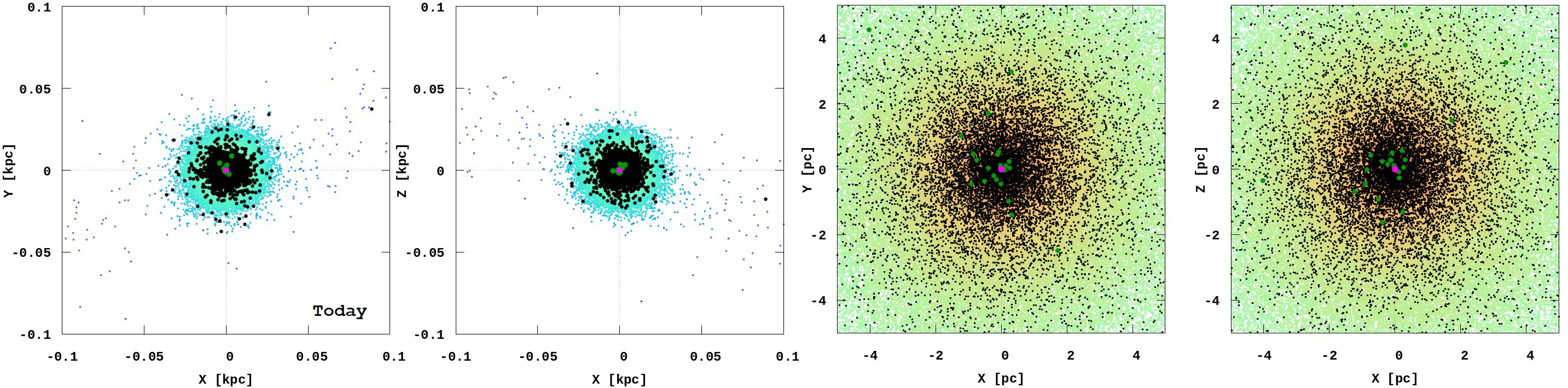}
\caption{The distribution of GC density, in the coordinate system connected with the cluster density centre, for central part of the GC (palette colour) for the circular type of orbit at different moments in time in TNG potential for mixed generations of stars {\tt g1+g2}. Black dots represent the distribution of the {\tt g2}, magenta dots -- BH, dark green dots -- NS only for second generation of stars {\tt g2}. The orbital and stellar evolution is presented in two planes -- ($X$, $Y$) and ($X$, $Z$) for lookback time $T$ = -5.984, -5.971, -5.947, -5.898, -5.800 Gyr and today, \textit{from left to right} 
The detailed central GC's area (5$\times$5 pc) are shown in \textit{two right panels}.}
\label{fig:cr-loc-tng}
\end{figure*}

\begin{figure*}[ht]
\centering
\includegraphics[width=0.93\linewidth]{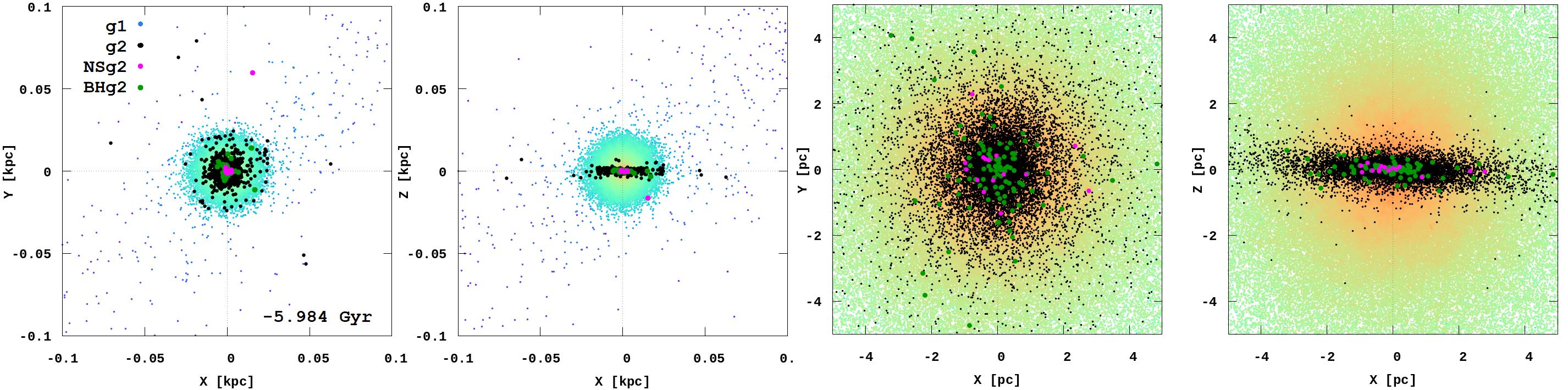}
\includegraphics[width=0.93\linewidth]{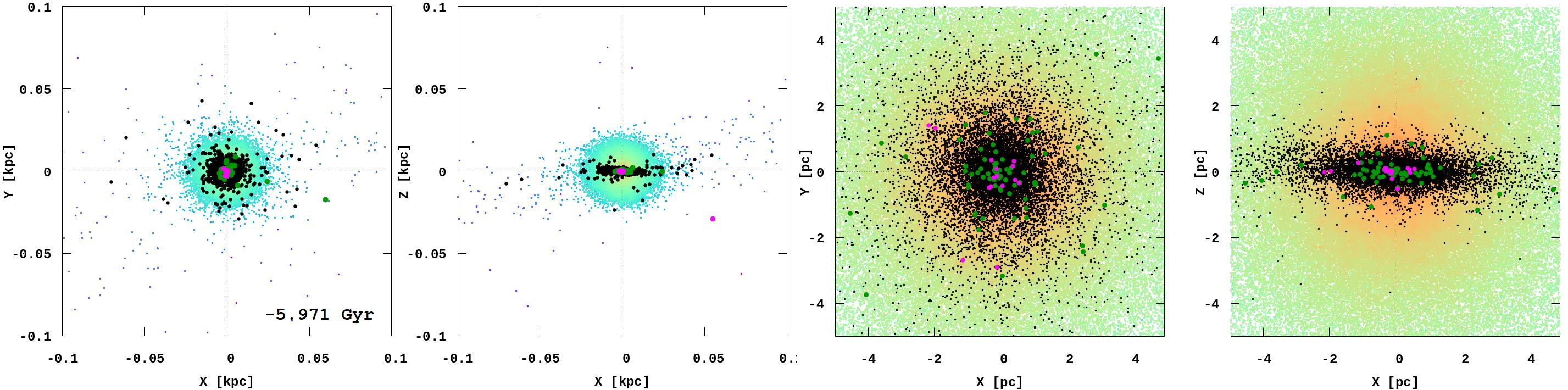}
\includegraphics[width=0.93\linewidth]{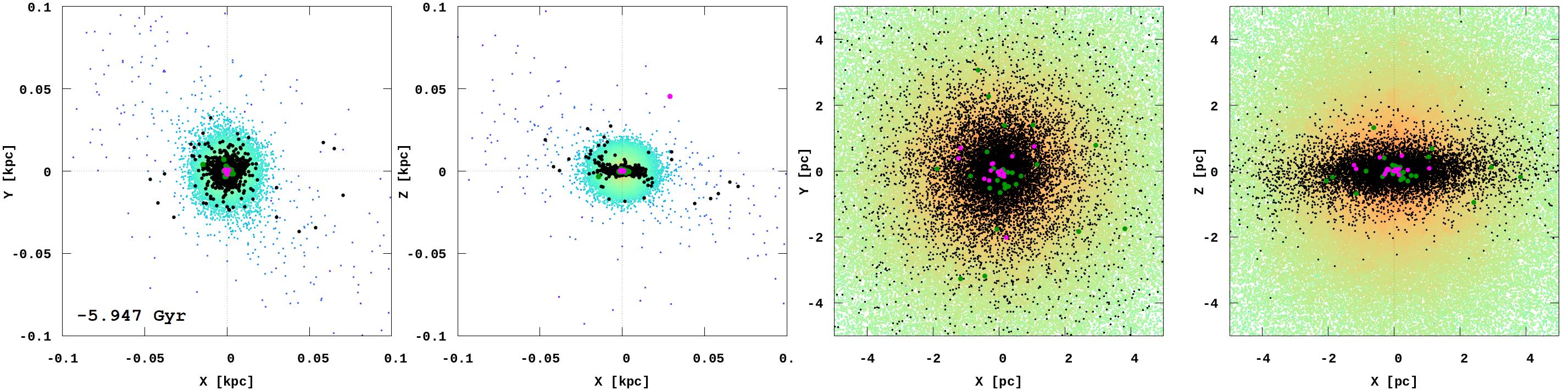}
\includegraphics[width=0.93\linewidth]{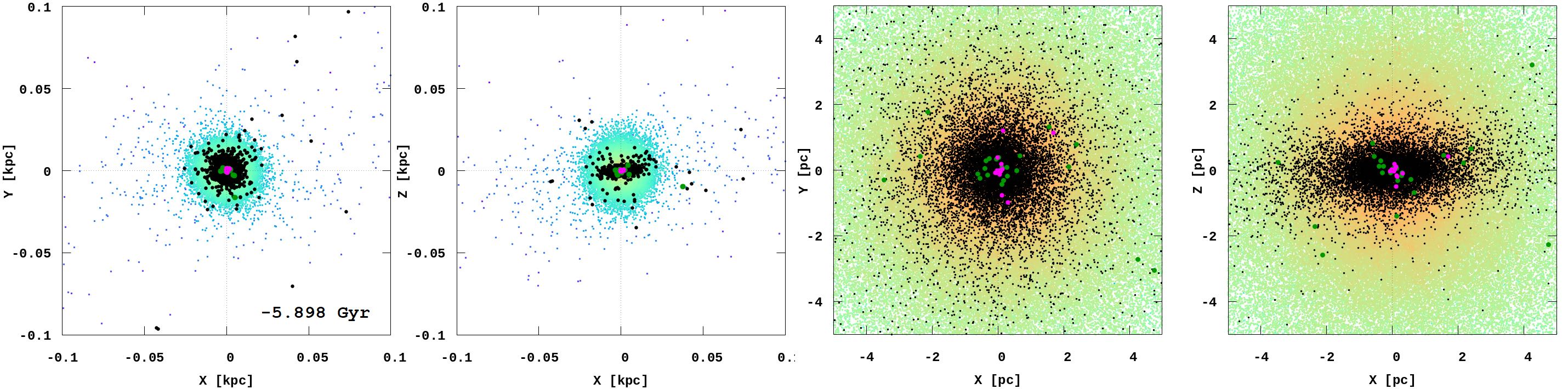}
\includegraphics[width=0.93\linewidth]{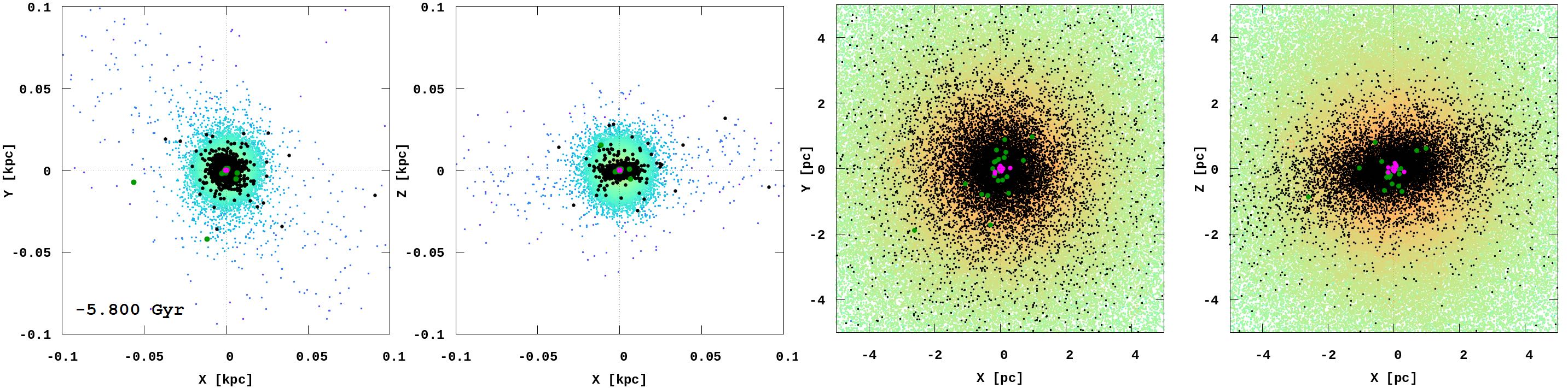}
\includegraphics[width=0.93\linewidth]{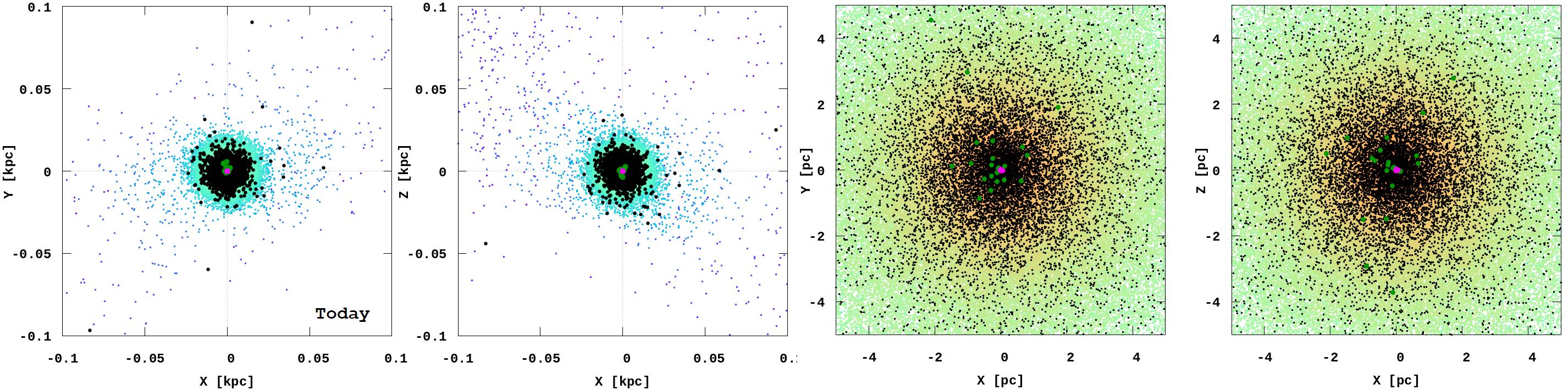}
\caption{Same as Fig. \ref{fig:cr-loc-tng}, but in the {\tt FIX} potential for a GC on a circular orbit.}
\label{fig:cr-loc-fixg}
\end{figure*}

\begin{figure*}[ht]
\centering
\includegraphics[width=0.93\linewidth]{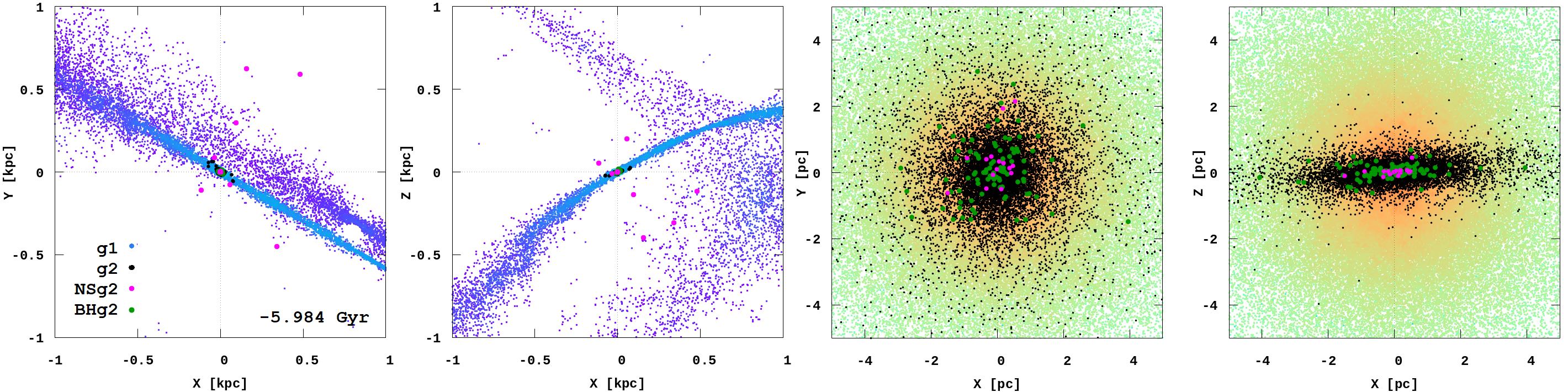}
\includegraphics[width=0.93\linewidth]{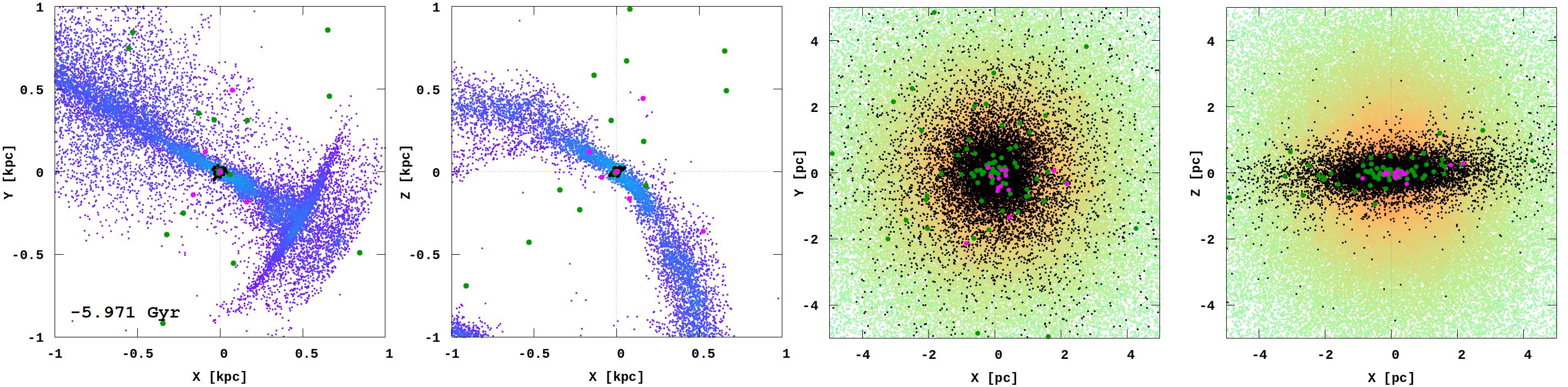}
\includegraphics[width=0.93\linewidth]{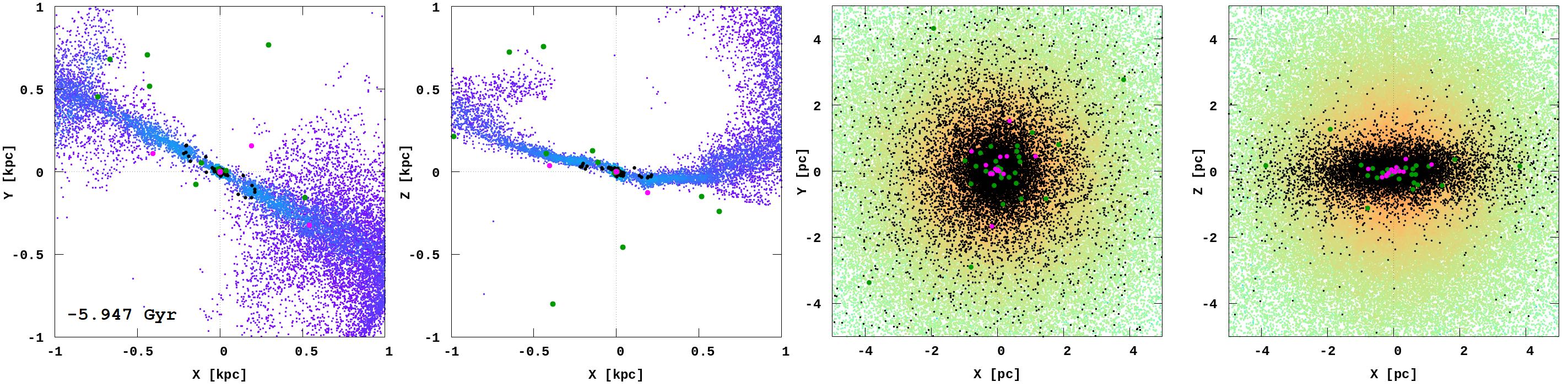}
\includegraphics[width=0.93\linewidth]{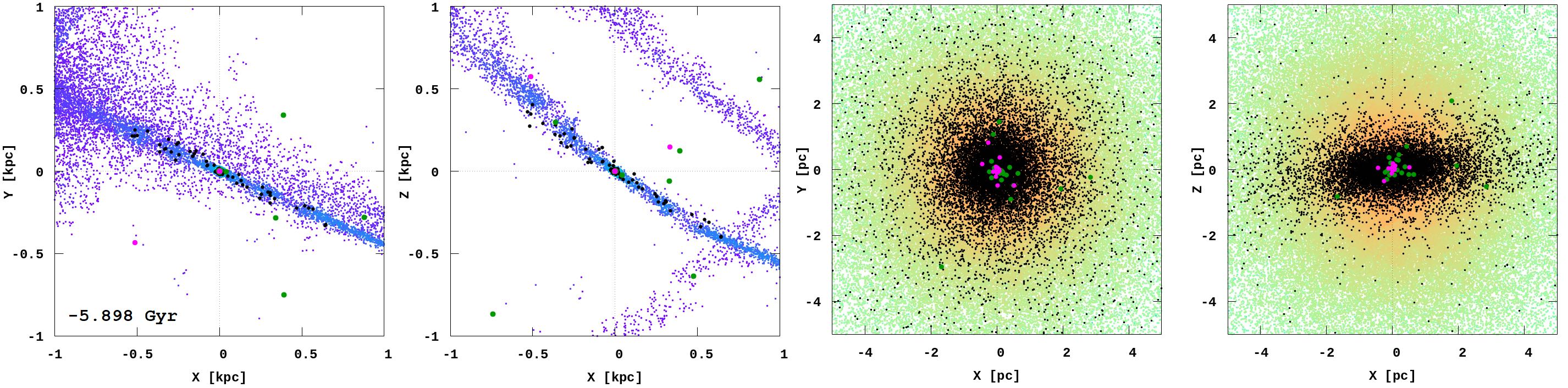}
\includegraphics[width=0.93\linewidth]{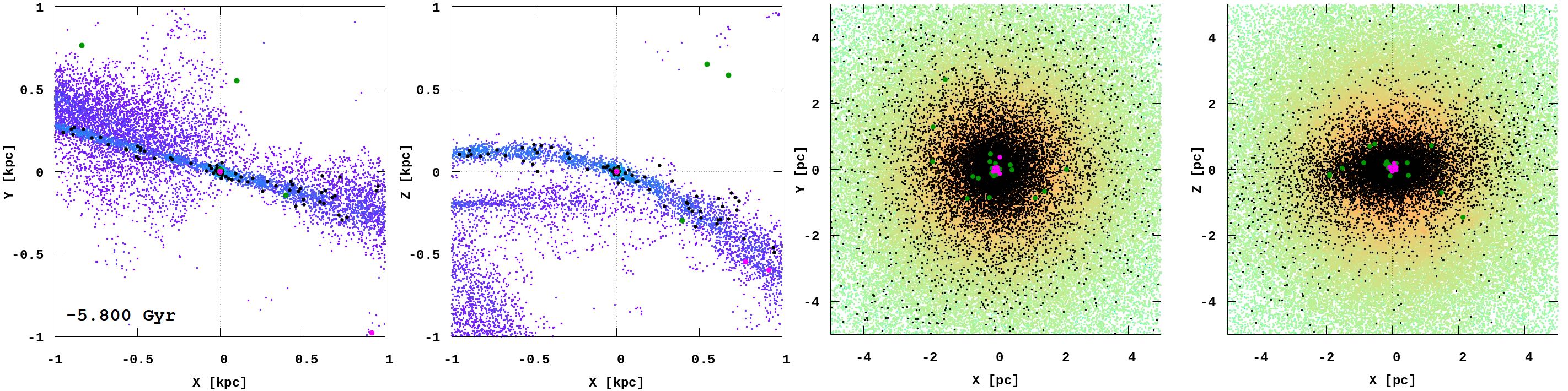}
\includegraphics[width=0.93\linewidth]{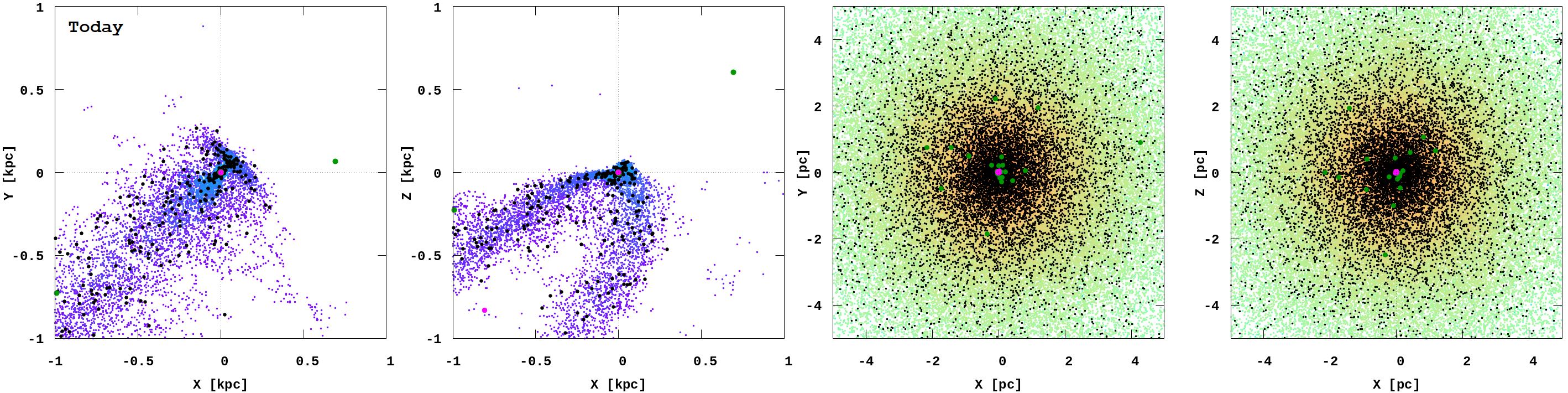}
\caption{Same as Fig. \ref{fig:cr-loc-tng}, but in the {\tt TNG} potential for a GC on a tube orbit.}
\label{fig:tb-loc-tng}
\end{figure*}

\begin{figure*}[ht]
\centering
\includegraphics[width=0.93\linewidth]{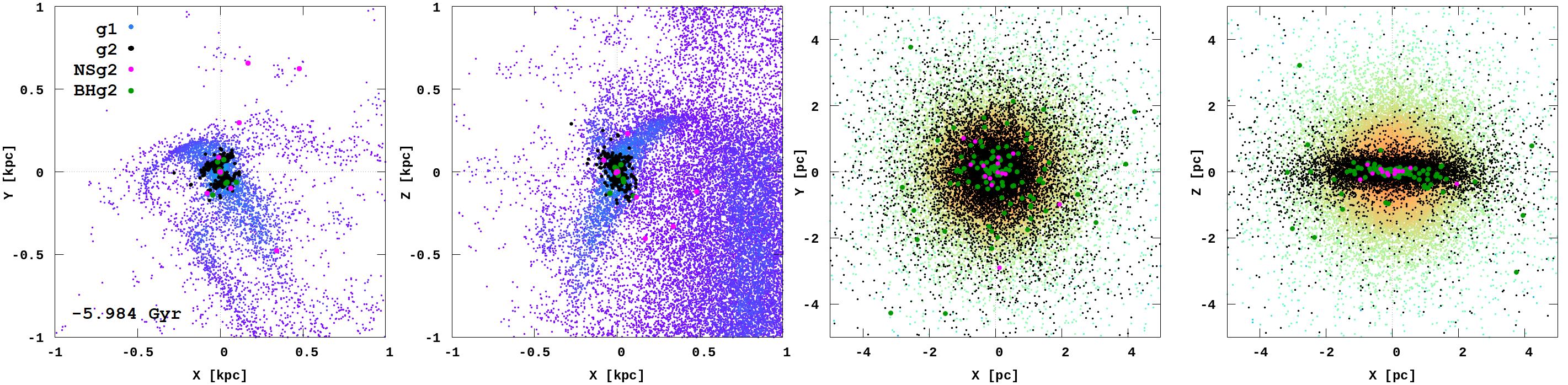}
\includegraphics[width=0.93\linewidth]{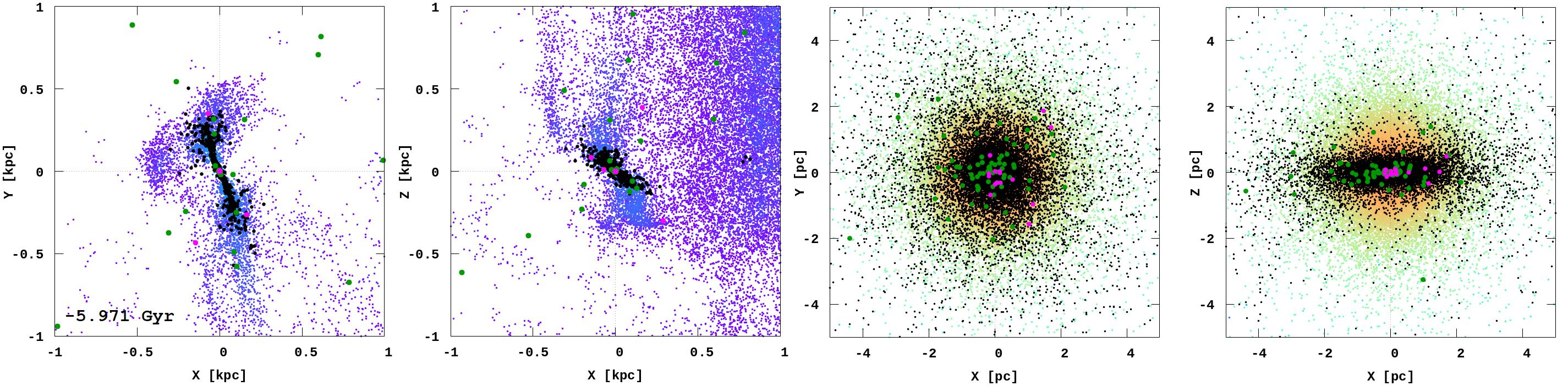}
\includegraphics[width=0.93\linewidth]{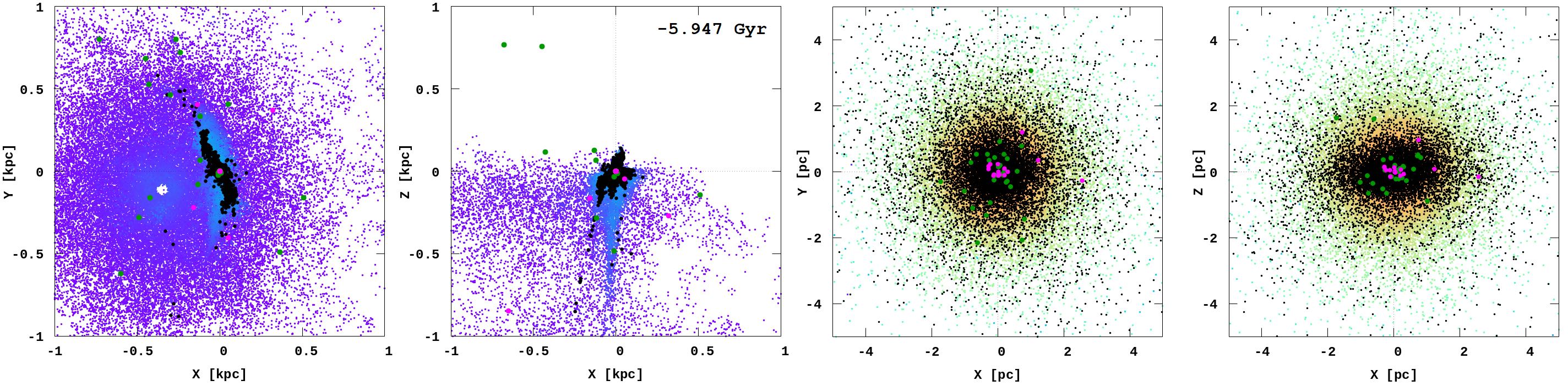}
\includegraphics[width=0.93\linewidth]{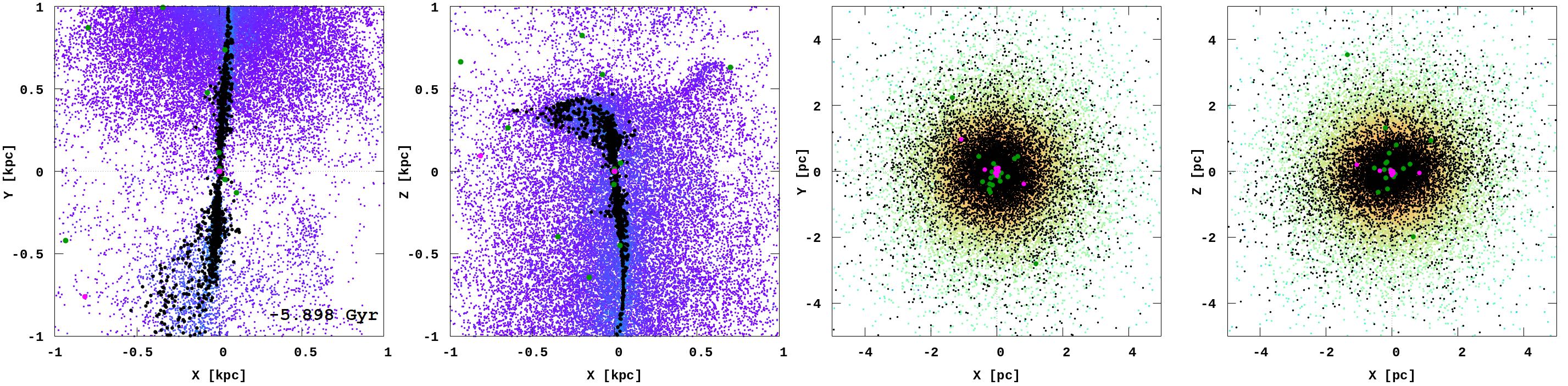}
\includegraphics[width=0.93\linewidth]{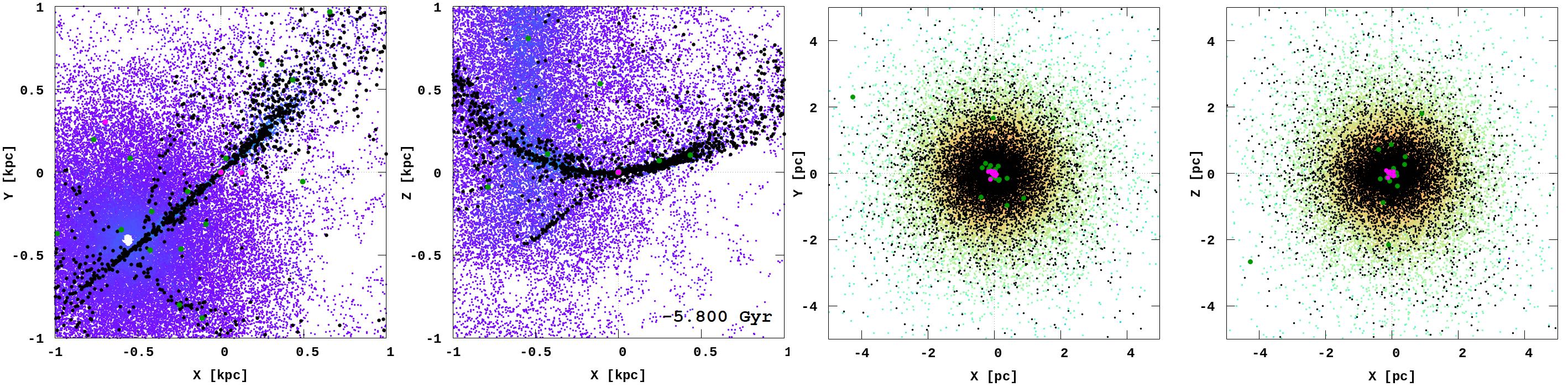}
\includegraphics[width=0.93\linewidth]{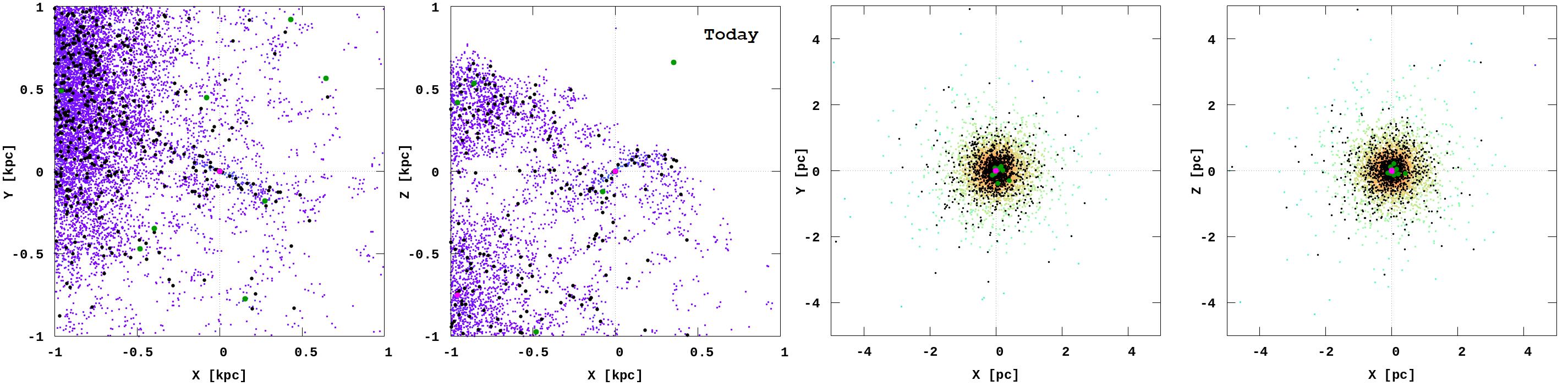}
\caption{Same as Fig. \ref{fig:cr-loc-tng}, but in the {\tt FIX} potential for a GC on a tube orbit.}
\label{fig:tb-loc-fix}
\end{figure*}

\begin{figure*}[ht]
\centering
\includegraphics[width=0.93\linewidth]{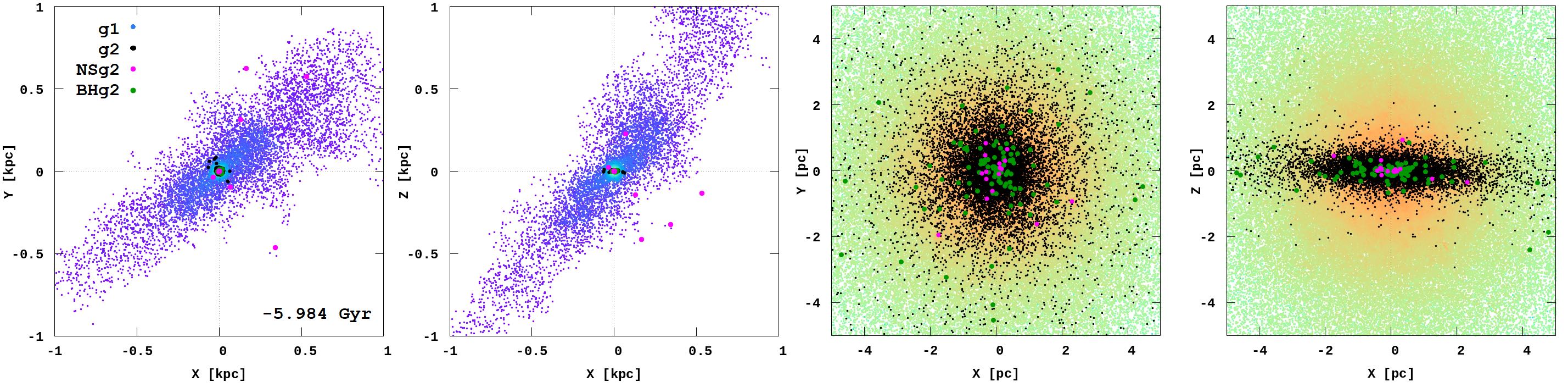}
\includegraphics[width=0.93\linewidth]{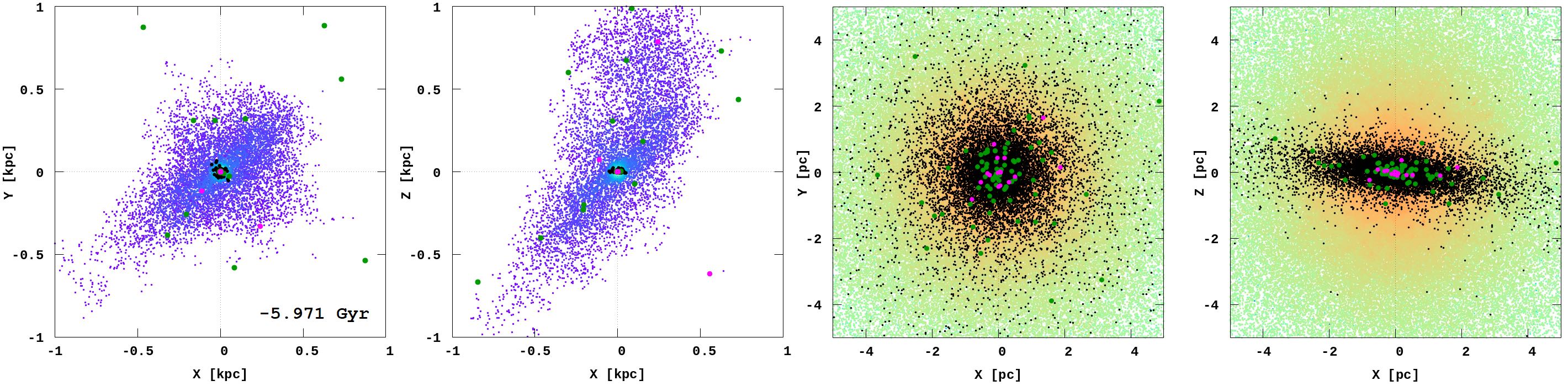}
\includegraphics[width=0.93\linewidth]{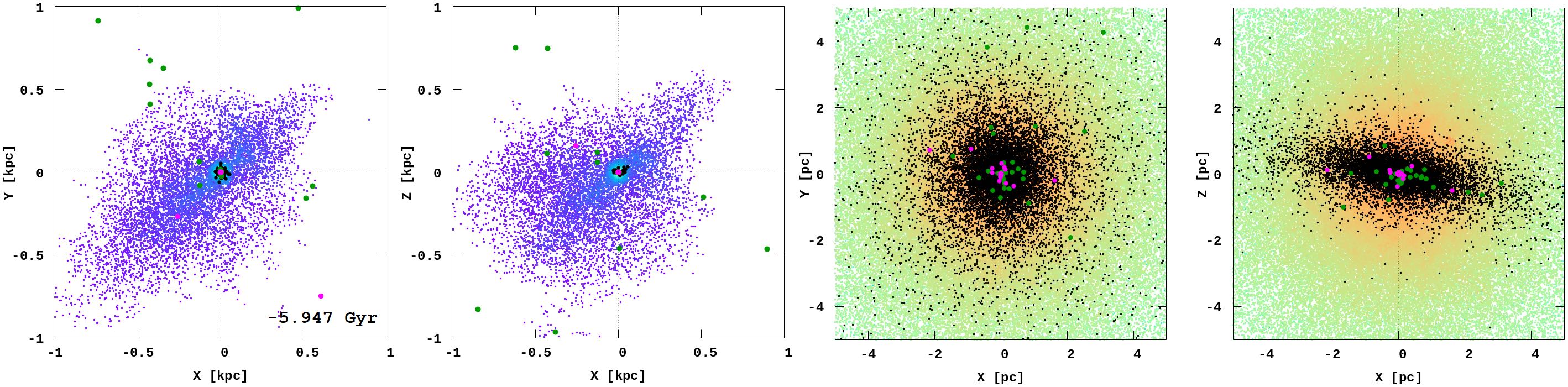}
\includegraphics[width=0.93\linewidth]{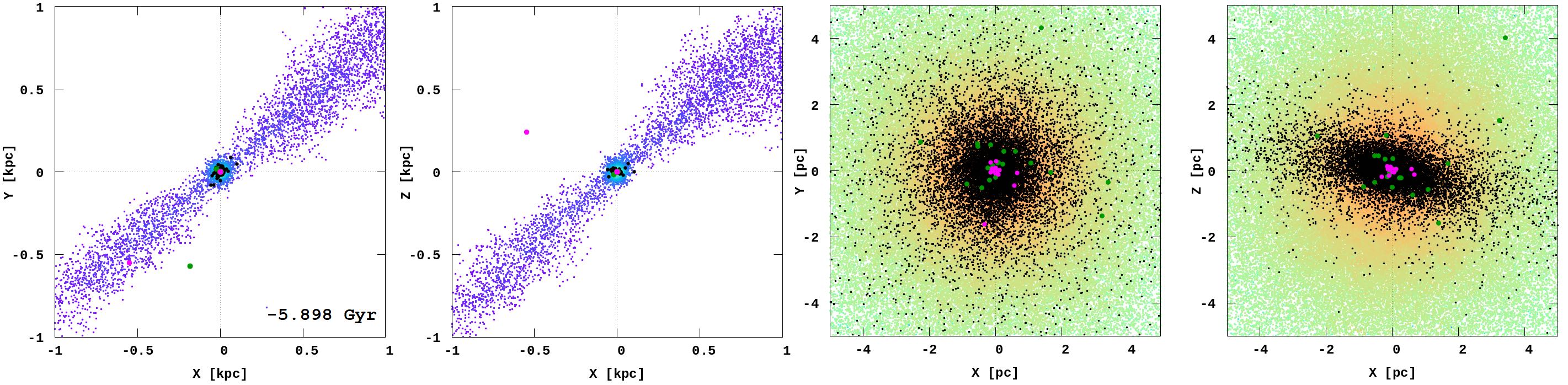}
\includegraphics[width=0.93\linewidth]{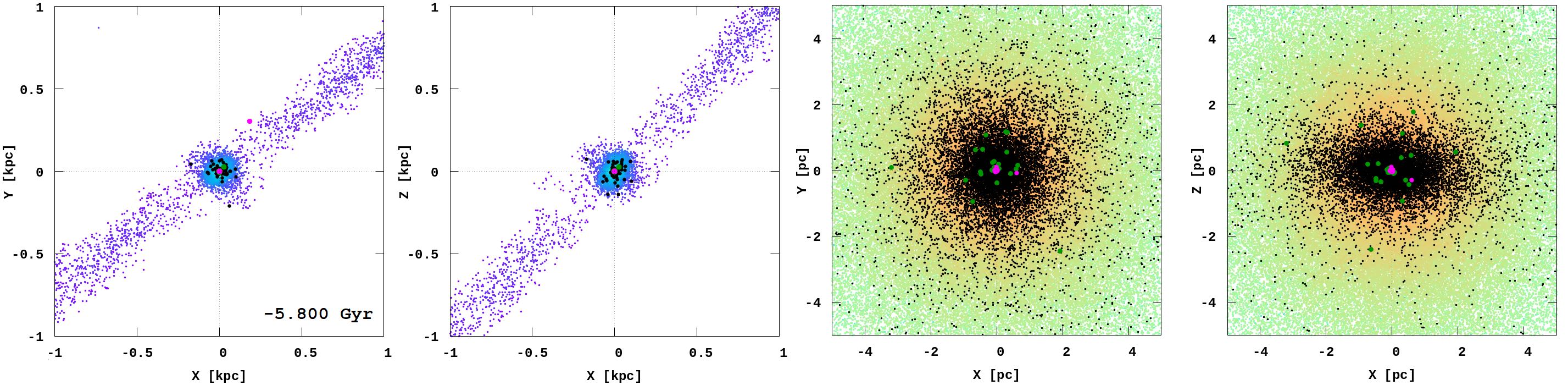}
\includegraphics[width=0.93\linewidth]{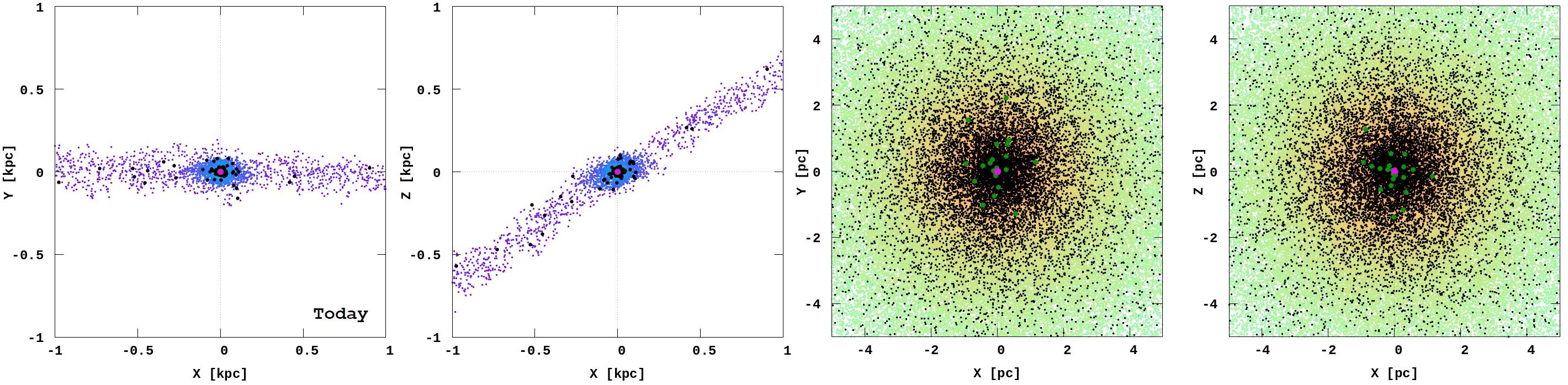}
\caption{Same as Fig. \ref{fig:cr-loc-tng}, but in {\tt TNG} potential for a GC on a long radial orbit.}
\label{fig:lr-loc-tng}
\end{figure*}

\begin{figure*}[ht]
\centering
\includegraphics[width=0.93\linewidth]{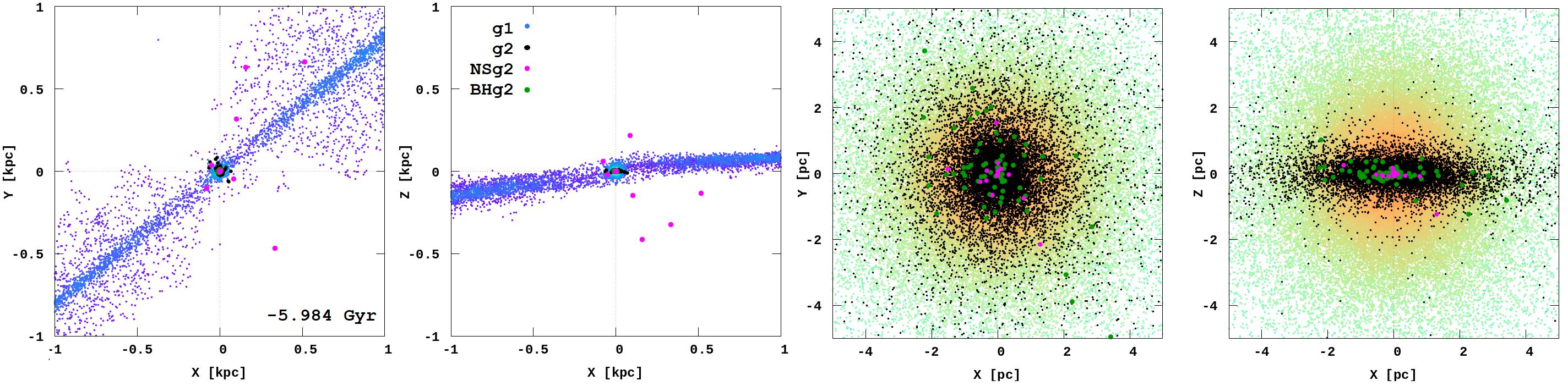}
\includegraphics[width=0.93\linewidth]{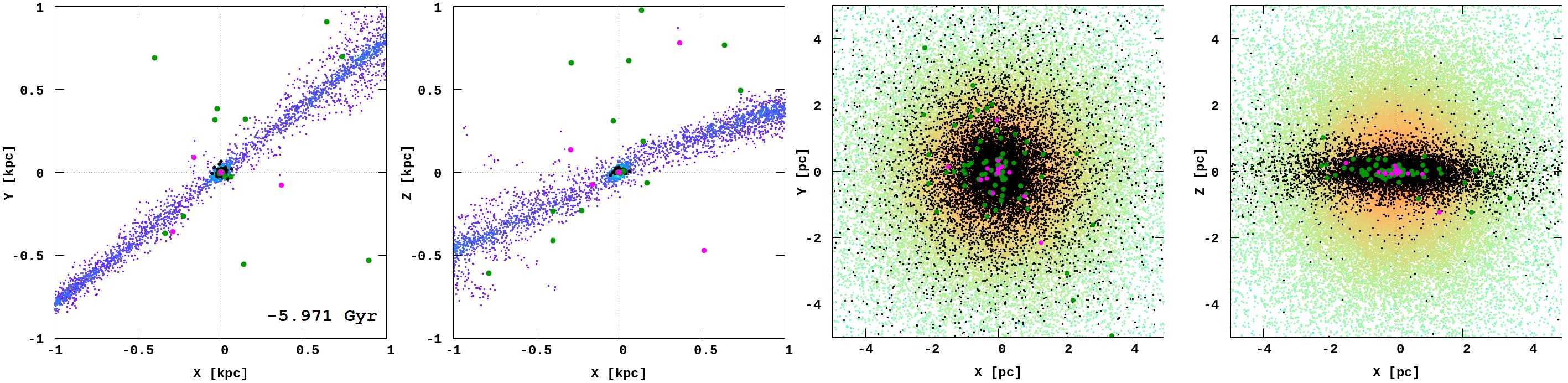}
\includegraphics[width=0.93\linewidth]{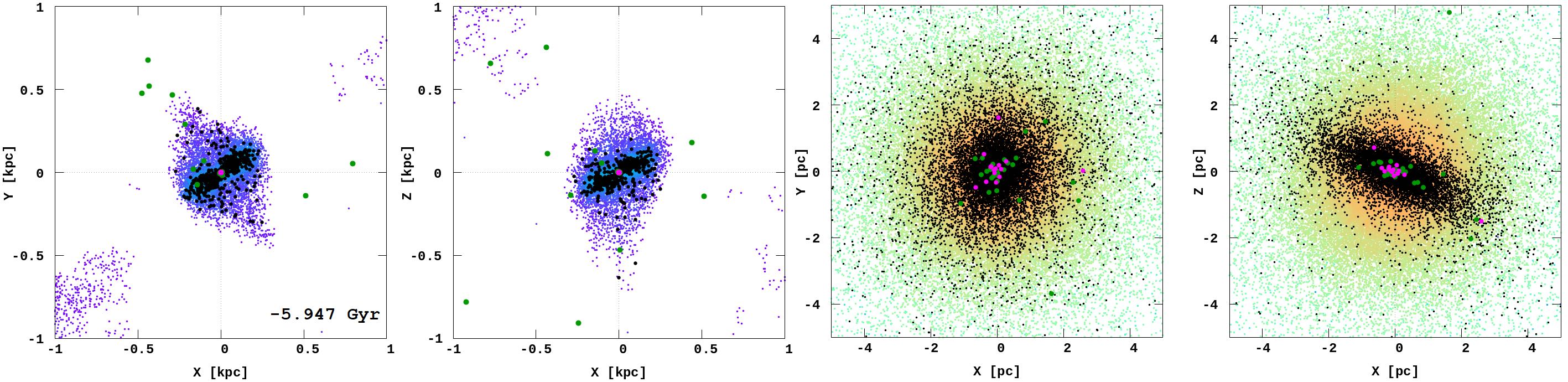}
\includegraphics[width=0.93\linewidth]{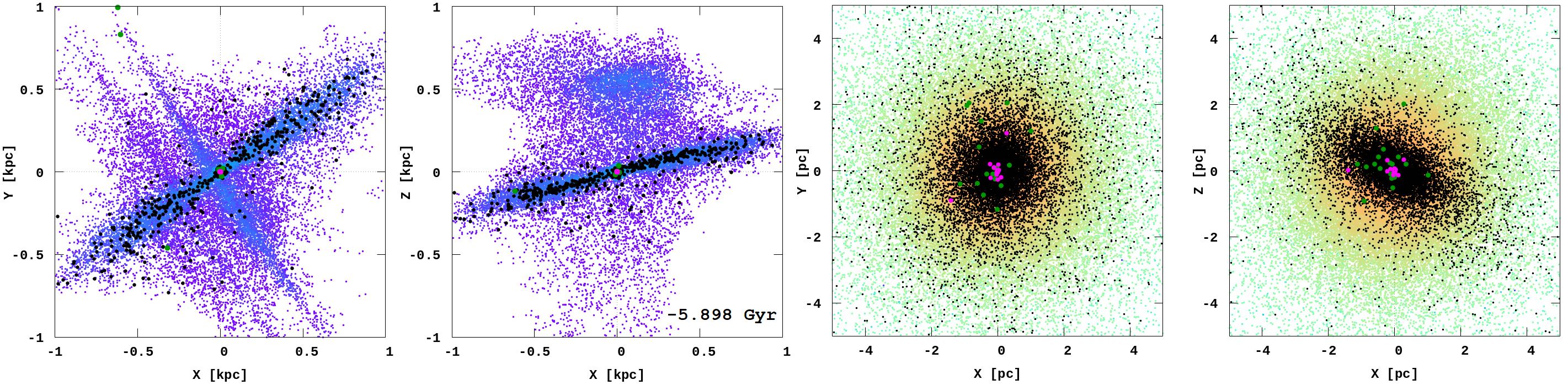}
\includegraphics[width=0.93\linewidth]{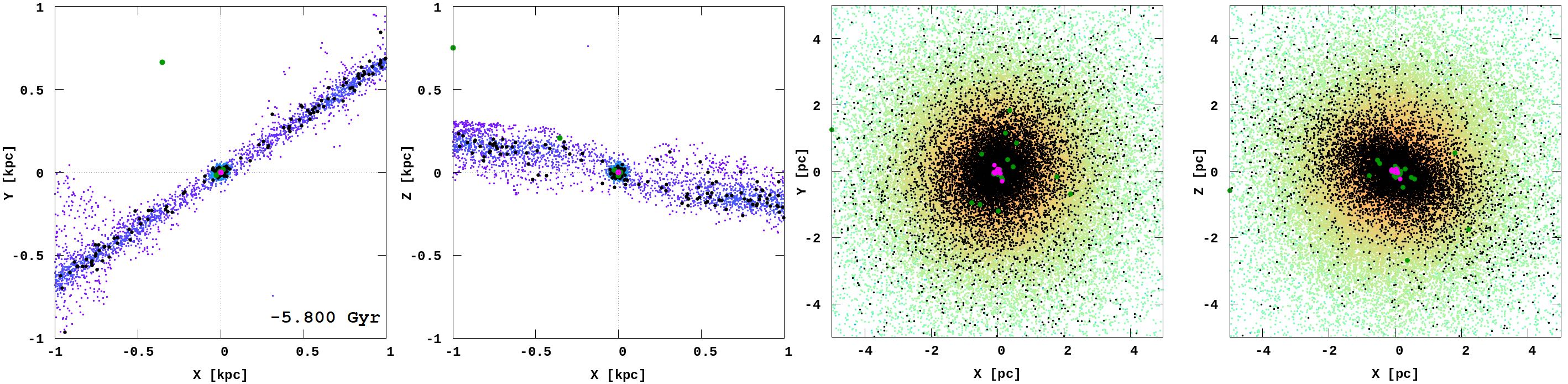}
\includegraphics[width=0.93\linewidth]{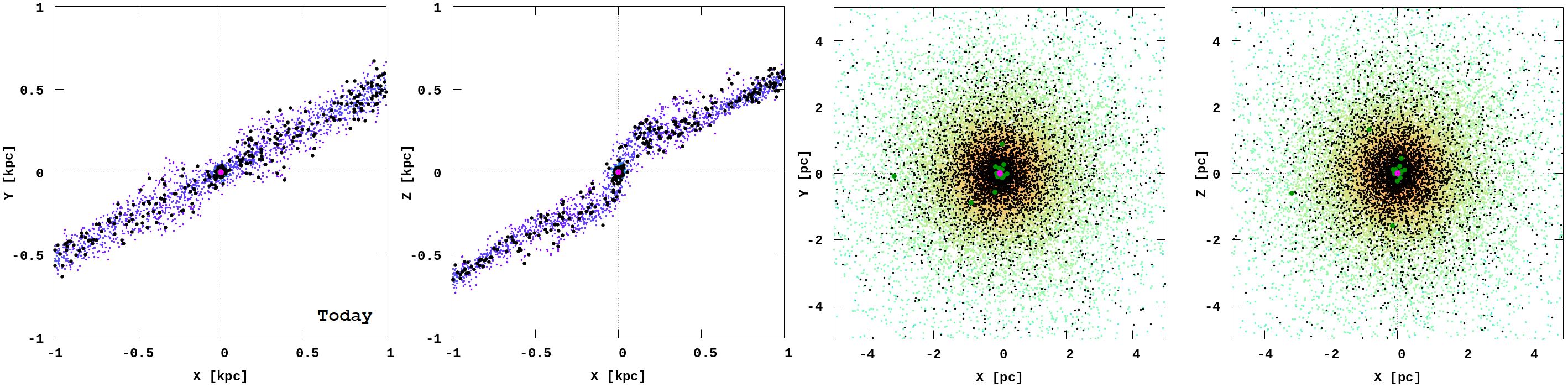}
\caption{Same as Fig. \ref{fig:cr-loc-tng}, but in the {\tt FIX} potential for a GC on a long radial orbit.}
\label{fig:lr-loc-fix}
\end{figure*}

In Figures \ref{fig:cr-loc-tng}, \ref{fig:cr-loc-fixg}, \ref{fig:tb-loc-tng}, \ref{fig:tb-loc-fix}, \ref{fig:lr-loc-tng} and \ref{fig:lr-loc-fix} we present the density distributions of {\tt g1} and {\tt g2} positions in the coordinate system connected with the GC's density centre. In \textit{two left panels} we demonstrate the central $\sim$kpc box. In \textit{two right panels} we demonstrate the very inner 5$\times$5 pc area of the GC's. The right panels clearly demonstrate the timescale (in order of $\sim$100 Myr) how quickly the initially disky distribution of the {\tt g2} getting more and more spherical. 

\clearpage


\end{appendix}

\end{document}